\documentclass[preprint,12pt]{elsarticle}

\usepackage{amsmath}
\usepackage{amssymb}
\usepackage{amsfonts}
\usepackage{amsthm}

\usepackage{tikz}
\usetikzlibrary{cd}
\usetikzlibrary{perspective,3d}
\tikzcdset{scale cd/.style={every label/.append style={scale=#1},
    cells={nodes={scale=#1}}}}

\usetikzlibrary{hobby}
\usetikzlibrary{arrows,decorations.markings,arrows.meta,patterns}
\tikzset
{
   ->-/.style={decoration={markings,mark=at position 0.5 with {\arrow{Straight Barb}}},
               postaction={decorate}}
}

\usepackage{caption}
\usepackage{subcaption}
\usepackage{physics}
\usepackage[margin=2cm]{geometry}

\newtheorem{lem}{Lemma}
\newtheorem{defn}{Definition}

\newcommand{\argmax}{\text{argmax}}

\usepackage{graphicx}
\usepackage[export]{adjustbox}

\usepackage{xcolor}
\newcommand{\NB}[1]{\textcolor{red}{NB: #1}}

\newcommand{\Z}{\mathbb{Z}}
\newcommand{\N}{\mathbb{N}}

\newcommand{\F}{\mathbb{F}}

\journal{Physica D}

\begin{document}

\begin{frontmatter}



\title{Automatic classification of magnetic field line topology by persistent homology}


\author[MSI]{N Bohlsen}
\author[RSPHYS]{V Robins}
\author[MSI]{M Hole}

\affiliation[MSI]{organization={Australian National University, Mathematical Sciences Institute},
            city={Canberra}, 
            postcode = {2601},
            state={ACT},
            country={Australia}}

\affiliation[RSPHYS]{organization={Australian National University, Research School of Physics},
            city={Canberra}, 
            postcode = {2601},
            state={ACT},
            country={Australia}}

\begin{abstract}
A method for the automatic classification of the orbits of magnetic field lines into topologically distinct classes using the Vietoris-Rips persistent homology is presented. The input to the method is the Poincare map orbits of field lines and the output is a separation into three classes: islands, chaotic layers, and invariant tori. The classification is tested numerically for the case of a toy model of a perturbed tokamak represented initially in its geometric coordinates. The persistent $H_1$ data is demonstrated to be sufficient to distinguish magnetic islands from the other orbits. When combined with persistent $H_0$ information, describing the average spacing between points on the Poincare section, the larger chaotic orbits can then be separated from very thin chaotic layers and invariant tori. It is then shown that if straight field line coordinates exist for a nearby integrable field configuration, the performance of the classification can be improved by transforming into this natural coordinate system. The focus is the application to toroidal magnetic confinement but the method is sufficiently general to apply to generic $1\frac{1}{2}$d Hamiltonian systems.
\end{abstract}




\end{frontmatter}





\section{Introduction}

\label{Chapter1}

Developments in controlled thermonuclear fusion have spurred substantial study into the dynamics of chaotic Hamiltonian chaos over the past several decades \cite{mackay1983renormalization,greene1979method,mackay1984transport,isodrasticFields}. This is partially because the charged particle motion in these machines is generically a chaotic Hamiltonian flow, but also because the magnetic field itself is also a chaotic Hamiltonian system. For the case of ideal tokamak devices, the field is an integrable system, however real tokamak devices are not perfectly axisymmetric, both because they are constructed with a finite number of toroidal field coils and due to the presence of error fields, and therefore confine their plasma with partially chaotic fields. Understanding the structure of these fields is key to predicting the mechanisms  and rate of particle escape from the machine \cite{de2001structure,mathias2017chaotic}. For example, magnetic field lines confined to toroidal surfaces pose a barrier to transport for particles orbiting in a tokamak \cite{joffrin2003internal}. Therefore by detecting these field lines it is possible to partition the domain of the tokamak into regions between which particle transport, and therefore energy transport, is suppressed and hence characterise this transport.

Separately, there also exists a renewed interest in the stellarator design concept for a toroidal magnetic plasma confinement device \cite{dinklage2018magnetic}. Unlike the tokamak design, stellarators are not symmetric in the toroidal direction. As a result the magnetic field lines in these machines are generically non-integrable and therefore contain nested flux surfaces, magnetic islands, and stochastic (chaotic) layers. The location and size of these islands and stochastic layers affects the transport of charged particles throughout the machine and hence its confinement characteristics. Studying one of these chaotic magnetic fields requires rendering and then analysing, largely by-hand, a Poincare section of many different field lines to see where islands exist in the field and what families of other islands accompany them. 

This poses a problem for the computer aided design of stellarators and real tokamaks. Using software such as the Stepped Pressure Equilibrium Code (SPEC), we can generate many more magnetic field configurations, than we can reasonably analyse by hand \cite{hudson2012computation}. To resolve this, we seek to develop software which can automatically characterise the magnetic field from its Poincare map. This paper presents an initial investigation into one method for achieving this using techniques from \textit{Topological Data Analysis} (TDA), chiefly \textit{Persistent Homology}.

Magnetic field line flow in a stellarator or tokamak is a $1\frac{1}{2}$d Hamiltonian system where the toroidal angle serves the role of time \cite{Morrison2000}. The field lines themselves can therefore be classified into the same standard categories as orbits of any $1\frac{1}{2}$d Hamiltonian systems. That is, as either invariant tori, island chains, or stochastic regions. These different classes of orbit are topologically distinct on a Poincare section. Invariant tori are homeomorphic to circles and island chains to finite unions of circles. Stochastic layers are positive measure fractal regions so are topologically distinct from tori and island chains, as measured by their  Hausdorff (fractal) dimensions and by the number of holes in the region \cite{umberger2020fat}. 

\begin{figure}
    \centering
        \includegraphics[width = 0.6\textwidth]{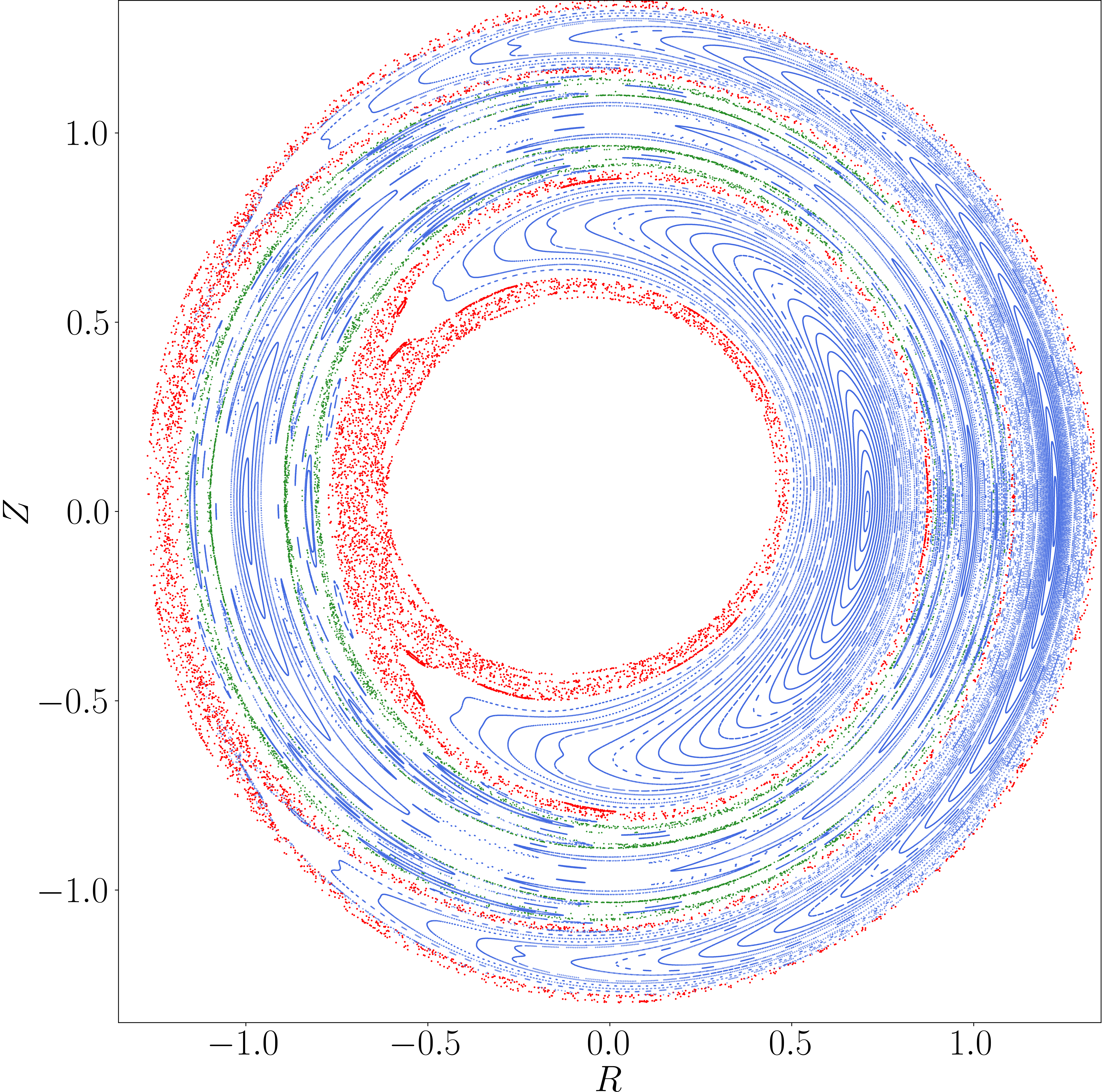}
    \caption{Orbit type classification determined by persistent homology. Orbits are generated by iterating the standard map for $100$ initial conditions with $k=0.8$. These orbits are then projected from the cylindrical $S^1\times[0,1]$ domain of the standard map to an annular disk. Islands are shown in blue, large stochastic orbits in red, and small (i.e. thin) stochastic layers and invariant tori are green. Detection of the orbit class was performed using our $TDA$ island detection scheme, and our $\langle d(PH_0)\rangle$ method for stochastic layer detection discussed in Sections \ref{EnclosureDefining} and \ref{StochLayersWithIslands} respectively.}
    \label{fig:standard_map_islands_coloured3}
\end{figure}

In this paper we describe a scheme for the  classification of magnetic field line orbits into one of the above three types. We achieve this by computing standard topological descriptors from an orbit and  use simple thresholds on these quantities to allocate the orbit type. 
We illustrate this automated classification scheme in Figure~\ref{fig:standard_map_islands_coloured3}, which presents the results of applying the method to a model Poincare map for which the radius and angle of the points are evolved by Chirikov's standard map \cite{CHIRIKOV1979263}
\begin{align}
    &p_{n+1} = p_n-\frac{k}{2\pi}\sin(2\pi q_n)\,,\nonumber\\
    &q_{n+1} = q_n+p_{n+1}\, \hspace{13mm} (\text{mod } 1)\label{stdmap}\,.
\end{align}



We observe that the islands and chaotic orbits have been separated by the topological descriptors and are therefore coloured differently. Note that the topological descriptors required for the above calculation are readily computable using open-source software on consumer hardware in a matter of minutes.





\subsection{Field line orbits}

Magnetic field lines are, mathematically speaking, just the integral curves of the magnetic field $\mathbf{B}$. That is, they are the solution curves of the ODE
\begin{equation}\label{FieldLineEquation}
    \dv{\mathbf{x}}{t}  = \mathbf{B}(\mathbf{x}(t))\,.
\end{equation}
Assuming a toroidal magnetic field configuration it is possible to construct a \textit{magnetic field line Hamiltonian} $\chi$ for which the solutions to \eqref{FieldLineEquation} satisfy
\begin{align}\label{Field Line Hamiltonian}
    &\dv{\psi}{\zeta} = -\pdv{\chi}{\theta}\,,\nonumber\\
    &\dv{\theta}{\zeta} = \pdv{\chi}{\psi}\,,
\end{align}
where $(\psi,\theta,\zeta)$ are  toroidal coordinates of a point $\mathbf{x}$ \cite{Morrison2000}. Note that this is not actually possible in general and for detailed discussion of conditions for the existence of the field line Hamiltonian, $\chi$, we refer to \cite{DUIGNAN2023133749}. However, we are only interested in configurations such that $\chi$ exists here.

Observe that \eqref{Field Line Hamiltonian} has the form of the Hamilton equations in $1\frac{1}{2}$d with the toroidal angle $\zeta$ acting as time. This is why magnetic field lines can be studied like a Hamiltonian system and why they exhibit the same island structures as non-integrable Hamiltonian systems. So, while we specifically focus on the case of magnetic field lines in this paper, the methods presented are sufficiently general as to apply to any Hamiltonian dynamical system.

\subsection{Related Work}

Research in computational topology and geometry has regularly been inspired by its applications to the study of dynamical systems and chaos. For example, in one of the earliest papers in the field Muldoon \textit{et al} investigated the extraction of the topological invariants of orbits from experimental time series \cite{muldoon1993topology}. This helped to introduce simplicial homology as a tool for the study of dynamics. In recent years there has been substantial development in this program. Tempelton and Khasawneh demonstrated that the shape of projections of chaotic trajectories to subspaces of the phase space appear geometrically distinct from those of ordered trajectories \cite{tempelman2020look}. They used this observation to construct a TDA based tool for the detection of chaotic trajectories by analysing the sub-level set persistent homology of kernel density estimates of their orbits. They demonstrate the application of their detection scheme for several dynamical systems but none of Hamiltonian type. Tymochko \textit{et al} demonstrated that Hopf bifurcations in one parameter families of dynamical systems can be detected with zig-zag persistent homology \cite{tymochko2020using}. Again, their analysis is concerned with dissipative systems not Hamiltonian ones. 

The most substantial contribution to the use of computational topology to the automated analysis of Hamiltonian systems is the dissertation of Yip \cite{YipDissertation}. Yip used Minimal Spanning Trees (MST)s to construct a similar topological classification of Hamiltonian orbits to the one we propose. However, MSTs only encode information about the connectivity of a dataset and are not sensitive to other topological features, such as homology. As a result they rely on standard chaos detection schemes to separate invariant tori, and chaotic layers, since both of these are connected sets and therefore not substantially distinguished by their MSTs. We will see later that our TDA based methods improve on this by encoding information about the number of internal holes in chaotic orbits.

TDA, or more specifically persistent homology, has been applied in contexts more closely related to the magnetic field trajectories which concern us here. Kramar \textit{et al} have demonstrated that the structures of convecting fluids can be computed, and the qualitative similarities of otherwise quantitatively distinct fluid flows can be observed in the topological data \cite{KRAMAR201682}. Similarly, it has been suggested that the lagrangian orbits of fluid packets can be classified by the topological structure of their intersections with a Poincare section and preliminary evidence has been created by Nunez \textit{et al} that this classification is computable with TDA \cite{nunez2022topological}. Lagrangian orbits of fluid flows are very similar to magnetic field lines, in the sense that they are also of Hamiltonian form when the fluid is steady state and incompressible. As such our work on field line classification is heavily inspired by the work of Nunez \textit{et al}. In a specifically plasma physics context, Banesh \textit{et al} demonstrated that TDA can be used to detect reconnection events in simulations of astrophysical magnetic fields \cite{banesh2020topological}. Their method does allow for the detection of topological change in the evolution of the field but it is restricted to the case of a planar magnetic field and hence is $2$d. It is also formally dissipative and therefore also not of Hamiltonian form. 

\subsection{Structure}

This paper is organised as follows: Section \ref{Chapter2} presents an informal introduction to the TDA methods which we apply. 
Section \ref{Chapter3} then presents our main results on automated classification of magnetic field line orbits and discusses limitations of the methodology. 
Section \ref{Chapter4} shows how these limitations can be relaxed in the specific case when the chaotic field is a perturbation of a known integrable field by transforming into the `natural' action-angle coordinate system. This also suggests a numerical method to compute a global measure of non-integrability of such a system. Section \ref{Chapter5} contains concluding remarks.



\section{Mathematical Background} 

\label{Chapter2}

\begin{figure}
    \centering
    \begin{subfigure}[b]{0.3\textwidth}
        \includegraphics[width = \textwidth]{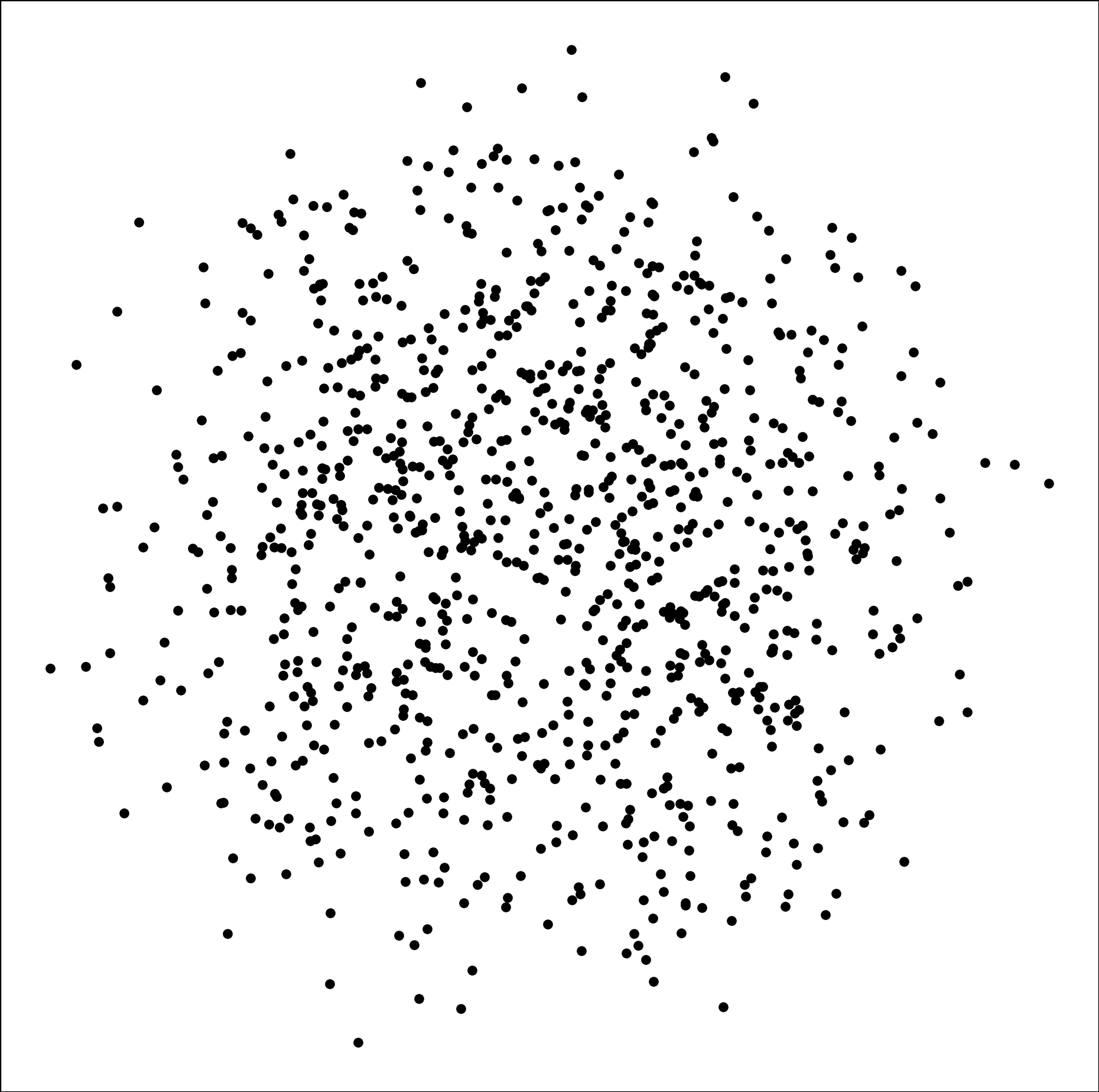}
    \end{subfigure}
    \begin{subfigure}[b]{0.3\textwidth}
        \includegraphics[width = \textwidth]{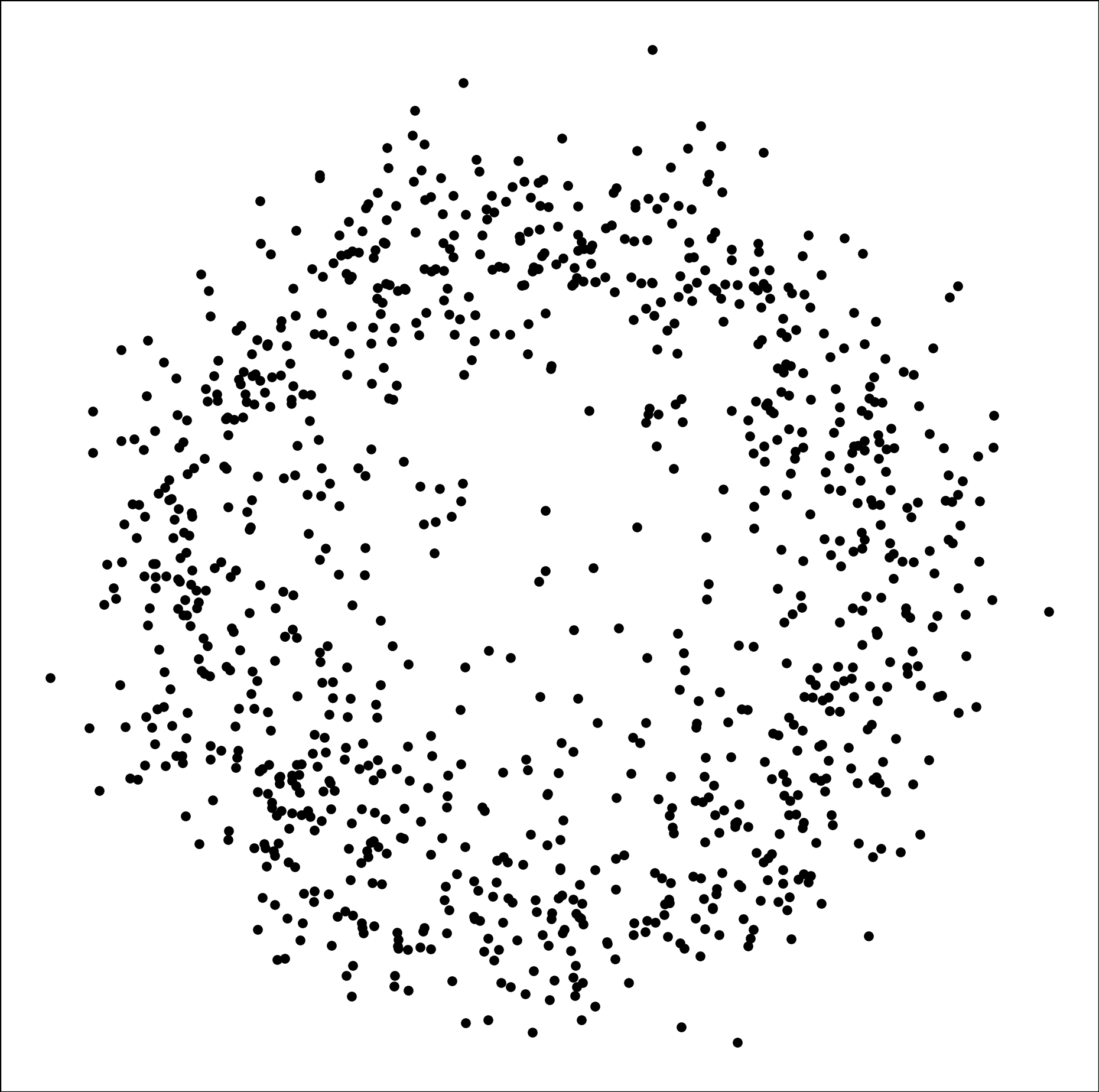}
    \end{subfigure}
    \caption{Two example point clouds}
    \label{fig:PointClouds}
\end{figure}

At its fundamental level, Topological Data Analysis (TDA) is concerned with quantifying the \textit{shape} of datasets. To see the meaning of this consider Figure \ref{fig:PointClouds} which presents images of two hypothetical datasets in the plane. While both datasets are roughly circular it is obvious to a person that they are qualitatively different because there is a \textit{hole} in the cloud on the right which is not present in the left one. TDA makes this notion mathematically rigorous so that a computer can detect this hole. Holes in spaces are topological features not geometric ones and so we would initially expect that they can tell us very little about the shape, that is geometry, of a dataset. However, by studying how the number of holes in our data changes with the length scale on which we examine the data we can determine geometric features as well as topological ones. 

In this section we provide an informal, and largely geometric, introduction to these ideas. Further, since their application to Hamiltonian mechanics is relatively recent our presentation is pedagogical in style and designed to ensure this paper is self-contained. For a more general and detailed introduction to TDA, we refer to any of the recent review articles on the subject,~\cite{chazal2021introduction,boissonnat2022topological,murugan2019introduction,10.1115/1.4055184}.


\subsection{Persistent homology}

The datasets we concern ourselves with here are all just a finite set of points $X$ which belong to a metric space $M$. We refer to such a set as a \textit{point cloud}. $X$ inherits the topology of the metric space but is also topologically trivial as it is just a set of disconnected points. So, if we want to study the topology of data we need to associate $X$ with a topological space which is non-trivial. The standard tool to accomplish this is the \textit{simplicial complex}.

\begin{defn}
Given a set, $X$, a simplicial complex, $K$, is a collection of finite subsets of $X$ that is closed under the subset relation $\leq$.  
 \begin{equation}
    \text{for all } \sigma \in K, \sigma'\leq \sigma \implies \sigma' \in K\,.
 \end{equation}
An element of $K$, $\sigma = \{p^0, p^1, \ldots, p^n\}$, is called an $n$-simplex and its subsets are called faces. The elements of $X$ are $0$-simplices called vertices. 
\end{defn}

A simplicial complex can be thought of as a higher-dimensional generalisation of a graph. 
The topological dimension of $K$ is given by the largest $r$ of its simplices. We call such a complex a simplicial $r$-complex. A graph is an example of a simplicial $1$-complex and a triangulation of a $2$-manifold is a simplicial $2$-complex. 
A graph is often visualised as a geometric object with vertices mapped to points and edges to line-segments. 
Similarly, a geometric realisation of a simplicial complex maps the elements of $X$ to points in a metric space, and an $r$-simplex to the convex hull of its vertices. 
Thus, $0$-simplices are points, $1$-simplices are line segments connecting two points, $2$-simplices are triangles, $3$-simplices are tetrahedra and so on.  
Precise definitions of simplicial complexes and their geometric realisations can be found in any text on computational or algebraic topology, see \cite{Hatcher,EDELSBRUNNER,GHRIST,Nakahara}.

If we can associate a simplicial complex to a dataset, then studying the topology of this complex provides information about the geometry of the dataset. This is a fundamental idea in TDA and it is how we will describe the shape of data. 

\subsubsection{Vietoris-Rips complexes}

If we have a point cloud $X$ in a metric space $M$ then we can construct a simplicial complex by connecting points which are near each other with respect to the distance on $M$. We can then add triangles to our complex as soon as all of their faces are also in the complex. This procedure forms what is called the Vietoris-Rips (VR) complex of the point cloud at a given diameter $\epsilon>0$ \cite{EDELSBRUNNER}. It is formally defined as follows . 

\begin{defn}
    Let $S$ refer to the set of all simplices whose vertex set is a subset of $X$. Then, for $\epsilon>0$ define the VR complex of $X$ as 
    \begin{equation}
        VR_\epsilon(X) = \{\sigma \in S| \, \text{diam}(\sigma) \leq \epsilon\}\,.
    \end{equation}
\end{defn}

\begin{figure}
    \centering
    \includegraphics[width = \textwidth]{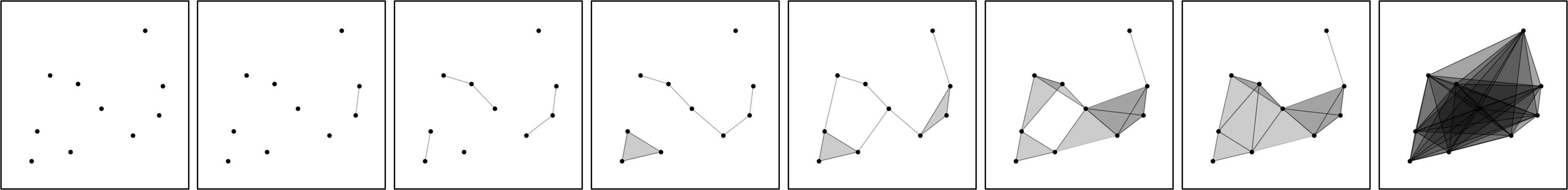}
    \caption{Image of Vietoris-Rips complexes of the same point cloud across a range of diameters showing only vertices, edges, and triangles. Note that the diameter increases from left to right.}
    \label{fig:RipsFiltration}
\end{figure}

Figure \ref{fig:RipsFiltration} presents images of the VR complex of a point cloud for a series of diameters $\epsilon$. For small diameters $VR_\epsilon(X) = X$ that is, only the actual points are included in the complex. However, for larger $\epsilon$ we see that edges and eventually triangles are added too. Also we see from this example that for the larger diameters the VR complex cannot be embedded in $\mathbb{R}^2$ without self-intersection. This is why we adopt the abstract, combinatorial definition of simplicial complex rather than a purely geometric one that would require the triangulation to be embedded.  

\subsubsection{Simplicial homology}

Computational topology as a subject is generally concerned with describing the topological properties of simplicial complexes. We focus on a famous algebraic topological invariant of spaces called homology. The fundamental idea of this construction is to associate to a topological space $S$ a family of abelian groups, called the homology groups $H_n(S)$, that encode the number of $n$-dimensional holes in the space. Here we briefly review the construction of simplicial homology.

In the following, assume that $K$ always refers to a simplicial complex. We will use the finite field $\F_2$ (integers with addition modulo 2) as our coefficient group, as this is standard in TDA applications. 

\begin{defn}
    Let the $n$-chain group $C_n(K)$ be the free abelian group generated by the $n$-simplices of $K$ with coefficients in the field $\F_2$.
\end{defn}

Note that by using a field as the coefficient group, $C_n(K)$ is actually a vector space with its basis being the $n$-simplices of $K$. Homology theory tells us that we can construct a chain complex 
\begin{equation}
    \cdots \rightarrow C_n(K) \overset{\partial_n}{\longrightarrow}C_{n-1}(K)\overset{\partial_{n-1}}{\longrightarrow}\cdots\,,
\end{equation}
where the maps are defined by summing the $(n-1)$-dimensional faces of a $n$-simplex $\sigma$. Formally: 

\begin{defn}
    We define a homomorphism from $n$-chains to $(n-1)$-chains called the boundary homomorphism $\partial_n:C_n(K)\rightarrow C_{n-1}(K)$. If $\sigma = \{p^0,p^1,\cdots, p^n\}$ is an $n$-simplex in $K$ the action of $\partial_n$ on $\sigma$ is given by the following sum over $(n-1)$-simplices
    \begin{equation}
        \partial_n \sigma = \sum_{i=0}^{n}\{p^0,\cdots, \hat{p}^i,\cdots, p^n\}\,,
    \end{equation}
    where the $\hat{p}^i$ indicates the removal of $p^i$ from $\{p^0,p^1,\cdots, p^n\}$. The definition of $\partial_n$ is then extended linearly over all $n$-chains.
\end{defn}

We then note that structures which geometrically represent loops will vanish under the action of the boundary operator. This leads us to the following definition
\begin{defn}
    If $c\in C_n(K)$ satisfies $\partial_n c=0$ then we call $c$ an $n$-cycle.
\end{defn}
The first key theorem of homology is that all boundaries are also cycles. That is
\begin{lem}
    If $c\in C_n(K)$ then $\partial_{n-1}\partial_n \sigma^n =0$, that is $\partial_{n-1}\partial_n=0$ for all $n\in \N$.
\end{lem}
The fact that all boundaries are cycles means we cannot use the kernel of the boundary map alone to detect holes in spaces. We instead define that a true hole in a space is represented by a cycle which is not a boundary. With the geometric interpretation, following the study of surfaces, being that such a cycle represents a hole because it describes a loop in the space which is not the boundary of a segment of surface. For example consider the loop around the hole in a doughnut, which is a cycle but that cycle does not contain any surface. Adopting this reasoning we define the following
\begin{defn}
    The degree-$n$ homology group of $K$ is defined as 
    \begin{equation}
        H_n(K) = \frac{\ker{\partial_n}}{\Im\partial_{n+1}}\,.
    \end{equation}
\end{defn}

As a quotient group, $H_n(K)$ is made of equivalence classes of cycles in $C_n(K)$ where the equivalence relation is that $c \sim c'$ if $c=c'+\partial_{n+1}d$ for some $d\in C_{n+1}(K)$. That is, classes are sets of cycles whose elements differ by the boundaries of $(n+1)$-chains. Each class in $H_n$ is associated geometrically to a single $n$-dimensional hole in the simplicial complex $K$.

The second key theorem of homology is that continuous maps between simplicial complexes lead to linear maps between homology groups. This result underlies the main constructions of TDA. 
Suppose we have simplicial complexes $K$ and $K'$ and a continuous map $\phi:K\rightarrow K'$. Then for any chain $c\in C_n(K)$ we define an induced map on chains $\phi_\#:C_n(K)\rightarrow C_n(K')$ by linearly extending the action of $\phi$, which is defined on simplices, over sums of simplices. It can be shown that $\phi_\#$ commutes with the boundary operator and as a result always maps cycles to cycles and boundaries to boundaries. For this reason $\phi_\#$ can be used to construct a map on the quotient group, that is the homology groups. If $[c]$ is an equivalence class in $H_n(K)$ described by the representative $c\in C_n(K)$, we can define a map $\phi_*:H_n(K)\rightarrow H_n(K')$ by the following
 \begin{equation}
     \phi_*([c]) = [\phi_\#(c)]\,.
 \end{equation}
 It follows that this map $\phi_*$ is a homomorphism between the homology groups and so we call it the induced homomorphism on homology. If we have a sequence of continuous maps $\phi,\psi,\ldots$ between several simplicial complexes $K,K',K'',\ldots$ that is
\begin{center}
\begin{tikzcd}
\cdots \arrow[r] & K \arrow[r, "\phi"] 
& K' \arrow[r, "\psi"] & K'' \arrow[r] & \cdots \,,
\end{tikzcd}
\end{center}
then there will exist a sequence of induced maps between the homology groups
\begin{center}
\begin{tikzcd}
\cdots \arrow[r] & H_n(K) \arrow[r, "\phi_*" ] & H_n(K') \arrow[r, "\psi_*" ] & H_n(K'') \arrow[r] & \cdots\,.
\end{tikzcd}
\end{center}
It is by studying such a sequence of homology groups that we will be able to determine how the topology of a dataset changes with scale.

\subsubsection{Filtrations and Persistent Homology}

We now describe a core tool of TDA called Persistent Homology~\cite{EDELSBRUNNER}. Suppose we have a finite sequence of simplicial complexes $K_0,K_1,\ldots K_m$ such that $K_i\subset K_j$ for all $i<j$. We can construct inclusion maps from one complex to another to form a sequence of inclusion maps
\begin{equation}
\begin{tikzcd}
0 \arrow[hookrightarrow]{r} & K_0 \arrow[hookrightarrow]{r} 
& K_1 \arrow[hookrightarrow]{r} & \cdots \arrow[hookrightarrow]{r} & K_m\,.
\end{tikzcd}
\end{equation}
Such a sequence is referred to as a \textit{filtration} over the set of complexes.

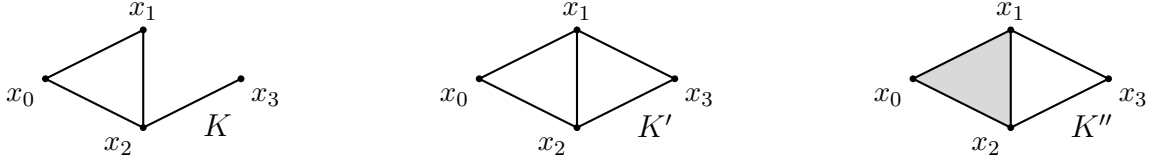
\begin{figure}
    \centering
    \begin{subfigure}[b]{0.32\textwidth}
        \centering
            \begin{tikzpicture}[scale=1.3] 
                \draw[thick, black] (0,0) -- (1,-0.5) -- (2,0);
                \draw[thick, black] (1,0.5) -- (0,0);
                \draw[thick, black] (1,-0.5)-- (1,0.5);
            
                \fill (0,0) circle (1pt) node[below left,node font=\small] {$x_0$};
                \fill (1,-0.5) circle (1pt) node[below left,node font=\small] {$x_2$};
                \fill (1,0.5) circle (1pt) node[above ,node font=\small] {$x_1$};
                \fill (2,0) circle (1pt) node[below right,node font=\small] {$x_3$};

                \fill (1.5,-0.25) node[below right]{$K$};
            \end{tikzpicture}
    \end{subfigure}
    \begin{subfigure}[b]{0.32\textwidth}
        \centering
            \begin{tikzpicture}[scale=1.3] 
                \draw[thick, black] (0,0) -- (1,-0.5) -- (2,0) -- (1,0.5) -- (0,0);
                \draw[thick, black] (1,-0.5)-- (1,0.5);
            
                \fill (0,0) circle (1pt) node[below left,node font=\small] {$x_0$};
                \fill (1,-0.5) circle (1pt) node[below left,node font=\small] {$x_2$};
                \fill (1,0.5) circle (1pt) node[above ,node font=\small] {$x_1$};
                \fill (2,0) circle (1pt) node[below right,node font=\small] {$x_3$};
                \fill (1.5,-0.25) node[below right]{$K'$};
            \end{tikzpicture}
    \end{subfigure}
        \begin{subfigure}[b]{0.32\textwidth}
            \centering
            \begin{tikzpicture}[scale=1.3] 
                \filldraw[thick, black, fill = gray!30] (0,0) -- (1,-0.5) -- (1,0.5) -- (0,0);
                \draw[thick,black]  (1,0.5) -- (2,0) -- (1,-0.5);
            
                \fill (0,0) circle (1pt) node[below left,node font=\small] {$x_0$};
                \fill (1,-0.5) circle (1pt) node[below left,node font=\small] {$x_2$};
                \fill (1,0.5) circle (1pt) node[above ,node font=\small] {$x_1$};
                \fill (2,0) circle (1pt) node[below right,node font=\small] {$x_3$};
                \fill (1.5,-0.25) node[below right]{$K''$};
            \end{tikzpicture}
    \end{subfigure}
    \caption{Diagram of three simplicial complexes. The grayed-in triangle indicates that $(x_0x_1x_2)$ is a simplex in $K''$.}
    \label{fig:examplecomplexes}
\end{figure}

We illustrate this in Figure \ref{fig:examplecomplexes} with three complexes that form a filtration $K\hookrightarrow K' \hookrightarrow K''$. Choosing representatives for the homology classes, we find the degree-1  homology groups of each complex: 
\begin{align*}
    &H_1(K) = \{ a[\{x_0,x_1\}+\{x_1,x_2\}+\{x_0,x_2\}] \,|\, a \in \F_2\}\,,\\
    &H_1(K') = \{ a[\{x_0,x_1\}+\{x_1,x_2\}+\{x_0,x_2\}]+b[\{x_1,x_3\}+\{x_2,x_3\}+\{x_1,x_2\}] \,|\, a,b \in \F_2\}\,,\\
    &H_1(K'') = \{ a[\{x_1,x_3\}+\{x_2,x_3\}+\{x_1,x_2\}] \,|\, a \in \F_2\}\,.
\end{align*}
We next find the image of each homology group under the maps induced by the above inclusions. For example, 
\begin{align*}
    {i_1}_*(H_1(K)) &= \{ a {i_1}_*([\{x_0,x_1\}+\{x_1,x_2\}+\{x_0,x_2\}]) \,|\, a \in \F_2\}\,,\\
    &= \{ a [i_1(\{x_0,x_1\}+\{x_1,x_2\}+\{x_0,x_2\})] \,|\, a \in \F_2\}\,,\\
    &= \{ a [\{x_0,x_1\}+\{x_1,x_2\}+\{x_0,x_2\}] \,|\, a \in \F_2\}\,,
\end{align*}
where in the final line $\{x_0,x_1\}+\{x_1,x_2\}+\{x_0,x_2\} \in C_1(K')$ instead of the $C_1(K)$ group in which it started. We can now make the  observation that ${i_1}_*(H_1(K)) \subset H_1(K')$ but $H_1(K')\neq {i_1}_*(H_1(K))$. 
That is, there are classes in $H_1(K')$ which are not in the image of $H_1(K)$ under ${i_1}_*$. 
This corresponds to the topological property that $K$ contains one hole (i.e., cycle) while $K'$ has two. In the language of persistent homology we would say the second $1$-cycle is \textit{born} when the $\{x_1,x_3\}$ edge is added to $K$.

In contrast, consider the inclusion-induced map on $H_1(K')$ which we compute as 
\begin{align*}
    {i_2}_*(H_1(K')) &= \{ a {i_2}_*([\{x_0,x_1\}+\{x_1,x_2\}+\{x_0,x_2\}])+b{i_2}_*([\{x_1,x_3\}+\{x_2,x_3\}+\{x_1,x_2\}]) \,|\, a, b \in \mathbb{Z}_2\}\,,\\
    &= \{ a [i_2(\{x_0,x_1\}+\{x_1,x_2\}+\{x_0,x_2\})]+b [i_2(\{x_1,x_3\}+\{x_2,x_3\}+\{x_1,x_2\})] \,|\, a, b \in \F_2\}\,,\\
    &= \{a [\partial_2 \{x_0,x_1\}]+b [\{x_1,x_3\}+\{x_2,x_3\}+\{x_1,x_2\}]  \,|\, a, b \in \F_2\}\,,\\
    &= \{b [\{x_1,x_3\}+\{x_2,x_3\}+\{x_1,x_2\}]  \,|\, b \in \F_2\}\,. 
\end{align*}
The third line holds because the $\{x_0,x_1,x_2\}$ simplex is in $C_2(K'')$. 
The fourth line holds because cycles in homology are equivalence classes with respect to the addition of boundaries. 
We see that $ {i_2}_*(H_1(K'))=H_1(K'')$ so no classes are born when the $\{x_0,x_1,x_2\}$ 2-simplex is added. 
However, 
\begin{equation}
    \ker{{i_2}_*} = \{ a[\{x_0,x_1\}+\{x_1,x_2\}+\{x_0,x_2\}] \,|\, a \in \F_2\}\cong \F_2\,.\\
\end{equation}
That is, a non-trivial class in $H_1(K')$ is sent to zero by ${i_2}_*$, and we say that a $1$-cycle \textit{dies} when the triangle $\{x_0,x_1,x_2\}$ is added to the complex. 
Note that a class can die without the selected representative being sent to zero when it merges with another class. The kernel of the induced map under inclusion will still be non-trivial in this case.

To make the example above into a general, formal theory we define the \textit{persistent homology groups}. 
\begin{defn}
    Given a filtration with $K_i\hookrightarrow K_j$ for $0\leq i\leq j\leq m$, let  $f^{i,j}_n:H_n(K_i)\rightarrow H_n(K_j)$ be the homomorphisms induced by inclusion. Then we define the $n$-th persistent homology groups as $H_n^{i,j} = \Im{f^{i,j}_n}$. 
\end{defn}
Observe that $H^{i,j}_n$ encodes those $n$-cycles from stage $K_i$ of the filtration that are still alive, or persist, in $K_j$. We now define the birth and death of homology classes as follows
\begin{defn}
    If $\alpha \in  H_n(K_i)$ but $\alpha \notin H_n^{i-1,i}$ we say that the class is \textit{born} at $K_i$. Similarly, we say this class $\alpha$ \textit{dies} entering $K_j$ if $f_n^{i,j-1}(\alpha) \notin H^{i-1,j-1}_n$ but $f_n^{i,j}(\alpha) \in H^{i-1,j}_n$.   The persistence of this class is defined to be $p(\alpha) = j-i$, the difference of the death and birth times.
\end{defn}
The definition of birth provided above follows our discussion from earlier, that a class is born when it is in the homology of a complex but not in the image of the homology of the previous complex in the filtration. The definition of death is somewhat subtle and includes both the possibility that classes are sent to zero as well as merged with older classes. For a detailed discussion we refer to Chapter 7 of \cite{EDELSBRUNNER}.

For a given filtration, we can compute the set of all birth and death times of degree-$n$ homology classes using matrix reduction algorithms. We refer to this set as the degree-$n$ Persistent Homology ($PH_n$) of the filtration. 
A key feature of this approach is that it is purely algebraic and does not require computing the homology of each complex in the filtration explicitly. For details regarding how to actually do the computation see \cite{EDELSBRUNNER}.

The \textit{modus operandi} of TDA is then as follows: we associate to our dataset a sequence of nested simplicial complexes; we construct a filtration from this sequence of complexes; we compute the PH for that filtration; and finally we study the PH itself to infer the shape of the dataset. 

\subsubsection{Persistence diagrams}

Consider Figure \ref{fig:RipsFiltration} again. Observe that each complex is a sub-complex of those to the right of it in the diagram. Generically we have that for a point cloud $X$, if $\epsilon<\epsilon'$ then
\begin{equation}
    VR_\epsilon(X) \subset VR_{\epsilon'}(X)\,.
\end{equation}

\begin{figure}
    \centering
    \includegraphics[width = \textwidth]{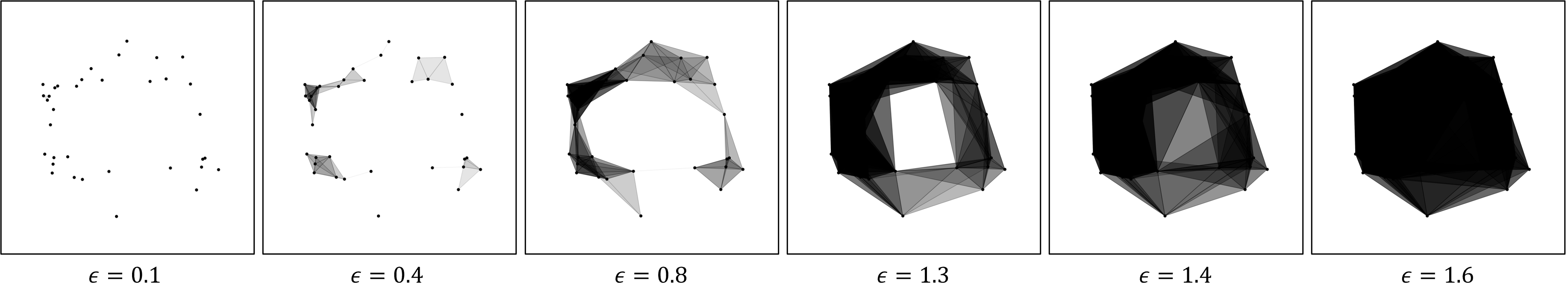}
    \caption{Image of Vietoris-Rips complexes of the point cloud sampled from the unit circle with normally distributed radial noise. The diameter ($\epsilon$) of each complex is indicated under each image.}
    \label{fig:RipsFiltration2}
\end{figure}

This implies that we can construct a filtration from the $VR$ complexes by the inclusion maps between complexes them. Hypothetically, this is a real valued filtration in the sense that we have a different VR complex for each $\epsilon\in \mathbb{R}^+$. However, if $X$ is a finite point cloud then the VR complex for each $\epsilon$ will not be unique and instead only changes at diameters which are lengths of edges of the complex. Therefore, we will have a series of intervals of the form $[\epsilon_i,\epsilon_j)$ over which the VR complex does not change and so we only need to consider a finite sequence of diameters $\epsilon_0 < \epsilon_1 <\ldots < \epsilon_m$ which we take as the lower bound of each of the intervals. Then our filtration is 
\begin{equation}
\begin{tikzcd}
0 \arrow[hookrightarrow]{r} & VR_{\epsilon_0}(X) \arrow[hookrightarrow]{r} 
& VR_{\epsilon_1}(X) \arrow[hookrightarrow]{r} & \cdots \arrow[hookrightarrow]{r} & VR_{\epsilon_m}(X)\,.
\end{tikzcd}
\end{equation}
We refer to this as the VR filtration and its persistent homology as the VR persistent homology of the point cloud $X$, and notate said $n$-th persistent homology as $PH_n(X)$.

In this document we will visualise $PH_n(X)$ using Persistence Diagrams (PD). The PD is a scatter plot of the birth-time and death-time pairs $(b,d)$ on the plane. Since all classes must die after they are born we have $d>b$ and so all persistence pairs appear above the line $y=x$. Note that the vertical distance between any point and this line is the persistence of the class. So classes which are very persistent appear well above the line. 

\begin{figure}
    \centering
    \includegraphics[width = 0.4\textwidth]{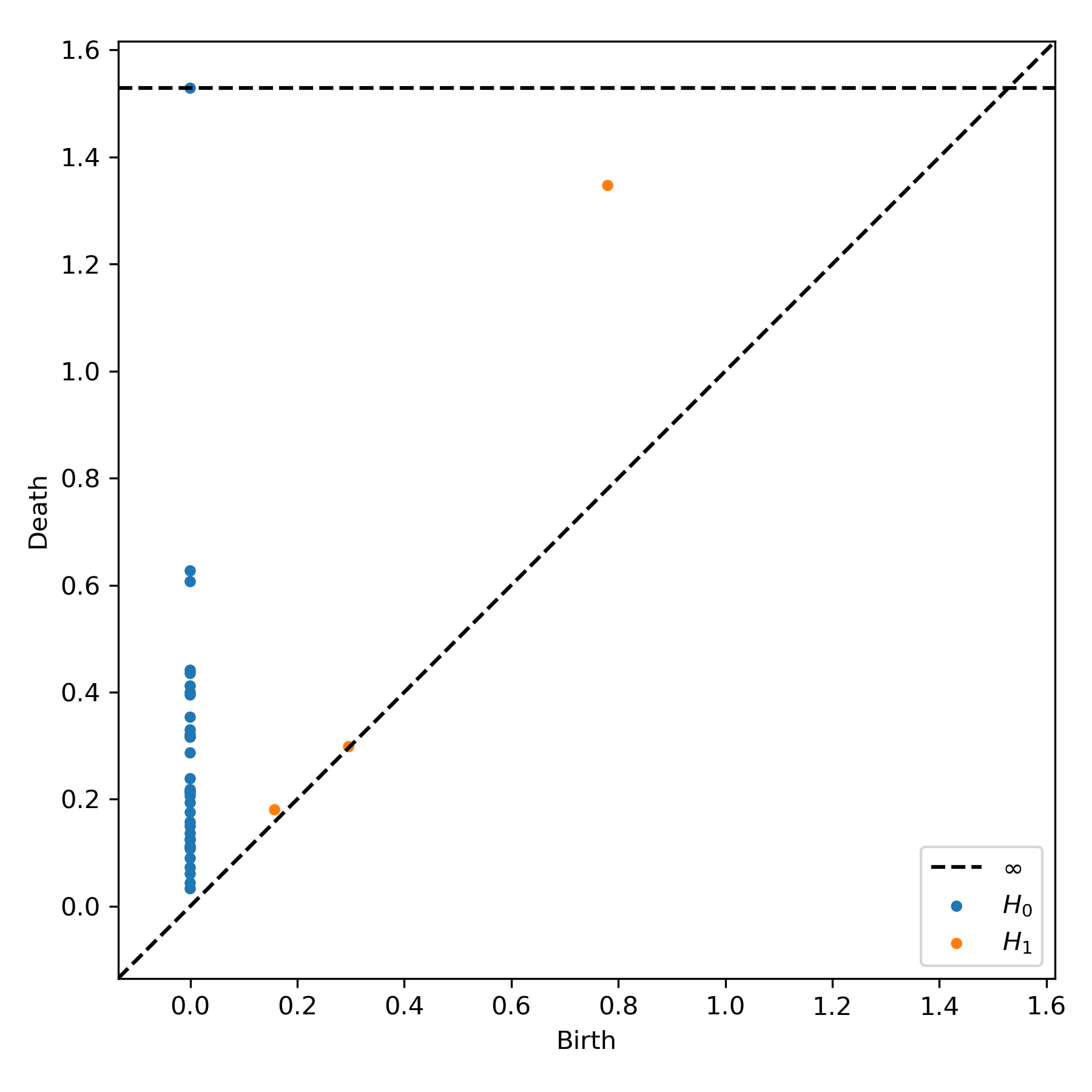}
    \caption{Persistence diagram of the VR filtration shown as Figure \ref{fig:RipsFiltration2}.}
    \label{fig:RipsFiltration2PD}
\end{figure}

Consider Figure \ref{fig:RipsFiltration2} which presents a sample of the complexes from the VR filtration of a point cloud which is approximately circular we will call $X$. For very small $\epsilon$ the complex is largely a collection of disconnected clusters. As $\epsilon$ is increased the clusters grow until there is only one connected component left. At $\epsilon\approx 0.8$ an edge is added which creates $1$-cycle in the VR complex corresponding to the hole in the center of the point cloud. At this point a degree-1 homology class is born. This class dies again at $\epsilon \approx 1.4$ when an edge and two triangles are added simultaneously and close the hole. For very large $\epsilon$ each point is connected to every other point and our render of the complex degenerates into a black mass.

To study the VR persistent homology of the $X$ cloud directly, we computed $PH(X)$ using the software program \verb!Ripser! \cite{Bauer2021Ripser}. This produces a persistence diagram which is presented as Figure \ref{fig:RipsFiltration2PD}. The classes from $PH_0(X)$ are shown as blue dots and those of $PH_1(X)$ as orange dots. We observe that there is one $H_1$ class well above the $y=x$ line, that is a highly persistent class. This corresponds to a loop in our dataset which persists across many scales and indicates the existence of a geometric hole in the dataset. It corresponds specifically to the fact that our dataset is approximately circular in shape. Note that the death time of the class $\epsilon_{\text{death}}\approx 1.4$ provides an estimate for the geometric size of the hole. The other classes in $PH_1$ are very close to the $y=x$ line and the convention in TDA is to attribute them to ``noise'' or more accurately ``sampling error''. 

\subsubsection{Multiplicative Persistence}

\begin{figure}
    \centering
    \begin{subfigure}[b]{0.49\textwidth}
        \includegraphics[width = \textwidth]{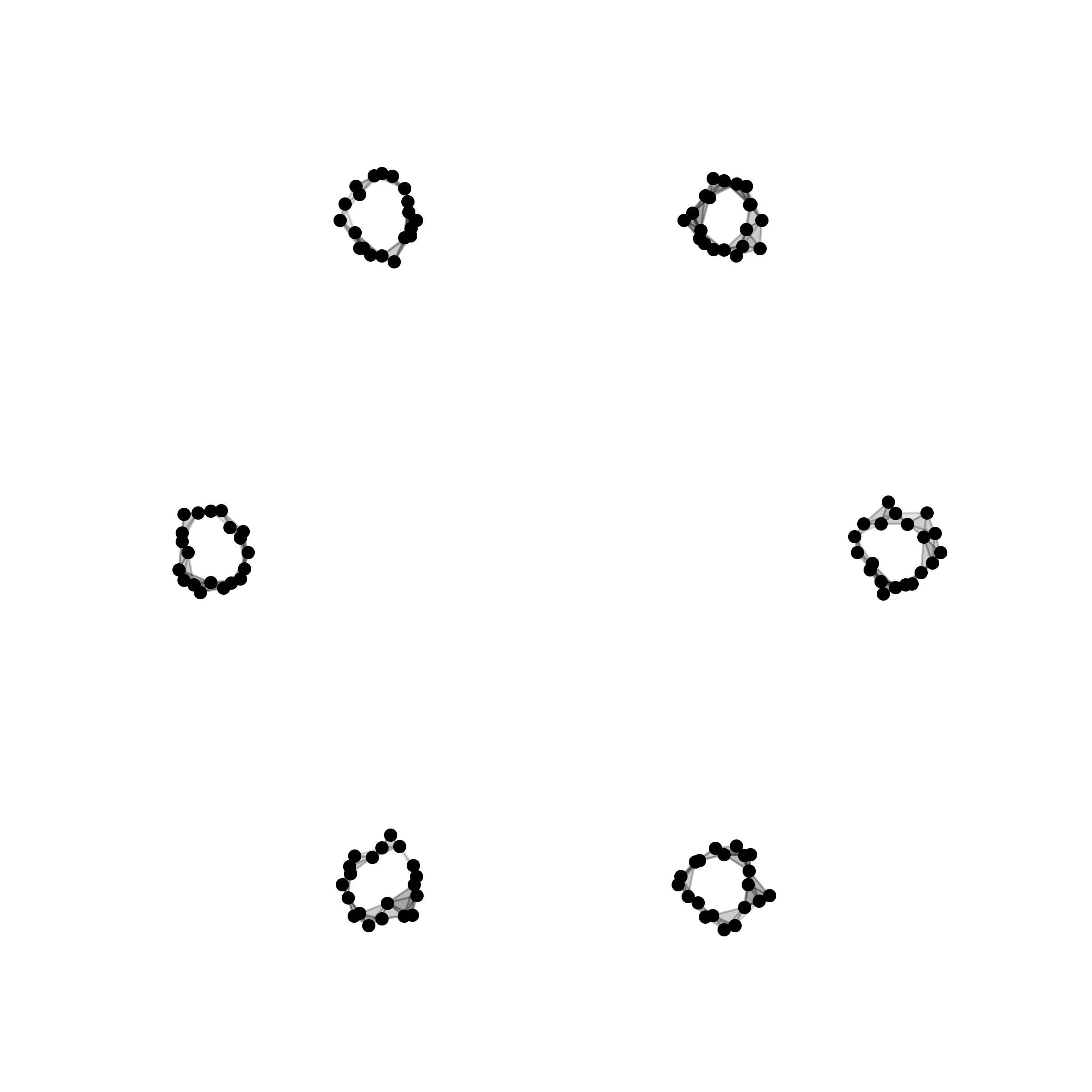}
        \subcaption{}
    \end{subfigure}
    \begin{subfigure}[b]{0.49\textwidth}
        \includegraphics[width = \textwidth]{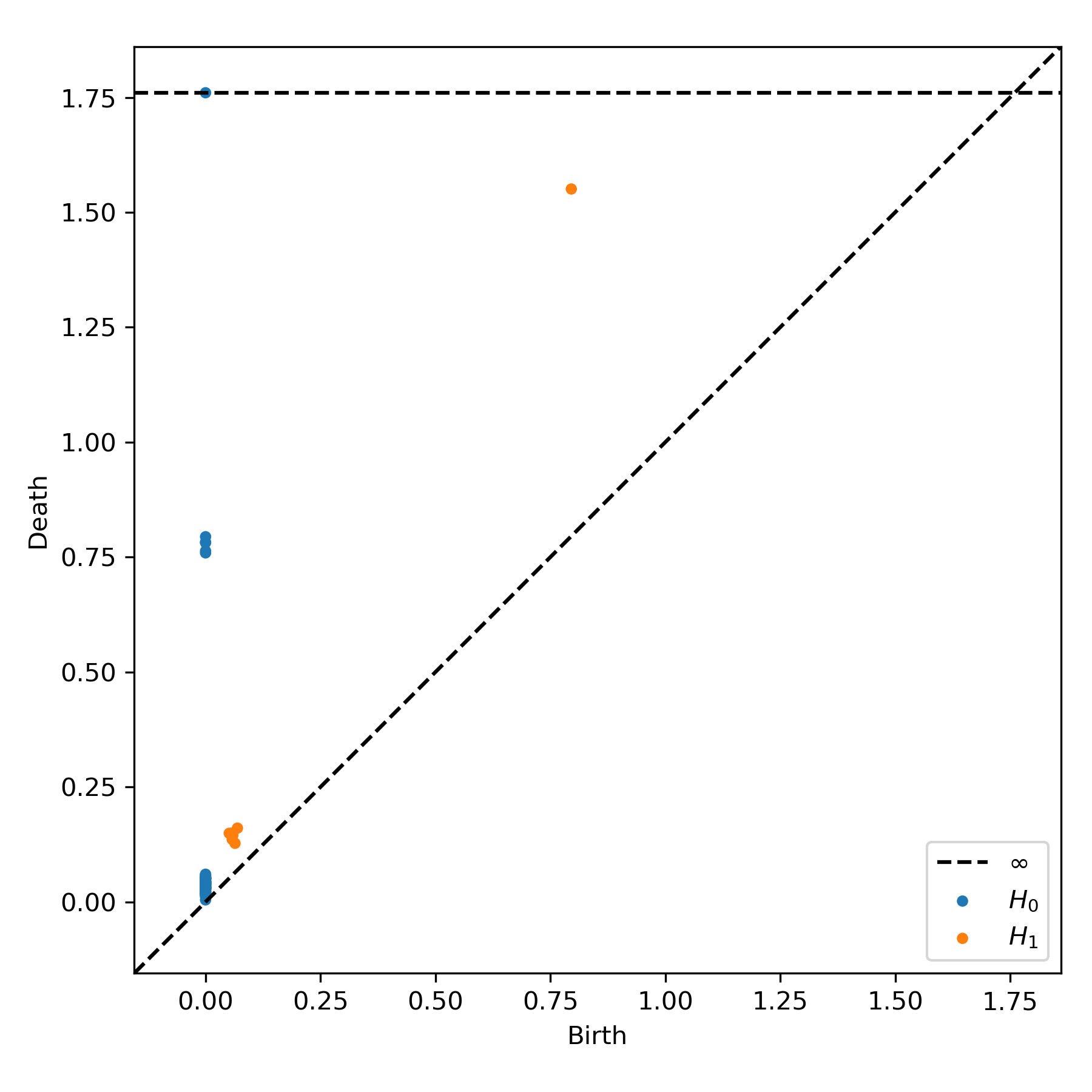}
        \subcaption{}
    \end{subfigure}
    \caption{Diameter $\epsilon = 0.1$ VR complex for a point cloud of small loop clusters and $PH(X)$ for said point cloud}
    \label{fig:loops}
\end{figure}

In TDA it is common to use the difference between the death and birth times of a class, defined above as the \textit{persistence}, to measure the statistical significance of said class. However, this can have disadvantages when we deal with point clouds and complexes with geometric structures on several length scales \cite{bobrowski2023universal}. Consider Figure \ref{fig:loops} which presents the $VR_{\epsilon= 0.1}(X)$ for a point cloud $X$, which is approximately an arrangement of six circles on the vertices of a regular hexagon, and its $PH(X)$. From the PD we note that there is one very persistent $H_1$ class and six $H_1$ classes with low persistence. The persistence of these classes is so small that without further investigation one might choose to consider them as less statistically significant and even ignore them in further analysis. Ignoring the low persistence $H_1$ classes we could conclude that our dataset is topologically close to a circle. The $H_0$ information tells us that the global circle is built from six tight clusters but nothing about those clusters. However, looking at the dataset geometrically we would argue that this is not representative. By completely ignoring the low-persistence $H_1$ classes any information about the small loops is lost. 

\begin{figure}
    \centering
    \includegraphics[width = 0.5\textwidth]{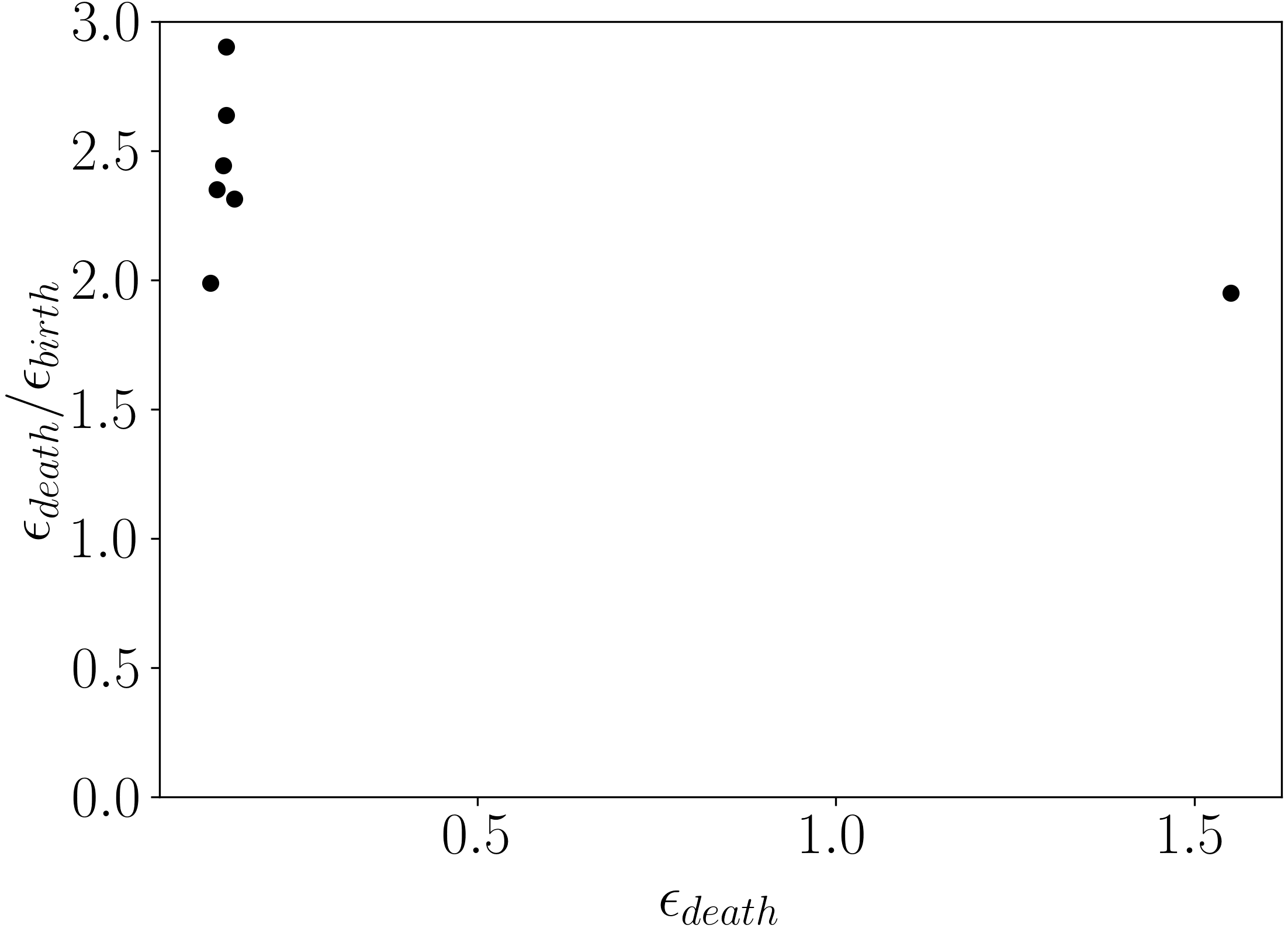}
    \caption{Scatter plot of the multiplicative persistence against the death time, that is the MPD, for each class in Figure \ref{fig:loops}. Collections of points arranged on vertical lines in the MPD correspond to families of geometric features with similar geometric size. }
    \label{fig:LoopsRelativerPersistence}
\end{figure}

The problem is that our dataset has topological structures on two geometrically distinct size scales. We will see in Section~\ref{Chapter3} that the same problem emerges when we study the Poincare maps of magnetic fields. Therefore, we would prefer to construct a tool which can measure the statistical significance of classes but without causing geometrically small features to be ignored. This can be achieved by looking at the ratio of the death and birth times instead of their difference. Figure~\ref{fig:LoopsRelativerPersistence} presents a scatter plot of the ratio of death time to birth time, that is $d/b$, versus $d$ for the $H_1$ classes in $PH_1(X)$. We see that the geometrically large loop, which has the greatest death time, has a very similar death time to birth time ratio as the six classes which die earlier. Hence, adopting $d/b$ as our measure of statistical significance we would conclude that the seven classes in $PH_1(X)$ are similarly statistically significant and hence recognise that the point cloud is not a single circle formed from six small clusters but instead is a single circle, or hexagaon, formed from six smaller circles.

In section~\ref{Chapter3} we will make repeated use of this alternative measure of class significance and so define it as follows.
\begin{defn}
    If $(b,d)$ are the birth and death times of a PH class respectively. Then $d-b$ is defined to be its additive persistence and $d/b$ its multiplicative persistence.
\end{defn}
Note that we will also use both the PDs and our $d/b$ scatter plots, which we refer to as Multiplicative Persistence Diagrams (MPDs), to visualise the topology of our point clouds and will refer to them generically as \textit{topological descriptors}.

\section{Classification of magnetic field lines by persistent homology}

 \label{Chapter3} 
 

Here we present our TDA approach to automated classification of magnetic field line orbits using VR persistent homology of the Poincare section of the field. We will begin by exploring the topological descriptors of selected field lines from a toy tokamak model to illustrate how the topological class of each field line is captured in its PD and MPD. The automatic classification scheme applied to our model is then demonstrated.

Computing the topology of a trajectory of a dynamical system using TDA requires that we have a finite representation for that trajectory. 
To obtain such a representation from a Poincare update map, we pick an initial point and then compute the trajectory for only a finite number of iterations in the forward time direction. 
For field lines this means we pick an initial point on the Poincare section $\Sigma$ and then integrate the field line flow, but stop integrating when the number of intersections between the field line and $\Sigma$ reaches a desired threshold. This gives a finite set of points in $\Sigma$ which admit the computation of the $PH$. 
 
\begin{defn}
    Suppose we have a magnetic field, or general dynamical system, and have selected a Poincare section $\Sigma$. Define $P:\Sigma\rightarrow\Sigma$ to be the Poincare update map induced on $\Sigma$ by field line flow. Given an integer $T$ and a point $x\in \Sigma$ we define the \textit{trajectory} of the point $x$ for finite time $T$ to be the set 
    \begin{equation}
        X_T(x) = \left\{ P^{(t)}(x),\text{ for } t \in 0,1,2,\ldots T\right\}\subset \Sigma\,.
    \end{equation}
\end{defn}

\subsection{Toy model for a perturbed tokamak field}

As our test magnetic field we use a toy model for a tokamak field formed. Specifically we adopt the vacuum field formed from a circular current loop, $I_{\phi}$ and an infinite straight line current, $I_z$. This model is scientifically problematic because the magnetic field is singular at the magnetic axis unlike a real tokamak. Nonetheless, the model is useful theoretically because it has the expected qualitative structure of a tokamak field --- that is of helical field lines confined to toroidal surfaces --- and can be analytically calculated making field line computation fast \cite{Morrison2000}. The exact analytic expressions for the required fields can be found in any standard text on electromagnetism, for example Section 5.5 of \cite{Jackson}. 

The tokamak field is modelled as 
\begin{equation}
    \textbf{B}(\textbf{x}) = \textbf{B}_t(\textbf{x})+\textbf{B}_p(\textbf{x})\,,
\end{equation}
where $\textbf{B}_t$ is the field generated by the straight line current $I_z$ and $\textbf{B}_p$ is the field generated by a circular current $I_\phi$. 
The labels $t$ and $p$ refer to the \textit{toroidal} and \textit{poloidal} components of the integrable field and are conventional in the discussion of the toroidal geometry of a tokamak. For the same reason we refer to the location of the current loop, $I_\phi$ as the \textit{magnetic axis}.


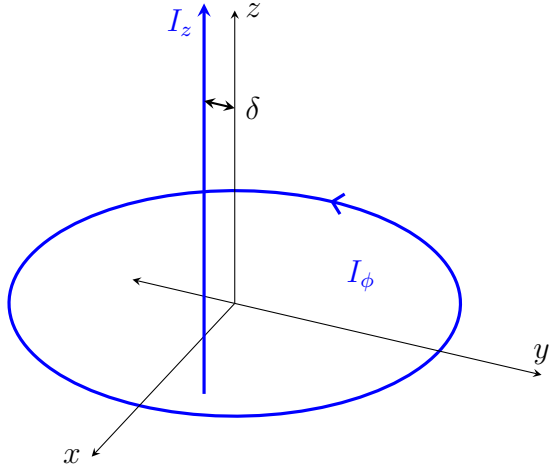
\begin{figure}
\centering
\begin{tikzpicture}[3d view={115}{30},>=stealth,scale=1.5]
    \coordinate (A) at (-0.75,0,0);
    \coordinate (B) at (0,0,0);
    \draw [very thick, blue, ->-] (B) ellipse (2 and 2);
    \fill (-0.5,1,0) node[above,blue] {$I_\phi$};

    \draw [->] (B) -- (3,0,0);
    \draw [->] (B) -- (0,3,0);
    \draw [->] (B) -- (0,-1,0);
    \fill (3,0,0) node[left] {$x$};
    \fill (0,3,0) node[above] {$y$};
    \draw [->] (B) -- (0,0,3);
    \fill (0,0,3) node[right] {$z$};
    \fill (0,-0.3,2.8) node[left,blue] {$I_z$};

    \draw [very thick, blue, ->] (0,-0.3,-1) -- (0,-0.3,0) -- (0,-0.3,3);

    \draw [thick,<->] (0,0,2) -- (0,-0.3,2);
    \fill (0,0,2) node[right] {$\delta$};
    
\end{tikzpicture}
\caption{Cartoon of the current configuration for our toy tokamak model. Note that, for our simulations $R=1$.}
\label{fig:ToyTokamakModel}
\end{figure}

We adopt the current geometry presented in Figure \ref{fig:ToyTokamakModel}. If we treated only the symmetric configuration presented on the left our magnetic field lines would be an integrable Hamiltonian system. That is, all field lines would lie on invariant circles in the Poincare section and therefore there would not be any different classes of field lines to classify \cite{Morrison2000}. To accurately represent the chaotic field line case, in which we are interested, we must explicitly break the symmetry of the setup. This is demonstrated in Figure \ref{fig:ToyTokamakModel} by the $\delta$ displacement of $I_z$ away from the $z$-axis. This perturbation breaks the symmetry of our setup and causes the field lines to become chaotic, generating the traditional island chains  of Hamiltonian chaos. 

Note that all Poincare sections presented below were computed for the case where 
\begin{equation}
    \frac{I_z}{I_\phi} = 1\,, \,   \ \frac{\delta}{R} = 0.005\,.
\end{equation}
It is only these ratios which meaningfully effect the field lines. This is because $\textbf{B}$ is linear in the current density and so increasing both currents by a uniform factor increases $\textbf{B}$ globally by the same factor but does not change the geometry of its field lines. Similarly, increasing both $\delta$ and $R$ by a constant factor corresponds only to rescaling the dimensions of length, a dilation transformation, and so similarly does not affect the geometry. We also adopt the half-plane defined by $\phi=0$ in cylindrical coordinates as our Poincare section $\Sigma$. This plane is shown explicitly in Figure \ref{fig:ToyTokamakModel}.


\subsection{VR persistent homology of magnetic field line orbits}

From the toy tokamak field we obtained a representative field line for each class, that is: an invariant torus, a single island, a multi-island chain, a large stochastic layer, and a thin stochastic layer. We computed the $X_{2000}$ for each field line and then the PD and MPD of the associated point cloud. The analysis of these topological descriptors follows.

\subsubsection{An invariant torus}

\begin{figure}
    \centering
        \includegraphics[width = 0.4\textwidth]{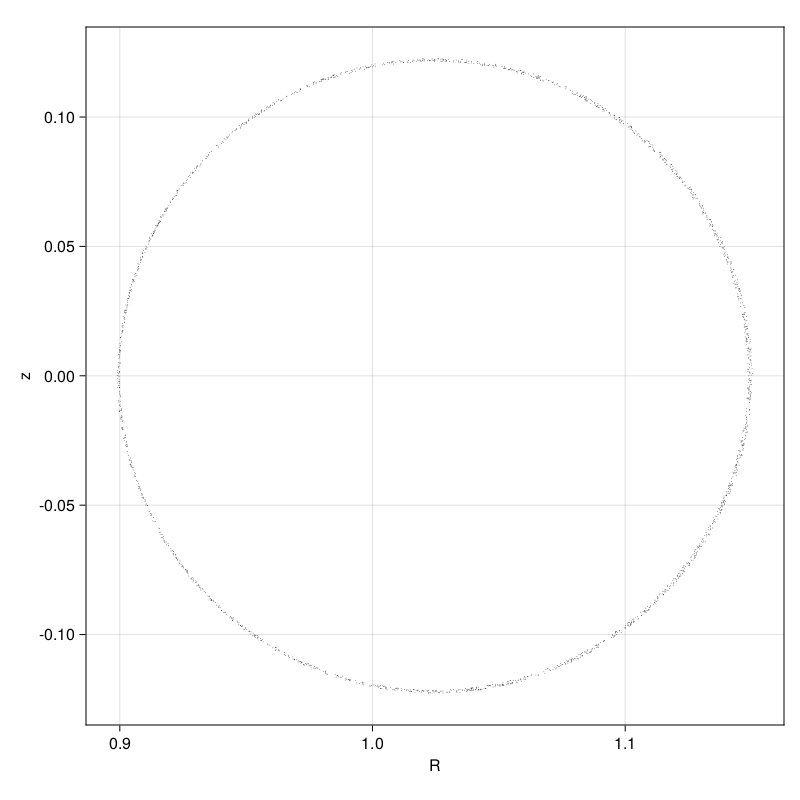}
    \caption{Point cloud of a field line on a invariant torus.}
    \label{fig:kamtorus_section}
\end{figure}

\begin{figure}
    \centering
    \begin{subfigure}[b]{0.39\textwidth}
        \centering
        \includegraphics[width = \textwidth]{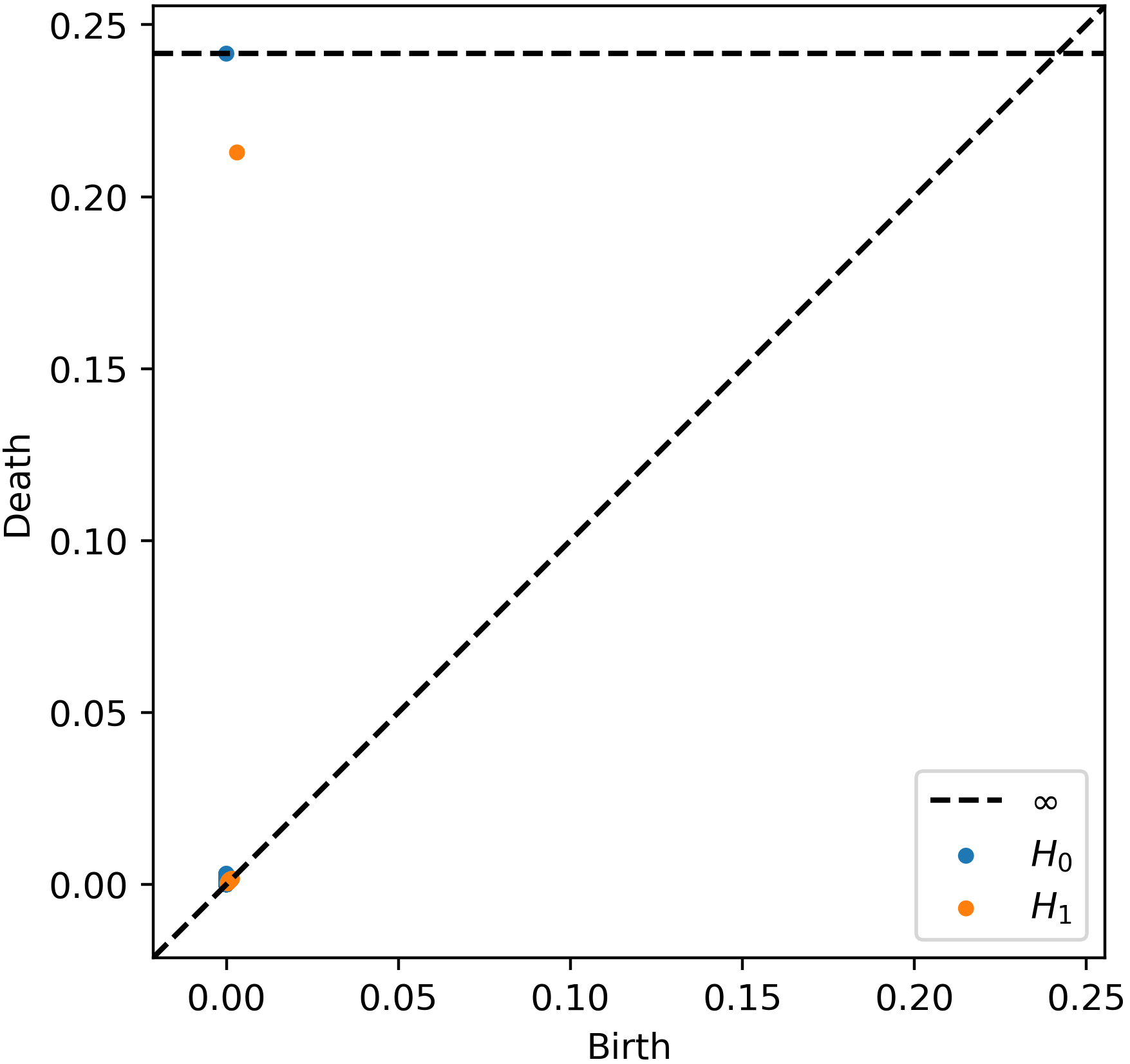}
        \subcaption{VR persistence diagrams.}
        \label{subfig:torusRips}
    \end{subfigure}
    \begin{subfigure}[b]{0.50\textwidth}
        \centering
        \includegraphics[width = \textwidth]{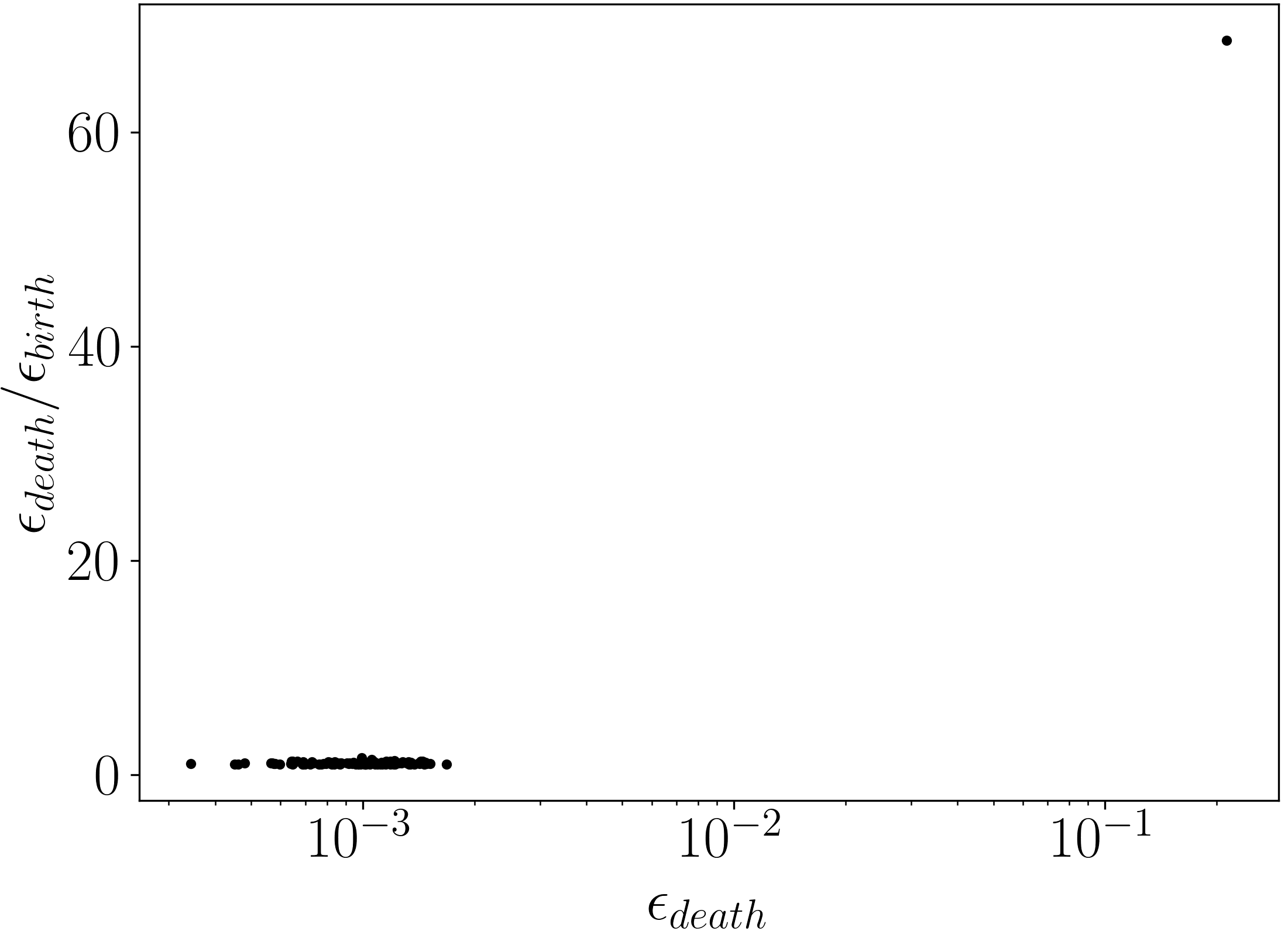}
        \subcaption{$PH_1$ class death to birth ratio.}
        \label{subfig:torusDB}
    \end{subfigure}
    \caption{Topological descriptors of the invariant
    torus shown in Figure \ref{fig:kamtorus_section}.}
    \label{fig:kamtorus_section_rips}
\end{figure}

The simplest field line orbit we can consider is that of an invariant torus. In the 3-d space in which the field lines live such an orbit densely covers a $2$-torus surrounding the magnetic axis. We therefore expect that after projecting to $\Sigma$ the orbit will form a topological circle containing the magnetic axis. Figure \ref{fig:kamtorus_section}, which presents the orbit of such a field line, confirms this expectation. Computing the $PH$ of this point cloud with the software \verb!Ripser! yields the PD presented as subfigure \ref{subfig:torusRips}. We observe a single $H_0$ class which exists for all diameters (The horizontal dotted line indicates that the class does not die by the end of the scan) corresponding to the fact that for $\epsilon$ larger than the diameter of the point cloud the VR complex is a connected simplicial complex. 

We also observe a very long lived $H_1$ class, which encodes the circular shape of the point cloud. A similar feature will appear in many of the cases which follow and its persistence is associated to whether our field line encloses the magnetic axis. Note that by \textit{encloses} here we are referring to whether or not the surface, or volume, filled by the field line separates the 3d space into two disconnected regions; one containing the magnetic axis and the other not. The invariant tori do separate space into two such regions and so we say they enclose the axis. In this case the point is very persistent, with a large death time to birth time ratio of $\epsilon_{\text{death}}/\epsilon_{\text{birth}}\approx 70$, and this suggests that it is highly likely that our field line encloses the magnetic axis, which it actually does in this case. 

\subsubsection{A magnetic island}

The next simplest type of field line orbit to consider after invariant tori are those associated to islands and island chains. The field lines of magnetic islands do not separate the magnetic axis from the rest of the space, or really the point at infinity. Their Poincare section may be made of several connected components and does not enclose the axis. For low order resonant perturbations, such as ours, the islands have a characteristic \textit{banana} shape \cite{Zaslavsky}.

\begin{figure}
    \centering
    \begin{subfigure}[b]{0.4\textwidth}
        \centering
        \includegraphics[width = 1.0\textwidth]{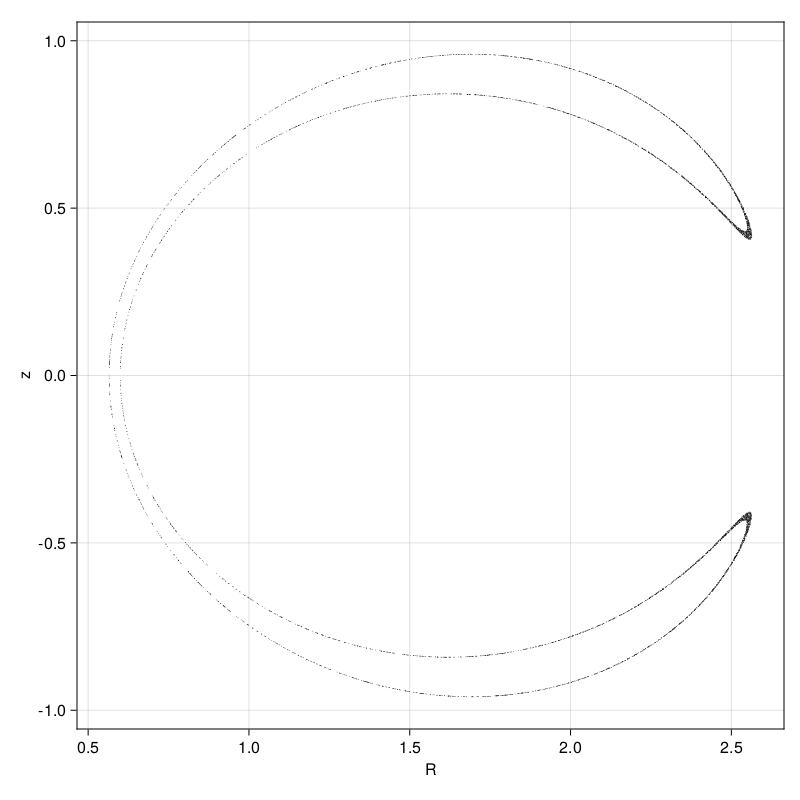}
        \subcaption{Poincare section.}
        \label{subfig:islandSection}
    \end{subfigure}
    \begin{subfigure}[b]{0.39\textwidth}
        \centering
                \includegraphics[width = \textwidth]{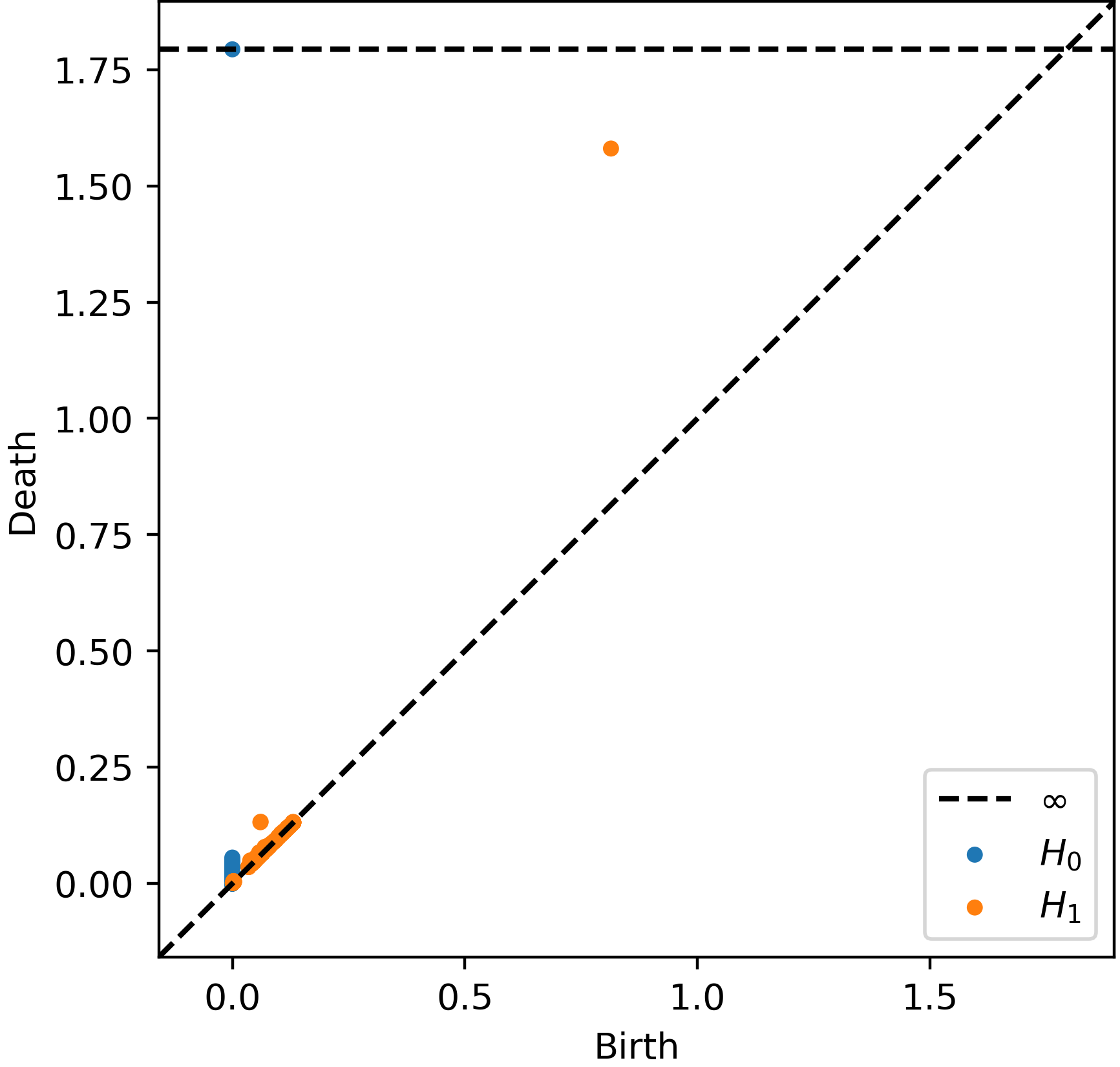}
        \subcaption{PD.}
        \label{subfig:islandRipsPH}
    \end{subfigure}
    \caption{Point cloud and $PD$ for an order one island.}
    \label{fig:island_section_and_rips}
\end{figure}

An example of such an island in our field is presented as subfigure \ref{subfig:islandSection}. The point cloud appears as an approximately closed loop which does not enclose the magnetic axis. The PD shown as subfigure \ref{subfig:islandRipsPH}, again includes the expected infinite lifetime $H_0$ class corresponding to connectivity. We can observe two interesting features in the $H_1$ PD: there is an $H_1$ class born at $\epsilon_{\text{birth}}\approx 0.7$ which is associated to the fact that our island nearly closes around the axis; and there appears to be another $H_1$ class born at $\epsilon_{\text{birth}}\approx0.1$ which we recognise is associated to the fact that our field line is a single closed island, however this class dies early $\epsilon_{\text{death}}\approx 0.15$ because the banana is a very narrow. These two features have very different additive persistence $\epsilon_{\text{death}}-\epsilon_{\text{birth}}$ but similar multiplicative persistence $\epsilon_{\text{death}}/\epsilon_{\text{birth}}$. This is a good example of why the multiplicative persistence is the preferred measure of the statistical significance of a topological feature when the shape in question contains structures on several length scales since it does not weight geometrically larger features more heavily \cite{bobrowski2023universal}.

\begin{figure}
    \centering
        \includegraphics[width = 0.7\textwidth]{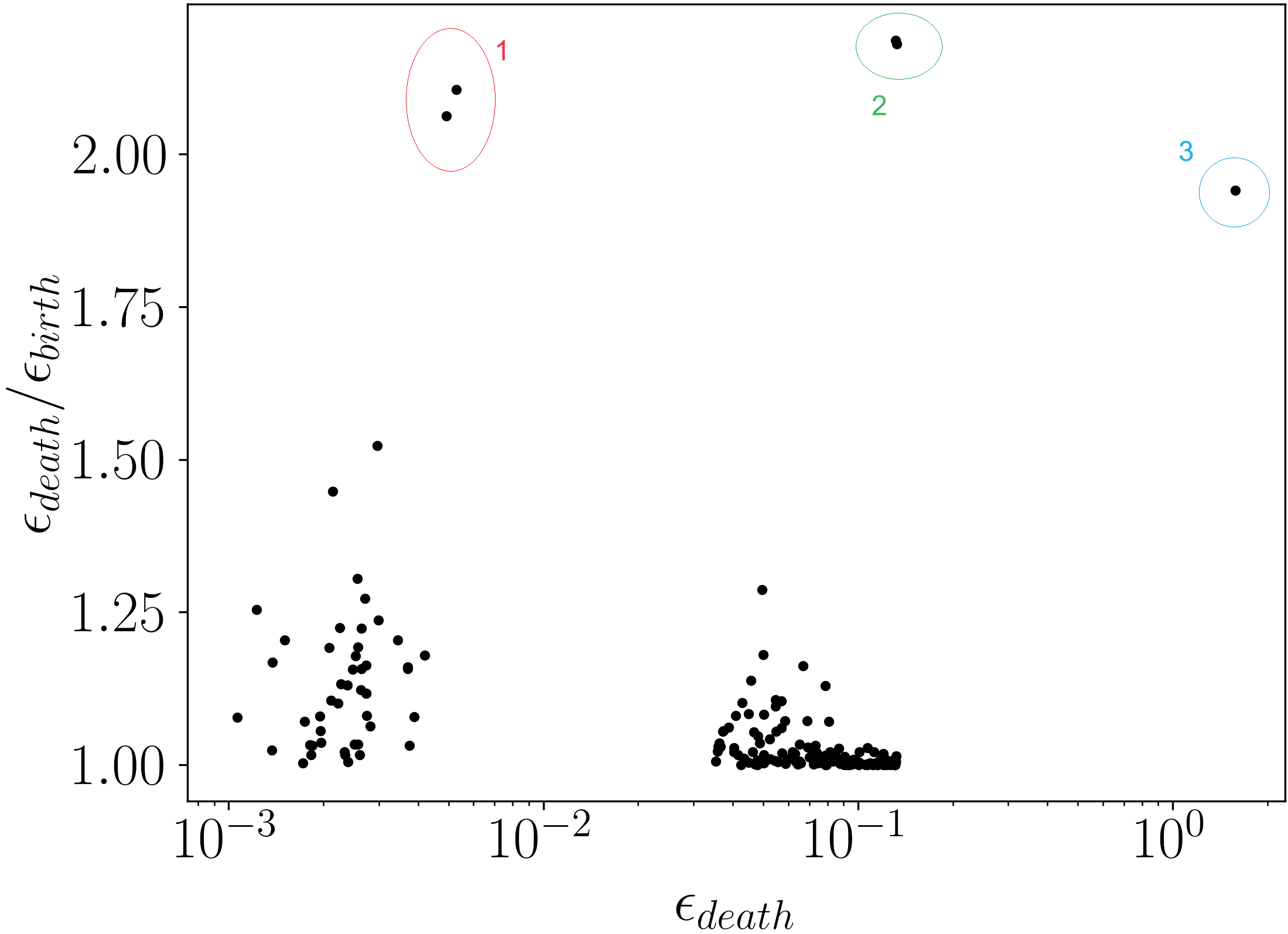}
    \caption{MPD of the island shown as subfigure \ref{subfig:islandSection}. The clusters $1,2,$ and $3$ are indicated in red, green, and blue respectively.} 
    \label{fig:island_section_bdrat}
\end{figure}

Consider Figure \ref{fig:island_section_bdrat} which presents the MPD for subfigure \ref{subfig:islandSection}. We observe three clusters of features with high multiplicative persistence at $\epsilon_{\text{death}} \approx 5(10^{-3})$, $10^{-2}$, and $2$. Which we will refer to as clusters $1$, $2$, and $3$ respectively. 

There are two classes in cluster $1$ and they are each associated to one of the two darker regions on the outboard side of the toy tokamak, that is, at the banana tips. The point cloud is \textit{thick}, fills a two-dimensional area, in these regions. The exact orbit of the field line would not be thick here but our computation of the field line orbit is not exact and so the integration error compounds at these points, where the field line changes direction quickly, and the field lines therefore start to fill out a 2d region. There exists small gaps between the points in the Poincare section at the filled in regions, allowing for the formation of closed loops in the VR complex, and this contributes a cluster of small diameter $H_1$ classes to $PH_1$ which we recognise as cluster $1$. 

Cluster $2$ is associated with the simple closed loop shape of the banana orbit but the reasonably low death time to birth time ratio of $\approx 2.2$ indicates that from only the specific point cloud examined we cannot be completely confident that the field line forms a simple closed curve. This is because near to the left edge, the Poincare section is somewhat sparse with an average distance between individual points on the order of the minimum width of the banana. This contrasts with the case of Figure \ref{fig:kamtorus_section} where the average spacing between points was several orders of magnitude less than the minimum width of the loop and therefore we were very confident that the trajectory is a simple closed curve. We would expect that if we followed the field line for longer and included more samples from the orbit the maximum multiplicative persistence of a class in cluster $2$ would increase and we would be more confident that the field line orbit is a closed loop in the Poincare section. 

Finally, cluster $3$ is associated with enclosure around the magnetic axis and the fact that its multiplicative persistence is lower than that of clusters $1$ and $2$ indicates that from the point cloud data we should not be confident that the field line surrounds the axis. It is good that we are not confident in this since the field line does not surround the axis.

\begin{figure}
    \centering
    \begin{subfigure}[b]{0.4\textwidth}
        \centering
        \includegraphics[width = 1.0\textwidth]{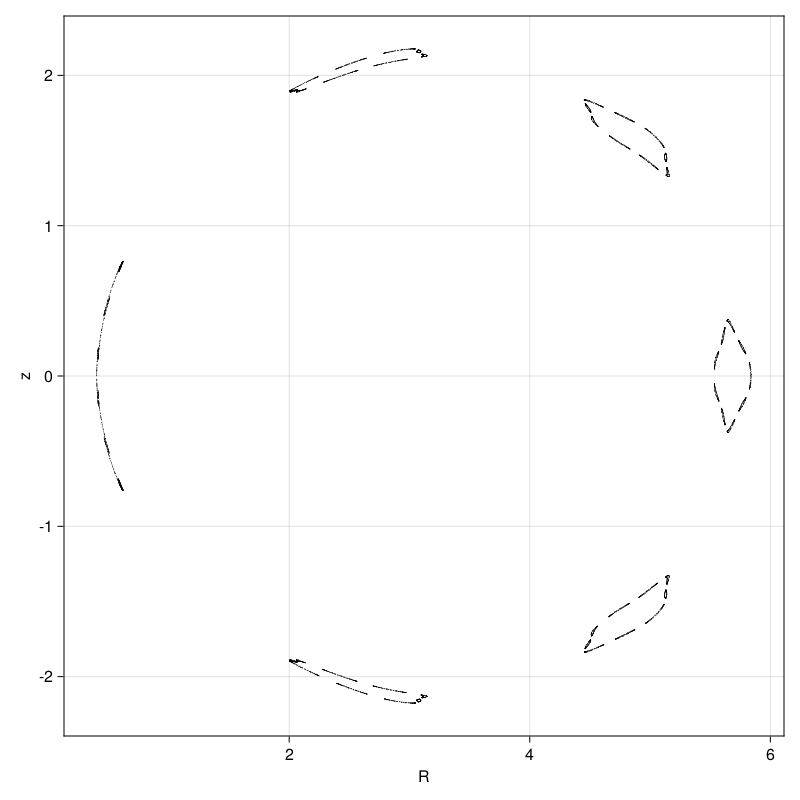}
        \subcaption{Poincare section.}
        \label{subfig:island_chainSection}
    \end{subfigure}
    \begin{subfigure}[b]{0.39\textwidth}
        \centering
                \includegraphics[width = \textwidth]{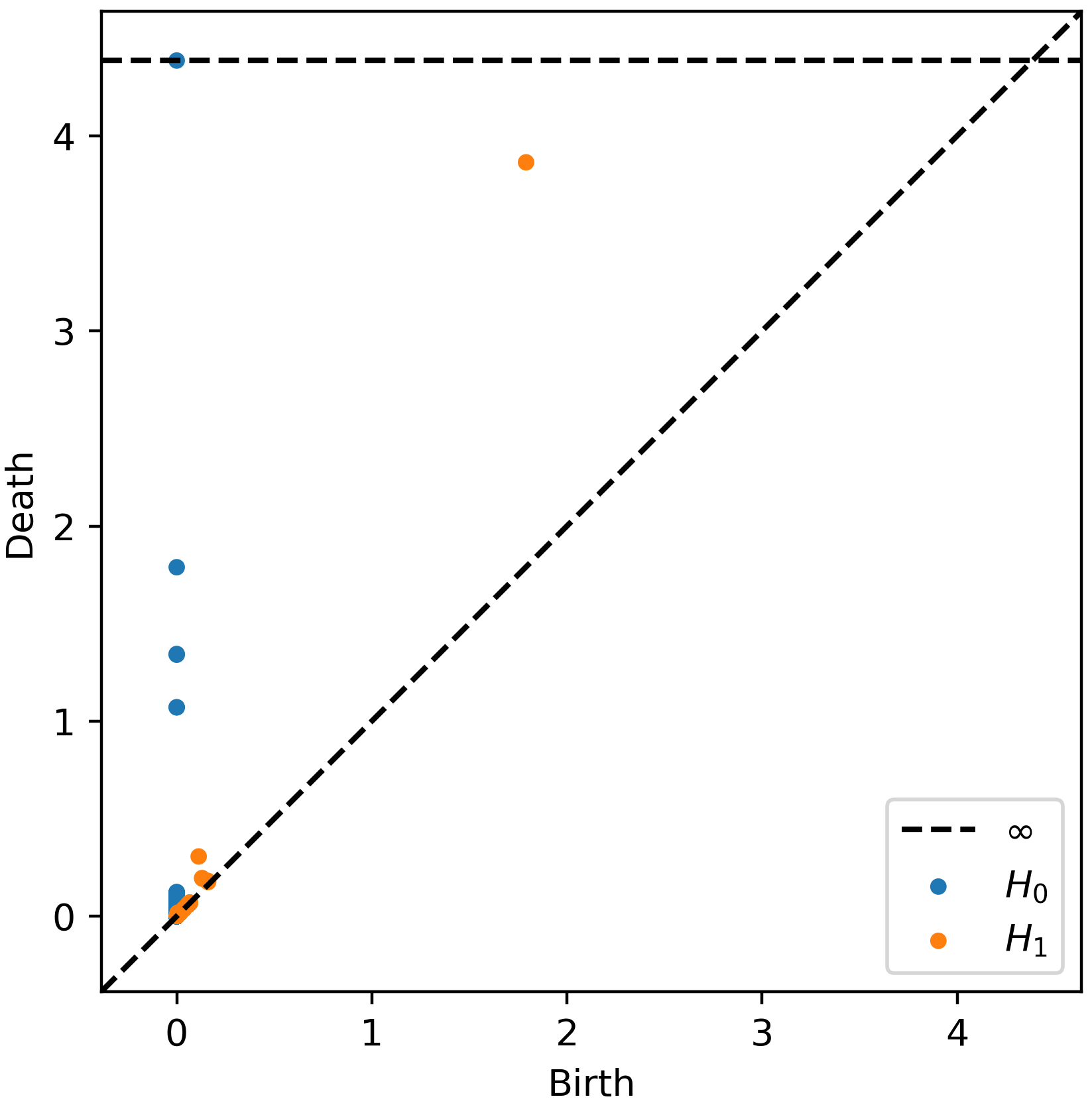}
        \subcaption{VR persistence diagrams.}
        \label{subfig:island_chainRips}
    \end{subfigure}
    \caption{Point cloud and persistent homology for an island chain.}
    \label{fig:island_chain_section_and_rips}
\end{figure}

\subsubsection{An island chain}

Next we will look at the case of the island chain presented as subfigure
\ref{subfig:island_chainSection}. We see that this is a chain of six major islands each of which is formed from many smaller islands. From the PD in \ref{subfig:island_chainRips} we observe what appears as four persistent $H_0$ classes. However, we would expect six persistent classes since there are six major islands in the orbit. In fact there are six persistence classes in \ref{subfig:island_chainRips} but two of them have been plotted under two of the others. This is because our orbit has a vertical parity symmetry and therefore the uppermost two islands are almost identical to the lowermost two and hence are associated to classes with nearly identical birth and death times. This is a downside of using PD as a visualisation technique, symmetries in the geometry of the point cloud will form coincident points on the diagram which then cannot be distinguished on inspection. 

\begin{figure}
    \centering
        \includegraphics[width = 0.6\textwidth]{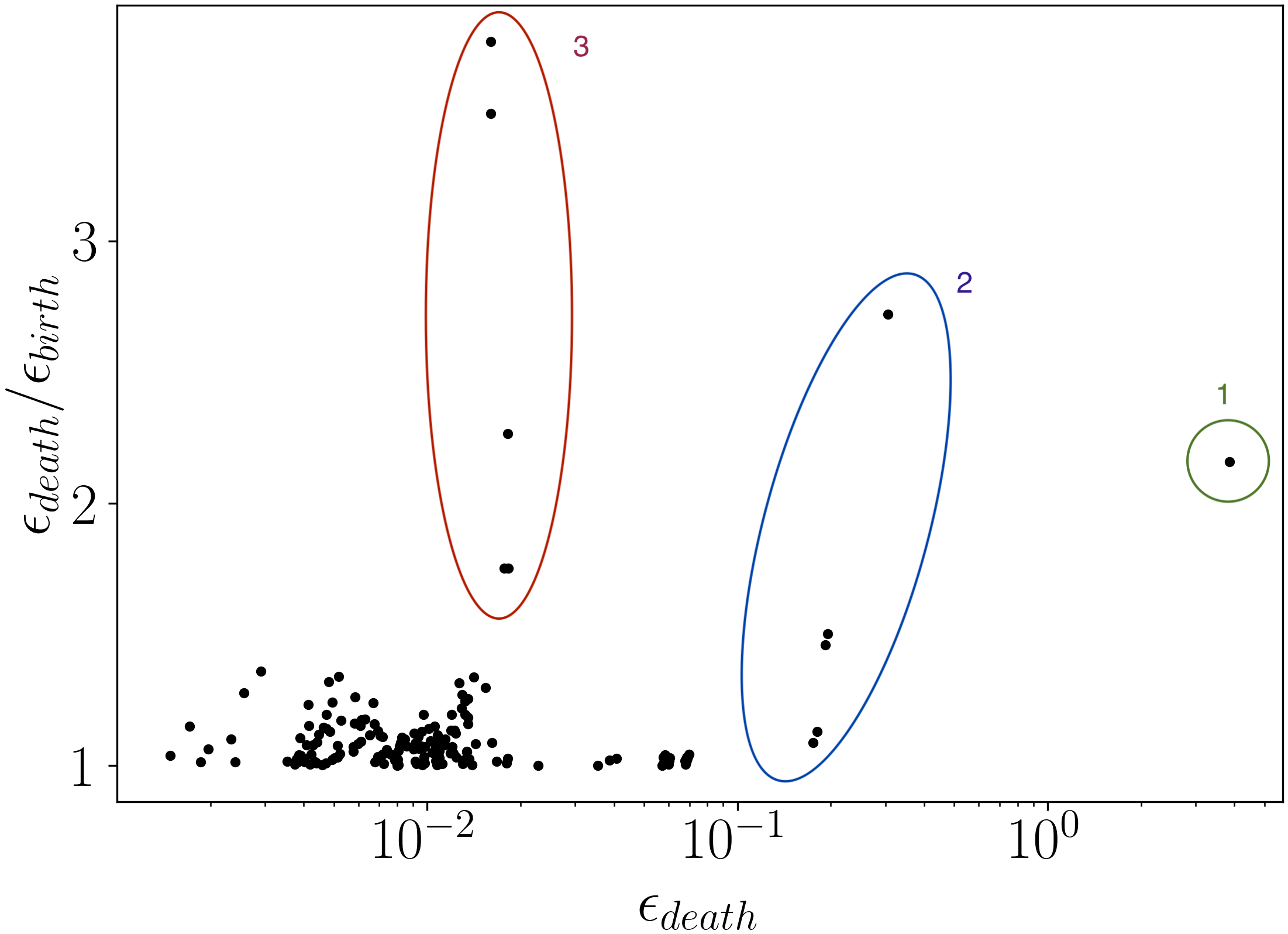}
    \caption{MPD of the island chain in subfigure \ref{subfig:island_chainRips}.}
    \label{fig:islandchain_section_bdrat}
\end{figure}

If we consider the MPD we obtain Figure \ref{fig:islandchain_section_bdrat} we observe three clusters of more persistent classes localised at different $\epsilon_{\text{death}}$, and hence corresponding to features on different geometric scales. Cluster $1$ is the point with largest $\epsilon_{\text{death}}$ and it corresponds, as before, to the enclosure of the magnetic axis. We recognise that the weak multiplicative persistence $\approx 2.25$ of this point corresponds to the fact that the island chain is composed of disconnected islands and so does not enclose the magnetic axis. 

Cluster $2$ contains five $H_1$ classes with $\epsilon_{\text{death}} \in (0.1,0.5)$. Each of these classes corresponds to a closed loop in the point cloud with a diameter in $(0.1,0.5)$. Observe from the Poincare section that the only structures in this radius range are the major islands. This indicates that each of the classes in the cluster corresponds to one of the major islands. The fact that there are five classes in the cluster indicates that the $PH_1$ has detected five major islands. However, we can see from inspection of the Poincare section that there are actually six major islands in the chain. So, we conclude that an island was not detected by the $H_1$ information. This should be expected because the island at smallest $R$ is so narrow as to appear on inspection as a straight line. The width of this island is less than the spacing between the sub-islands which comprise it and consequently a closed loop in the VR complex is never formed during the filtration, hence no class appears in the $H_1$ associated to this island. A class associated to the narrow island does appear in the $H_0$ persistence diagram though, as was noted above. We will see in Section \ref{Chapter4} that this problem can be mitigated by transforming the point cloud into straight field line coordinates.

Finally cluster $3$ is associated to structures on smaller scales. Specifically, the smaller islands which comprise the major islands. Some of these minor islands are circular enough to generate $H_1$ classes forming the small scale cluster of high multiplicative persistence islands observed.

\subsubsection{Thin stochastic layers}

\begin{figure}
    \begin{subfigure}[b]{0.45\textwidth}
        \centering
    \includegraphics[width = \textwidth]{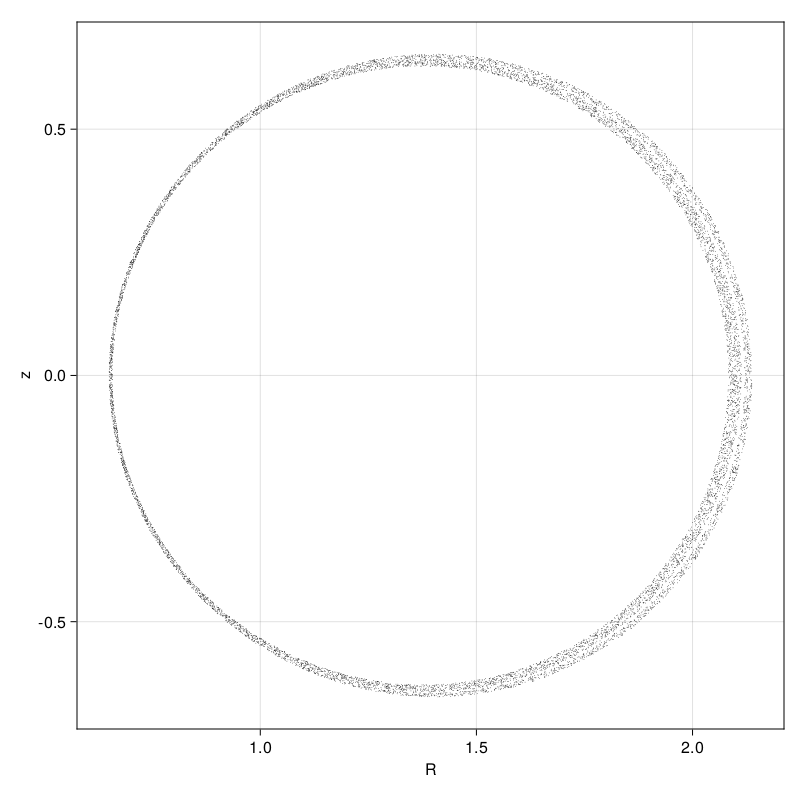}
    \subcaption{A thin stochastic layer.}
    \label{subfig:stochastic_thinSection}
    \end{subfigure}
    \centering
    \begin{subfigure}[b]{0.45\textwidth}
        \centering
        \includegraphics[width = \textwidth]{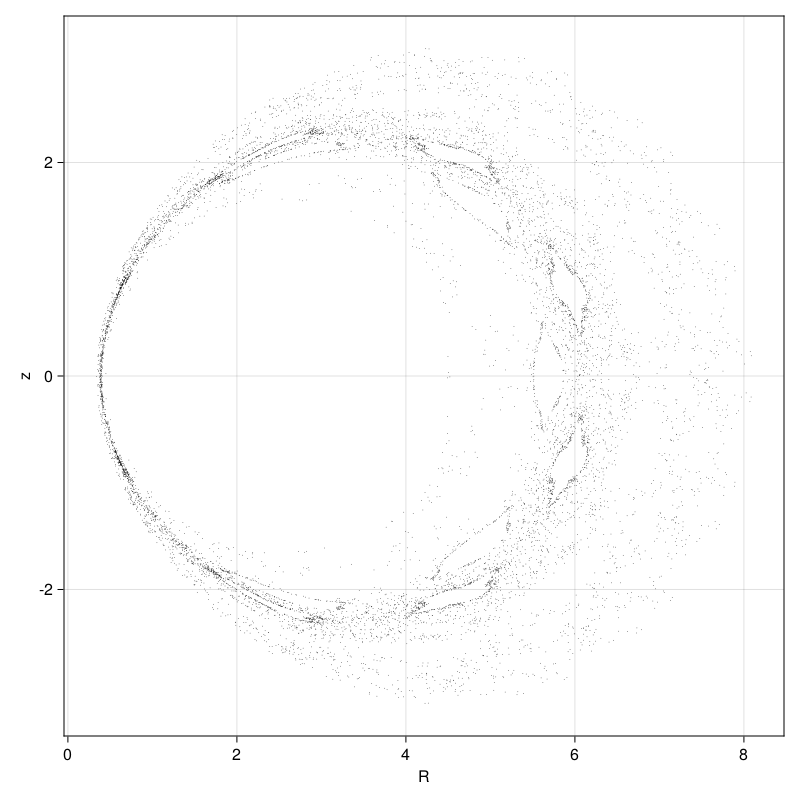}
        \caption{A thick stochastic region}
        \label{fig:stochastic_section}
    \end{subfigure}
    \caption{Orbits in stochastic regions.}
    \label{fig:stochastic_thin_section_and_thick_section}
\end{figure}

\begin{figure}
    \begin{subfigure}[b]{0.39\textwidth}
        \centering
        \includegraphics[width = \textwidth]{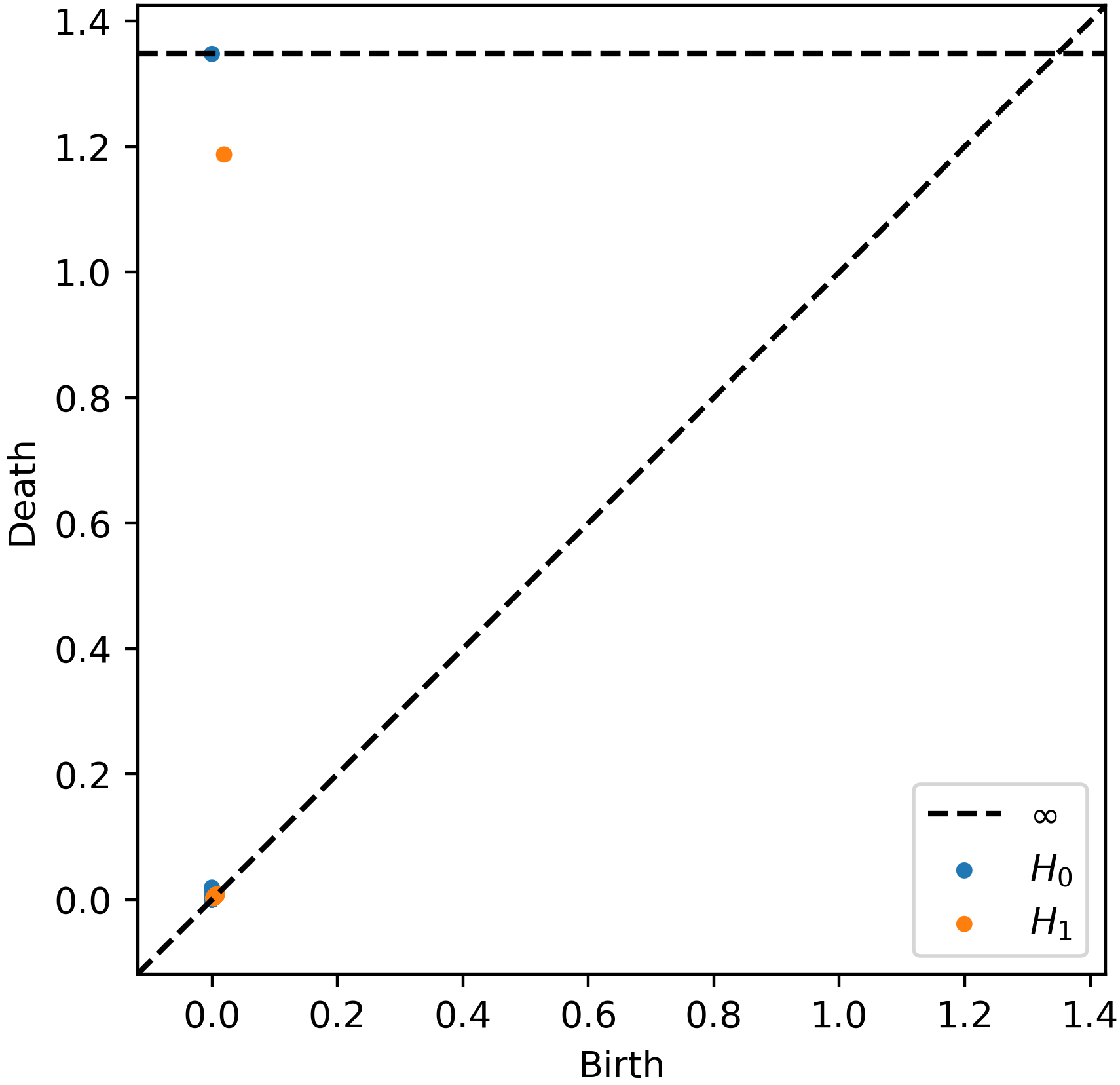}
        \subcaption{PD}
        \label{subfig:stochastic_thinRips}
    \end{subfigure}
    \centering
    \begin{subfigure}[b]{0.5\textwidth}
        \centering
        \includegraphics[width = \textwidth]{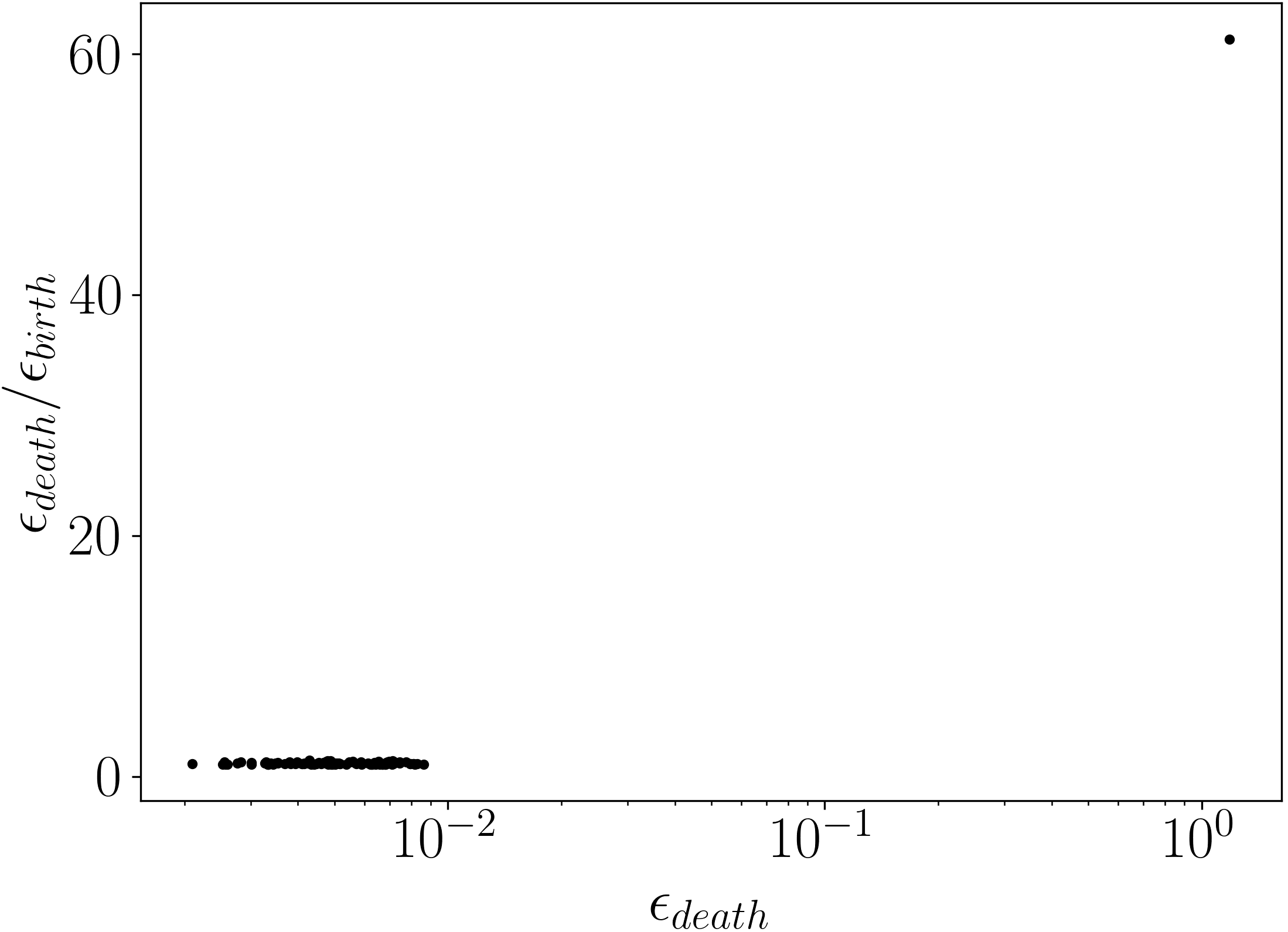}
        \caption{MPD}
    \label{fig:stochastic_thin_section_bdrat}
    \end{subfigure}
    \caption{Topological descriptors for subfigure \ref{subfig:stochastic_thinSection}.}
    \label{fig:stochastic_thin_section_and_rips}
\end{figure}

We now look at a case in which the topological descriptors struggle to delineate between two classes of structure, specifically a thin stochastic layer. These layers form between invariant tori and are very hard to distinguish from them both on inspection and numerically. Consider subfigure \ref{subfig:stochastic_thinSection} which presents one of these stochastic layers. This particular layer is bounded by invariant tori and actually contains a very thin banana island inside it. Subfigure \ref{subfig:stochastic_thinRips} presents the PD for this layer and it is on inspection very similar to that of \ref{subfig:torusRips}. Similarly, the MPD as shown in Figure \ref{fig:stochastic_thin_section_bdrat} contains only a cluster of classes very close to zero below $\epsilon_{\text{death}} = 10^{-2}$ and a single class with multiplicative persistence $\approx 60$ corresponding to the trajectory enclosing the axis. This is again very similar to the invariant torus case. It is evident that the VR persistent homology is not capable of clearly distinguishing a thin stochastic layer from an invariant torus and other methods are needed for this problem such as measurements of the Lyapunov exponent of the orbit, or the WBA chaos detection procedure discussed in \cite{SandersAndMeiss}. This demonstrates that our homological approach to classification is complementary to these more classical methods. In the sense that where they can, given long enough orbits, generically distinguish any periodic or quasi-periodic orbit from the aperiodic chaotic ones, our homological approach cannot. In contrast the classical chaos detection schemes cannot discern a difference between an invariant torus and an island chain, since they are both quasi-periodic, or a thin and thick stochastic layer, which are both aperiodic, but our homological approach can. Therefore, to create the most general and accurate automatic orbit classifier possible we would want to integrate both approaches. 

\subsubsection{Large stochastic region}\label{LargeStochasticExample}

For our final example we consider the case of a large stochastic region with many internal islands. Such an orbit is presented as subfigure \ref{fig:stochastic_section}. We observe that the orbit encloses the magnetic axis and hence we should expect, as above, that the last $H_1$ class to die will have a high multiplicative persistence. Since there are many islands enclosed by the stochastic region we should also expect that there will be many $H_1$ classes detected but few persistent $H_0$ classes. This is because the stochastic region itself is connected but not simply connected. That is, there are holes in the stochastic region, the islands, which ensure the first homology is nontrivial.

\begin{figure}[t]
    \centering
    \begin{subfigure}[b]{0.39\textwidth}
        \centering
        \includegraphics[width = \textwidth]{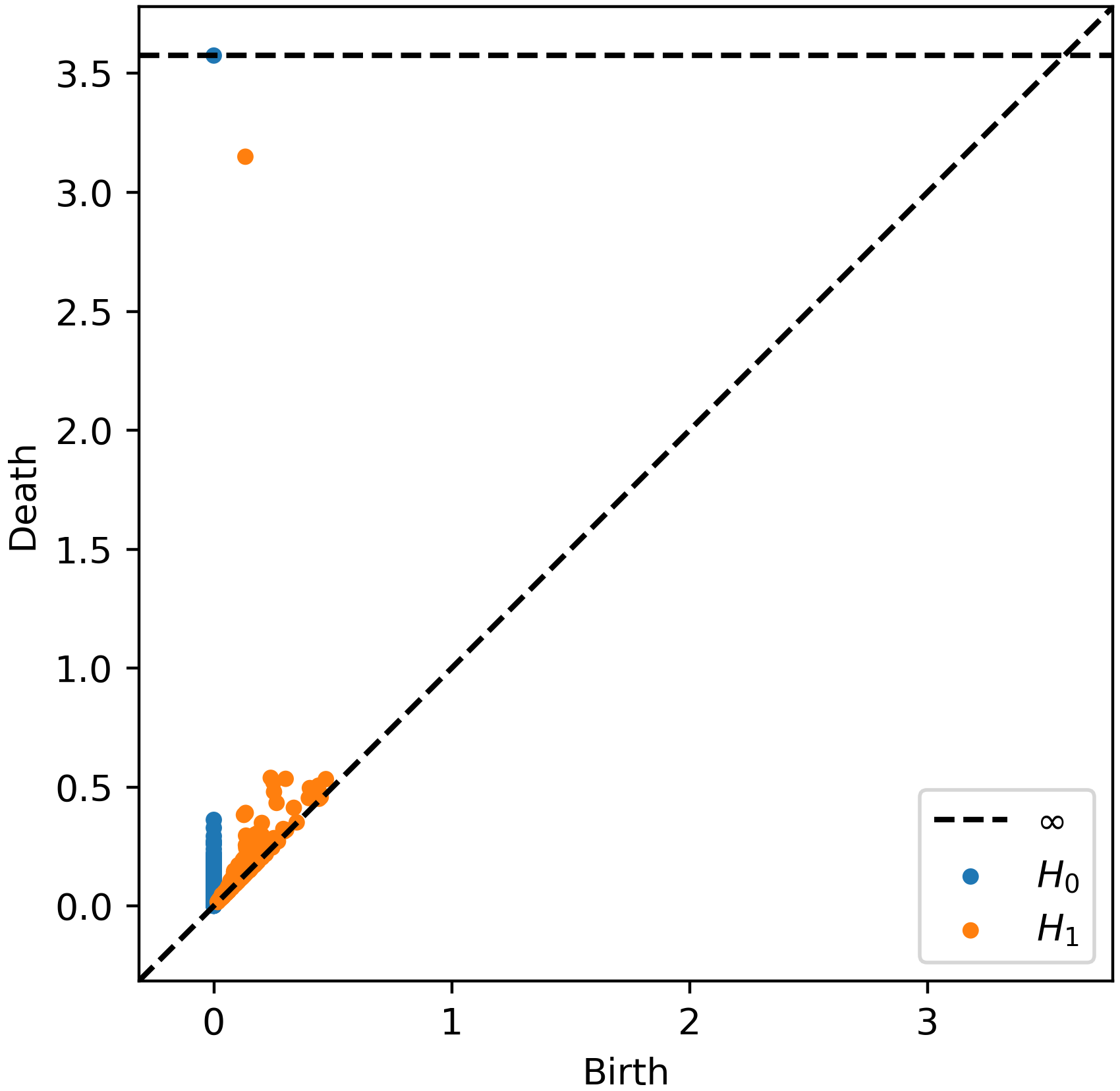}
        \subcaption{PD}
        \label{subfig:stochasticRipsPH}
    \end{subfigure}
    \begin{subfigure}[b]{0.50\textwidth}
        \centering
        \includegraphics[width = \textwidth]{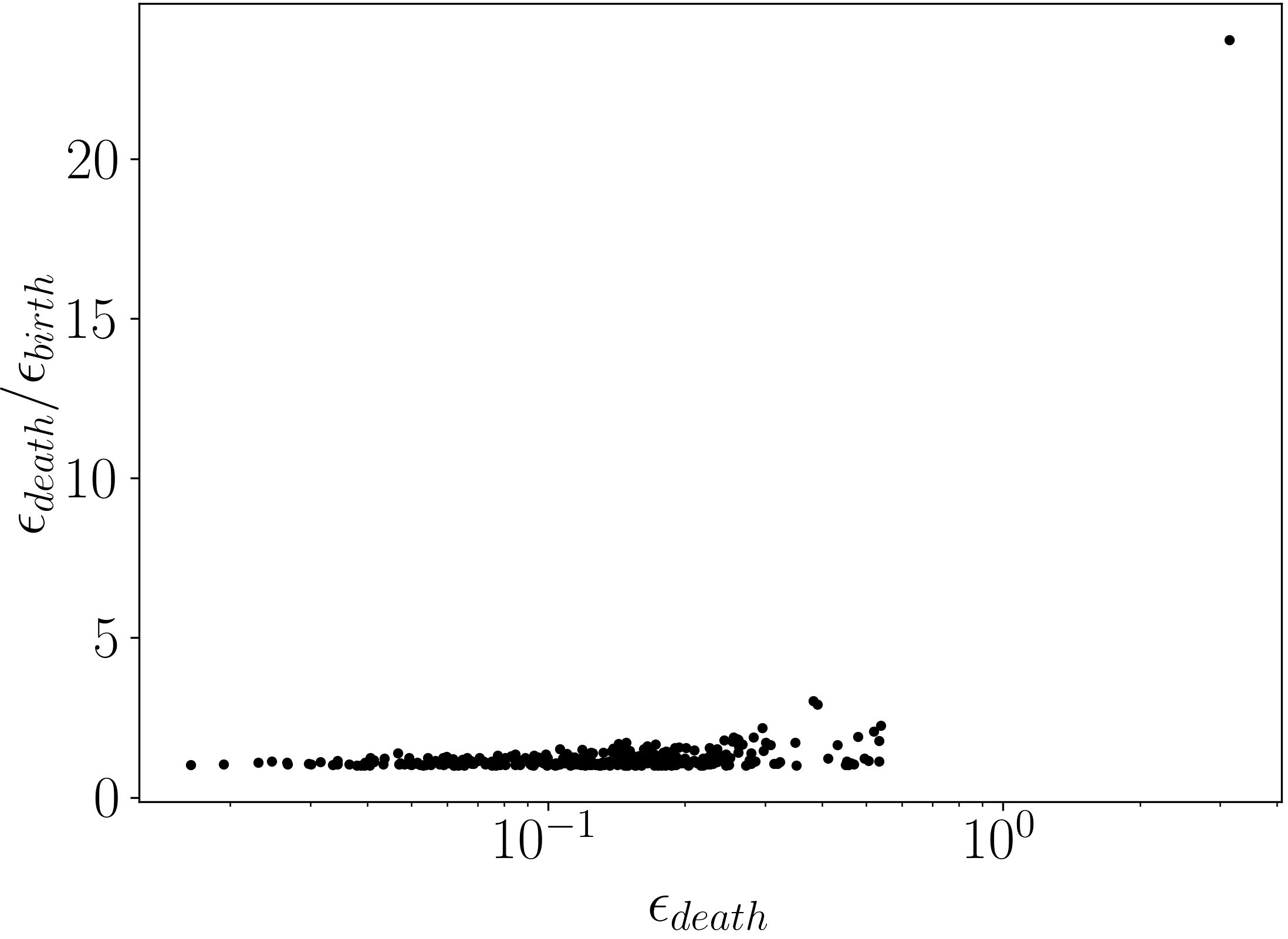}
        \subcaption{MPD}
    \end{subfigure}
    \caption{Topological descriptors for subfigure \ref{fig:stochastic_section} . }
    \label{fig:stochastic_section_rips}
\end{figure}

Looking at the topological descriptors, which are presented as Figure \ref{fig:stochastic_section_rips}, we see that they follow the expected behaviour noted above. We see the $H_1$ class we expect with high multiplicative persistence agreeing with our first hypothesis. Also, there is only one persistent $H_0$ class, corresponding to global connectivity, however we see many $H_1$ classes with death time above $\epsilon_{\text{death}} = 10^{-1}$ - recalling that for our invariant torus and thin stochastic layer there were no $H_1$ classes above this range - and multiplicative persistence of $\approx 2$ like those found in the island chains earlier. These $H_1$ classes are evidently associated to the many internal islands, agreeing with our second hypothesis.  

\subsection{An automated classification procedure}

Having analysed the $PH$ formation of different classes of magnetic field line orbit and connected geometric and topological features of the orbits to features in the persistent homology itself, we can proceed with developing a list of criteria which will allow us to automatically identify the class of an orbit only from its persistent homology. We will present methods based upon the observations made above for both distinguishing islands and islands chains from invariant tori and stochastic regions, and then for distinguishing different order island chains. We will also present the results of applying our classification criteria to the case of our toy tokamak model. 

\subsubsection{Distinguishing islands}\label{EnclosureDefining}

Recall that in the above section we observed that for all orbits except island chains the last to die $H_1$ class has a high multiplicative persistence. We related this to the observation that the islands are the only orbit to not form complete circles containing the magnetic axis in the Poincare section. This indicates that we can distinguish the Poincare section of an island chain from others by computing the multiplicative persistence of the last $H_1$ class to die. If $\epsilon_{\text{death}}/\epsilon_{\text{birth}} < \text{threshold}$ for said class, and a selected threshold, then we claim that the orbit corresponds to an island chain. 

We define a function $c_l$ mapping point clouds $X\subset \Sigma$ to the class in $PH_1(X)$ with the largest death time. That is we define it by
\begin{equation}
    c_l(X) = \underset{c\in PH_1(X) }{\argmax}\epsilon_{\text{death}}(c)\,.
\end{equation}
The multiplicative persistence of this class defines a number we will refer to as the \textit{enclosure}, $e$, of the  point cloud. That is we define
\begin{equation}
    e(X) = \frac{\epsilon_\text{death}(c_l(X))}{\epsilon_\text{birth}(c_l(X))}\,.
\end{equation}
To determine whether the orbit of a point belongs to the class of \textit{islands} or not we will check if the enclosure of the trajectory $X_T(x)$ is small, that is less than a chosen threshold. When it is small we declare the trajectory to lie in an island. 

Selecting a real number $e_\text{thresh}>0$, we define a set of points in $\Sigma$ called $IS(e_\text{thresh})$ by
\begin{equation}
    IS(e_\text{thresh}) = \left\{ x\in \Sigma |\, \ e(X_T(x))<e_\text{thresh} \right\}\,.
\end{equation}
We claim that if $e_\text{thresh}$ is selected correctly, $IS(e_\text{thresh})$ will be an acceptable approximation to the points whose orbits are islands. 

\begin{figure}
    \centering
        \includegraphics[width = 0.6\textwidth]{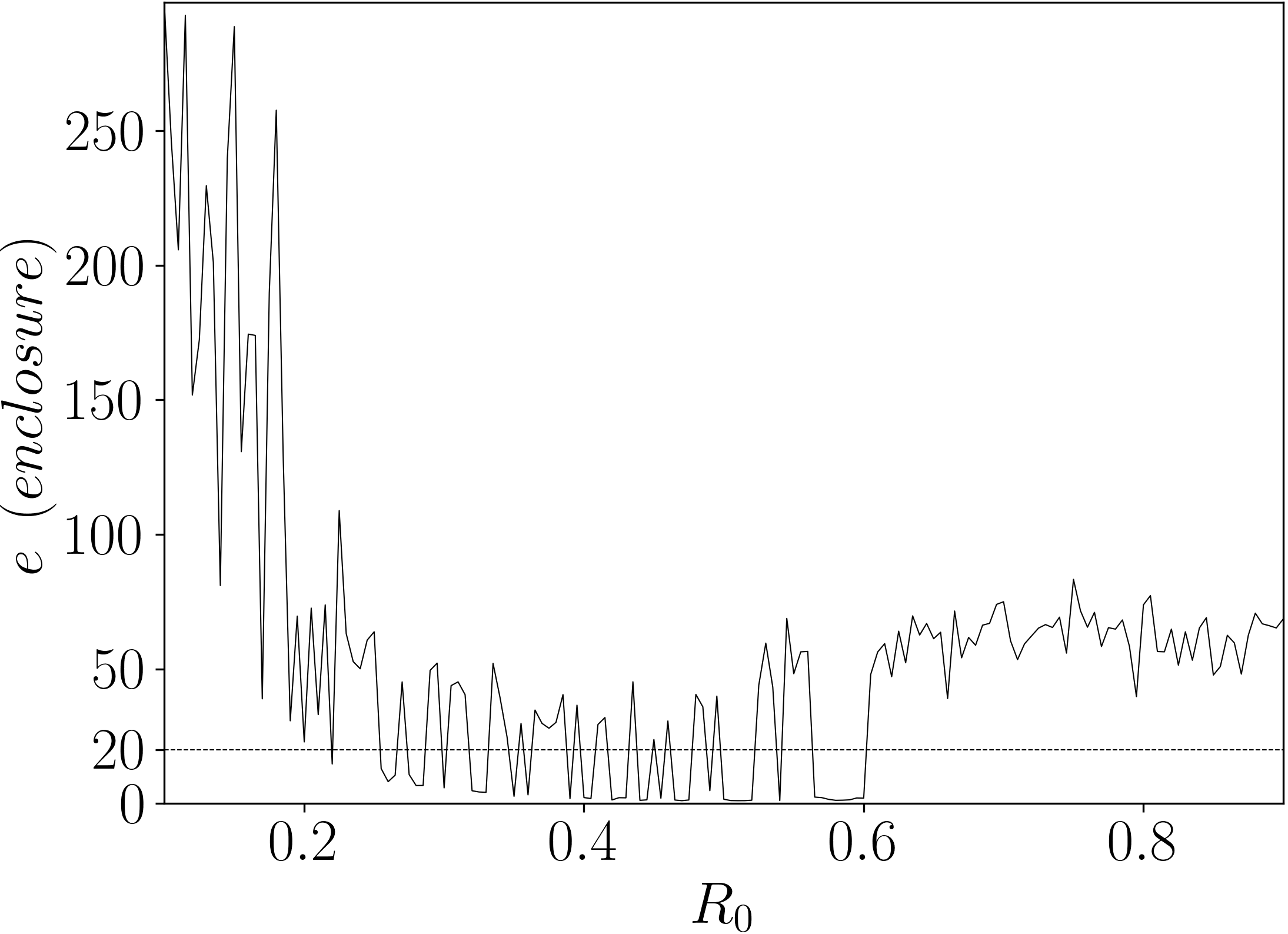}
    \caption{Enclosure as a function of the initial point $(R,Z)=(R_0,0)$ of the field line.}
    \label{fig:db_rat_radius}
\end{figure}

To demonstrate that this is a feasible concept we computed the enclosure $e(X_T(R,Z))$ for $T=2000$ point trajectories generated from $160$ initial points $(R,Z) = (R_0,0)$ with $R_0 \in [0.1,0.9]$. This set of initial conditions spans the left side of our toy tokamak. The plot of the calculated enclosure is presented as Figure \ref{fig:db_rat_radius}. We observe that the enclosure fluctuates as we vary $R_0$ but the values are all either above $e = 20$ or near zero. This indicates that if we choose our threshold at $\approx 20$ we can separate the trajectories into two classes which we recognise will be \textit{islands} and \textit{not islands}. Note that for $R_0>0.6$ the enclosure of all field lines appears to be large and so we would conclude that there are no magnetic islands near to the axis. This is expected as the magnetic axis is an elliptic fixed point of the Poincare map and so the field lines form an elliptic island around it comprised of invariant tori separated by thin stochastic layers, both of which have, as observed above, high enclosure. 

\begin{figure}
    \centering
        \includegraphics[width = 0.6\textwidth]{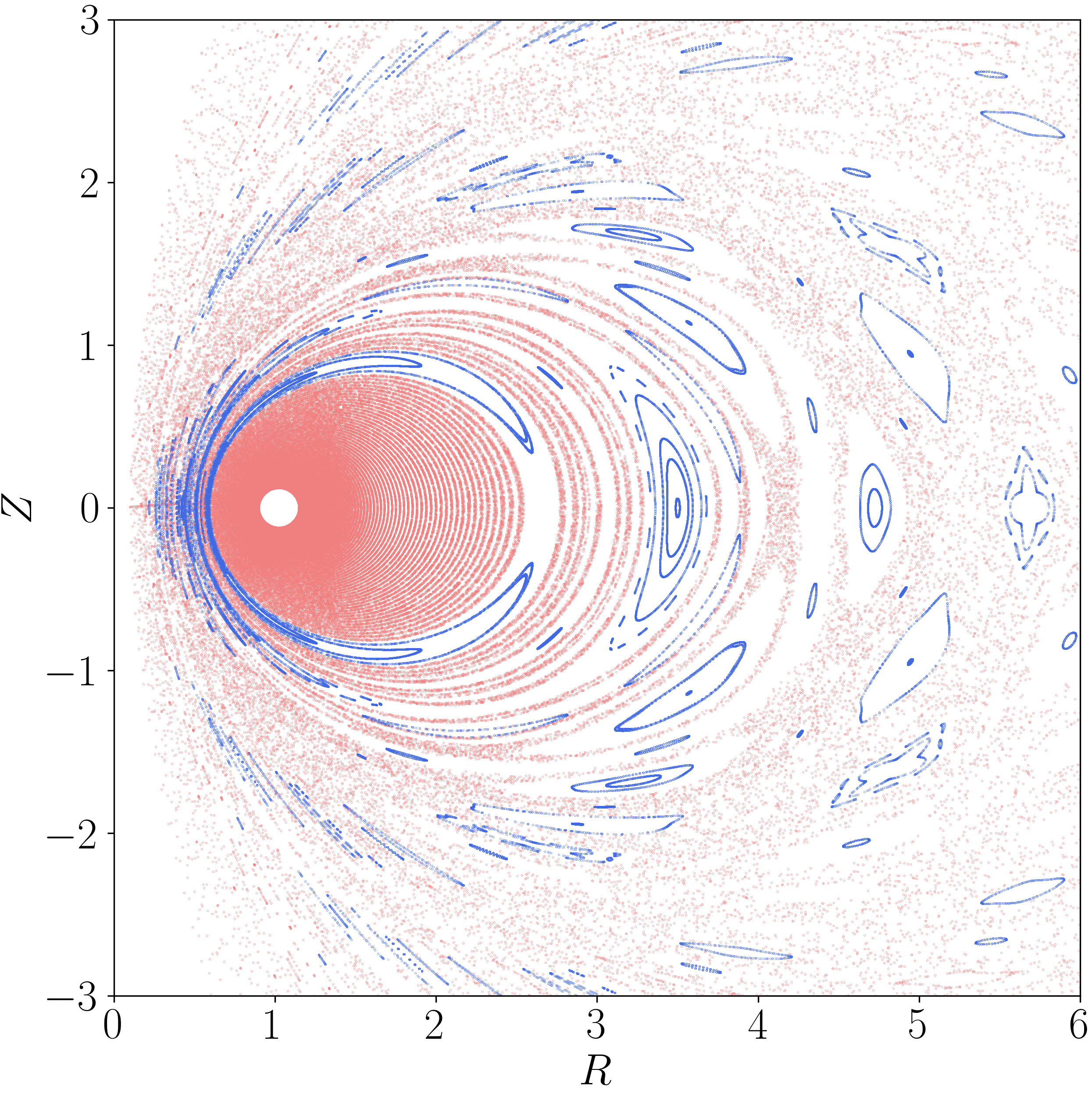}
    \caption{Orbit class map of islands detected by the condition $e_{thresh} = 20$. Islands are shown in blue and other classes in red.}
    \label{fig:toy_tokamak_islands_coloured1}
\end{figure}

To check that our classification scheme can correctly identify islands, at least up to the level of inspection, we now set our $e_\text{thresh} = 20$ and separate the initial conditions into those that belong to $IS(20)$ and those that do not. We now construct an image presenting the trajectories of all of our initial conditions. We colour the initial conditions in $IS(20)$ blue and the others in red. This produces the image presented as Figure \ref{fig:toy_tokamak_islands_coloured1}. Looking at this picture it appears, at least on inspection, that our classification scheme has acceptably separated the islands from the stochastic layers and invariant tori. 

\subsubsection{An approach to island counting}

Having demonstrated that we can identify the islands and island chains by their enclosure we now look at counting the number of major islands in each identified chain. We could achieve this by counting the number of $H_1$ classes. We get better results however by counting the number of $H_0$ classes. This is because we know that for an island chain, each connected component, and hence $H_0$ class, should correspond to a single island in the chain. 

\begin{figure}
    \centering
        \includegraphics[width = 0.65\textwidth]{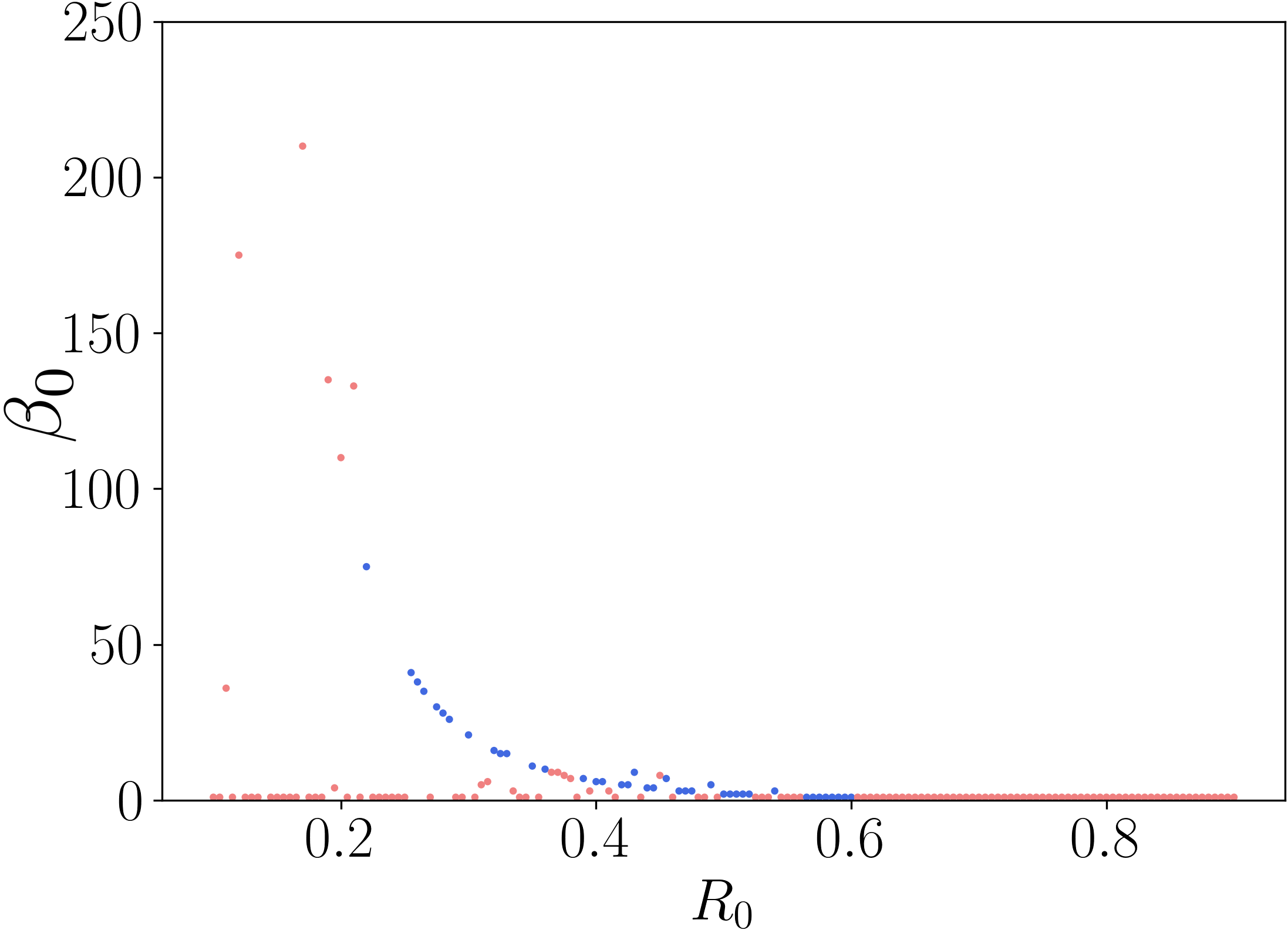}
    \caption{Scatter plot of the detected number of connected components in each orbit as a function of the initial radius for the case of $\epsilon_\text{thresh}  = 0.25$}
    \label{fig:betti0_radius}
\end{figure}

To count islands we cannot just compute the total number of $H_0$ classes in $PH_0(X_T(x))$. This is because $X_T(x)$ is, ignoring the case of periodic trajectories, a set of $T+1$ disconnected points and each of which provides a class to $PH_0(X_T(x))$ so the the total number of classes is always $T+1$. Instead we will compute the number of classes with absolute persistence greater than a chosen threshold. Specifically for a chosen $\epsilon_\text{thresh}>0$ define the $0$-th approximate Betti number of a point cloud $X$ as 
\begin{equation}
    \beta_0(X)  = \#\{ c \in PH_0(X) | \, \epsilon_\text{death}(c) > \epsilon_\text{thresh} \}\,.
\end{equation}

We can then compute $\beta_0(X_T(x))$ for each of our initial conditions along the $Z=0$ axis defined in the previous section. The result of this calculation for $\epsilon_{\text{thresh}}=0.25$ is presented as Figure \ref{fig:betti0_radius}. We have coloured points on this scatter plot blue if the underlying trajectory is classified as an island chain by its enclosure value, and red if not. We observe that the $\beta_0$ values for the islands appear to mostly lie on  a curve which increases as $R_0$ approaches zero. The $\beta_0$ of non-islands is either close to zero, for thin stochastic layers, or effectively random for large stochastic layers. A few red dots at small $R_0$ seem to lie on the blue curve. This is a result of a misclassification error in the enclosure based island detection scheme. Specifically, island chains with a very high number of major islands will be identified as invariant tori. 

Misclassification occurs when the ratio between the effective diameter of the island chain and the distance between adjacent major islands is greater than $e_{\text{thresh}}$. This allows us to estimate at what island count we should begin to see misclassification. As a rough approximation take that the point clouds are circles of radius $R$. Then let $n$ be the major island number and assume that the gaps between the islands are equal in size to the island themselves. This assumption enforces a \textit{worst case scenario} where the islands are already close to enclosing the axis. Then we can estimate the average distance between islands as of order $2\pi R/2n$. Misclassification will occur when the diameter $2R$ divided by this distance is greater than $e_{\text{thresh}}$
\begin{equation}
    \frac{2R}{2\pi R/2n} > e_\text{thresh}\,,
\end{equation}

\begin{equation}
    n >\frac{\pi}{2} e_{\text{thresh}} \approx \frac{3}{2} e_{\text{thresh}}\,.
\end{equation}
For the $e_\text{thresh} = 20$ value adopted earlier we have $n>30$. In practice this will be an underestimate of the island number at which misclassification occurs because field line orbits have circumferences larger than $2\pi R$ since they are not circular. This agrees with the observation that, in Figure \ref{fig:betti0_radius}, all red points which lie on the blue island curve have $\beta_0\approx n$ well above $30$. 

\begin{figure}
    \centering
    \begin{subfigure}[b]{0.49\textwidth}
        \centering
        \includegraphics[width = \textwidth]{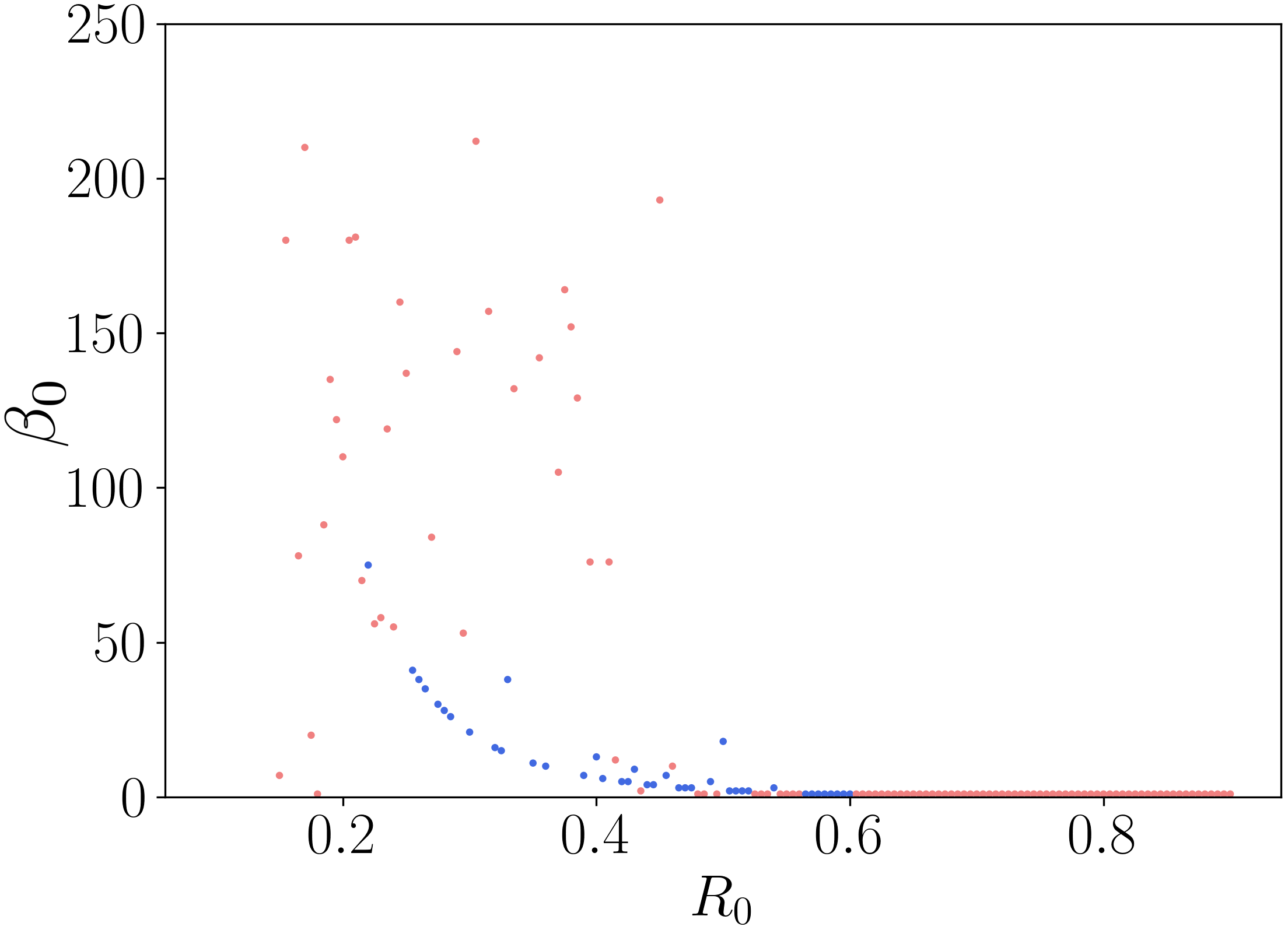}
        \subcaption{Case with $\epsilon_\text{thresh}  = 0.1$}
    \end{subfigure}
    \begin{subfigure}[b]{0.49\textwidth}
        \centering
        \includegraphics[width = \textwidth]{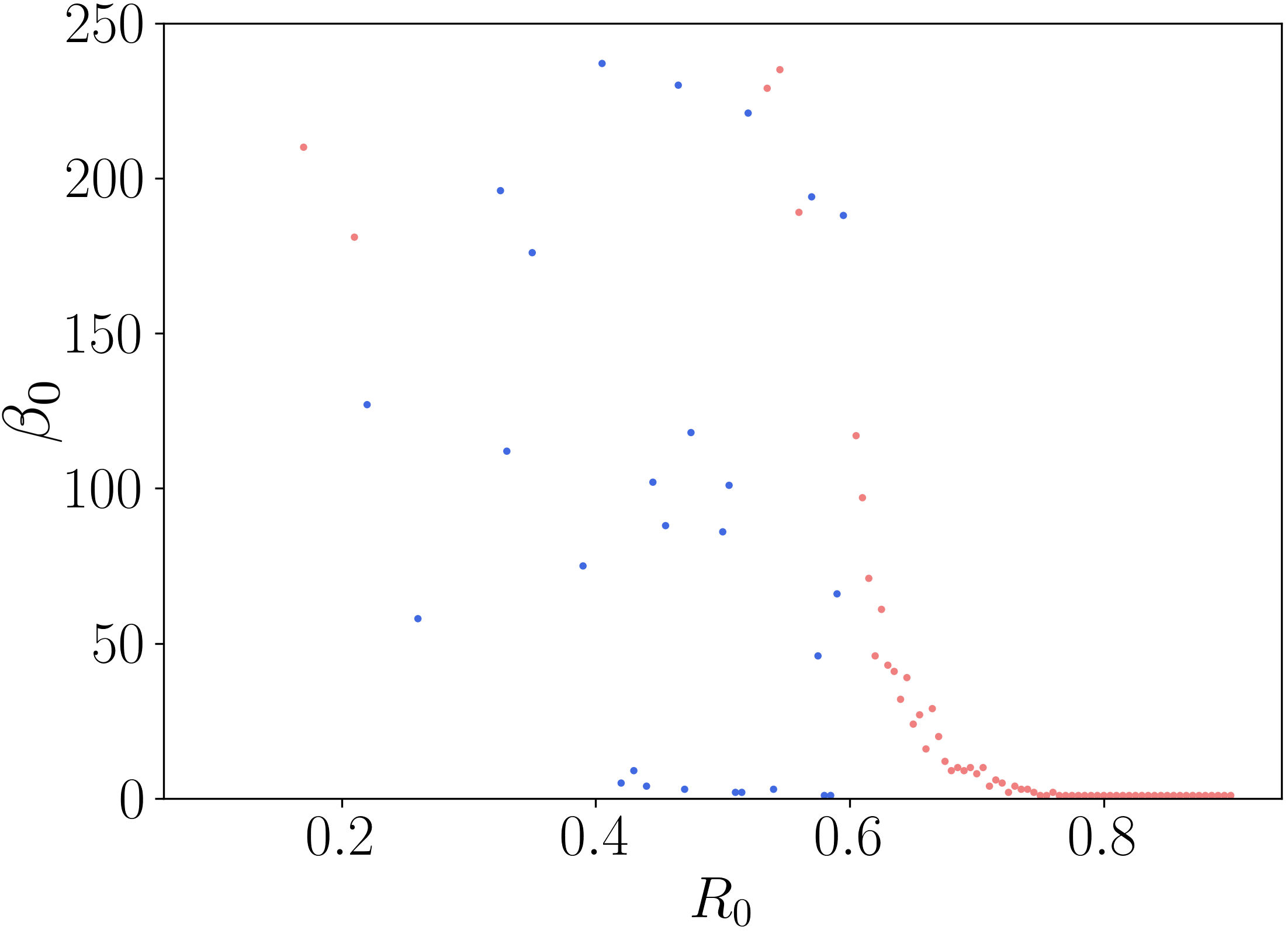}
        \subcaption{Case with $\epsilon_\text{thresh}  = 0.01$}
    \end{subfigure}
    \caption{Comparison of the Betti number plots with varying $\epsilon_{\text{thresh}}$. The curve of (a), which uses a higher threshold, is consistent with that of Figure \ref{fig:betti0_radius} while there is no curve emerging in (b) due to its threshold being to low to capture the island count correctly.}
    \label{fig:betti0_radius2}
\end{figure}

We now ask the question of what happens when we lessen the threshold $\epsilon_{\text{thresh}}$. Consider Figure \ref{fig:betti0_radius2} which presents the same plot of $\beta_0$ for two smaller values of the threshold. We observe that for $\epsilon_{\text{thresh}}=0.1$ the spread of Betti numbers of stochastic layers, the red points, increases dramatically but the detected number of components in the island orbits, the blue dots, is consistent with Figure \ref{fig:betti0_radius}. This suggests that when restricted to the case of orbits which are known to form island chains the detected Betti number is relatively insensitive to the specific choice of threshold. For an even lower threshold of $\epsilon_{\text{thresh}}=0.01$ the island curve disappears and the invariant tori near the magnetic axis, where $R_0\approx 0.7$, are detected as containing multiple components. This is inaccurate and suggest that the threshold has been selected too small. Clearly then as long as $\epsilon_{\text{thresh}}$ is chosen sufficiently large its precise value is not important.

\subsubsection{Finding stochastic layers filled with islands}\label{StochLayersWithIslands}

We saw above that we could discern the topological difference between island chains and stochastic layers using TDA but we also know that there are several different types of stochastic field lines, and we would prefer to also be able to distinguish them. This can be done and we demonstrate how in this subsection.

\begin{figure}
    \centering
    \begin{subfigure}[b]{0.49\textwidth}
        \centering
        \includegraphics[width = \textwidth]{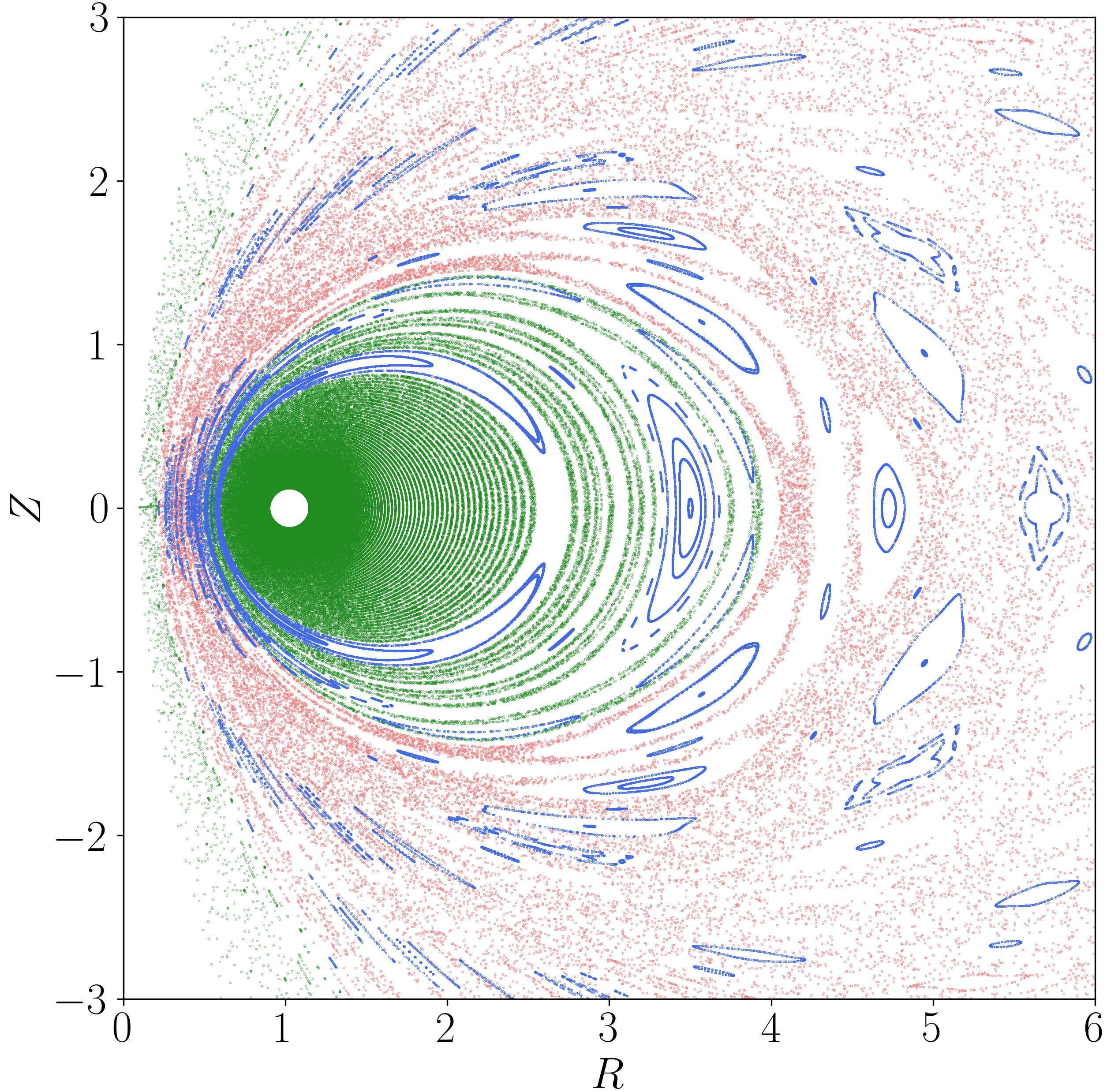}
        \caption{$\epsilon_{\text{thresh}}=10^{-1}$}
    \label{fig:toy_tokamak_islands_coloured2}
    \end{subfigure}
    \begin{subfigure}[b]{0.49\textwidth}
        \centering
        \includegraphics[width = \textwidth]{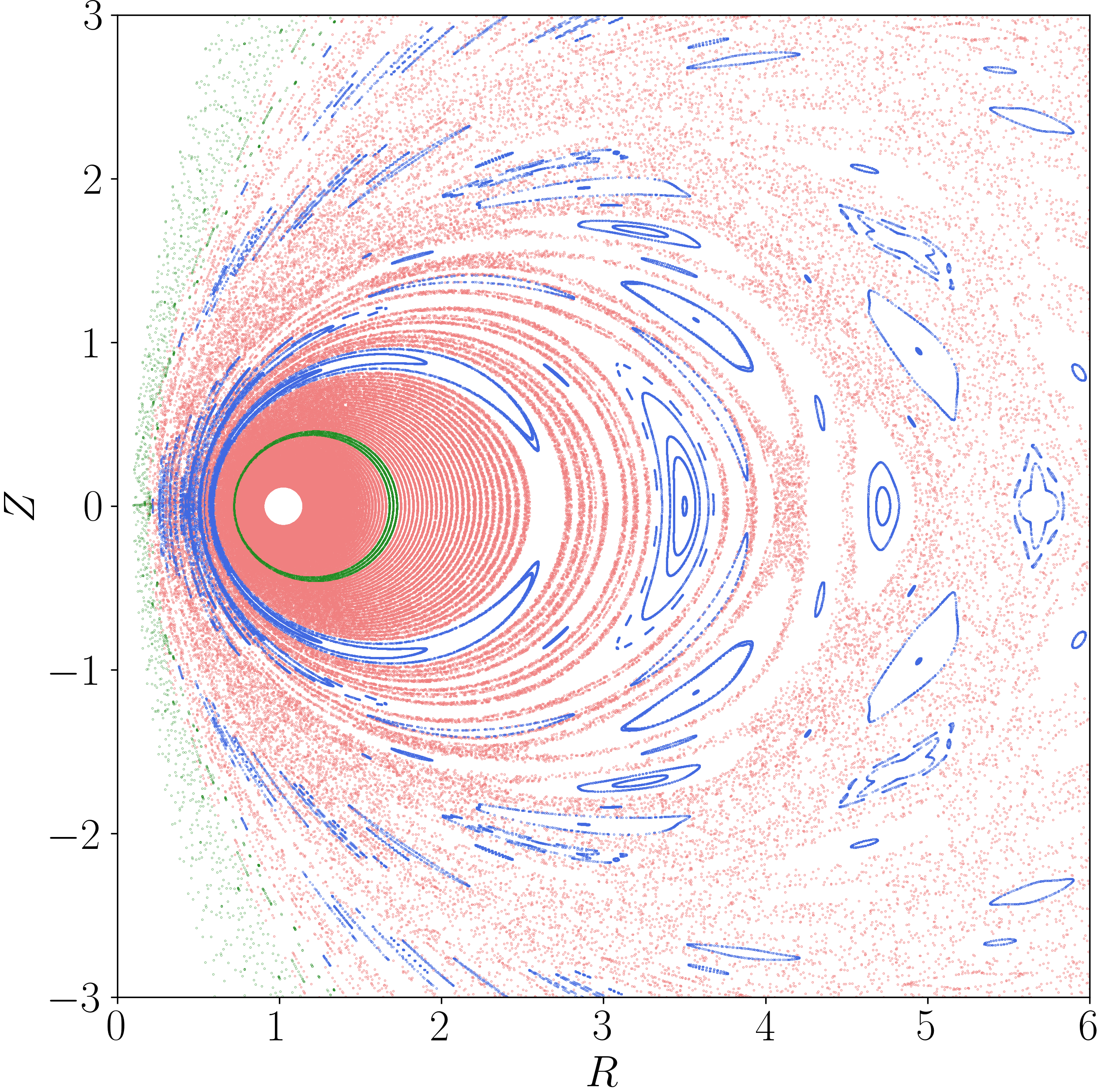}
    \caption{$\epsilon_\text{thresh}=\langle d(PH_0)\rangle$}
    \label{fig:toy_tokamak_islands_coloured3}
    \end{subfigure}
    \caption{Orbit class maps for the two different thresholding procedures. Blue, red, and green indicate islands, large stochastic layers, and thin stochastic layers and tori respectively. Note that in  subfigure \ref{fig:toy_tokamak_islands_coloured3}, only the orbits in the neighborhood of an invariant torus are green.}
\end{figure}



Recall from Section \ref{LargeStochasticExample} that $PH_1$ for a large stochastic trajectory, as opposed to a thin stochastic layer, contained many $H_1$ classes with death time $\epsilon_{\text{death}} \approx 10^{-1}$ well above the approximate spacing between points. This can be seen by analysing the persistence diagram presented as subfigure \ref{subfig:stochasticRipsPH} and noting that there exists many $H_1$ classes which die above the blue cluster which describes the deaths of $H_0$ classes. For all stochastic layers and invariant tori, at least one class must exist above this threshold, with death time of order $1$, the diameter of the surfaces of magnetic flux. So we can conclude that if the $PH_1$ contains more than one $H_1$ class which dies after $10^{-1}$ then the point cloud in question is most likely not a thin stochastic layer and hence is a large stochastic layer. Adopting this line of reasoning we construct a classification scheme as follows.

We define a function $ihc:\Sigma \cross \mathbb{R}^+\rightarrow \mathbb{N}$, the \textit{internal hole count}, as the number of $H_1$ classes in the $PH_1$ of the trajectory generated by the input point. That is we define
\begin{equation}
    ihc(x,\epsilon_{\text{thresh}}) = \# \{ c \in PH_1(X_T(x))| \, \ \epsilon_{\text{death}}(c) >\epsilon_{\text{thresh}} \}\,.
\end{equation}
Then the set of points lying on large stochastic layers is defined to be 
\begin{equation}
    LS(e_{\text{thresh}},\epsilon_{\text{thresh}}) = \{x \in \Sigma |\, \ x \notin IS(e_{\text{thresh}})\, \ \text{and}\, \ ihc(x, \epsilon_{\text{thresh}}) > 1\}\,.
\end{equation}

We now produce a new version of the orbit class map from Figure \ref{fig:toy_tokamak_islands_coloured1} but with the orbits whose initial point is in $LS$ coloured differently. This is presented as subfigure \ref{fig:toy_tokamak_islands_coloured2} where the points in $LS$ are presented in red and other stochastic layers are presented in green. We observe the expected behaviour that the large stochastic layers form a bounded layer between thin stochastic layers which lie near to the magnetic axis, $R_0 =1$ and invariant tori near to what would serve as the effective boundary of a machine $R_0 < 0.1$. We see that the trajectories which are coloured green form disconnected thin layers each separated either by a invariant torus or an island chain. This confirms that the $ihc$ metric is capable of distinguishing the large stochastic layers from thin layers and invariant tori. 

The method described above works but does not come equipped with a method for determining the best $\epsilon_\text{thresh}$. The threshold can be seen as a free parameter which can be adjusted by a user to select for thinner and thinner stochastic layers. However, we would prefer an automated procedure to obtain it. One possible approach is to determine $\epsilon_\text{thresh}$ using other information contained in the $PH$ of the orbit. We noted above that to distinguish stochastic layers we should search for $H_1$ classes representing holes which are larger than the spacing between points. This spacing can be estimated as the mean of the death times in $PH_0$, excluding the class which does not die. We refer to this mean as $\langle d(PH_0)\rangle$. Note that the death times of $H_0$ classes are edge lengths of a minimal spanning tree and their mean overestimates the average nearest neighbor distance. Thresholding with the $\langle d(PH_0)\rangle$ gives an automated procedure to generate $\epsilon_\text{thresh}$ but comes at the cost that the threshold will be different for each orbit, although this may be advantageous since the varying spacing of the points in different orbits is being compensated for. 

The results of constructing an orbit class map using $\langle d(PH_0)\rangle$ as $\epsilon_\text{thresh}$ is shown in subfigure \ref{fig:toy_tokamak_islands_coloured3}. Many thin stochastic layers have been reclassified as large stochastic layers and the only remaining green point clouds form a very thin layer around what is actually the only true invariant torus in the set of orbits included in the diagram. Hence we observe that by using $\langle d(PH_0)\rangle$ as a threshold we can partially distinguish invariant tori from thin stochastic layers by their persistent homology. However, this result is very preliminary and our confidence in this particular identification scheme is low.

\section{Persistent homology in straight field line coordinates}

 \label{Chapter4} 

\subsection{An under-counting problem}

The advantage of our homological approach to automated field line classification when compared to the previous graph theoretic approaches \cite{YipDissertation} is the ability for us to count internal holes in chaotic layers as statistically significant classes in $PH_1$. This allows us to infer the existence of an island chain without finding an orbit which is actually within the island chain. This is not possible when only considering MSTs, which measure only the connectivity of the point cloud, because the chaotic orbits are connected. 

\begin{figure}
    \centering
        \includegraphics[width = 0.5\textwidth]{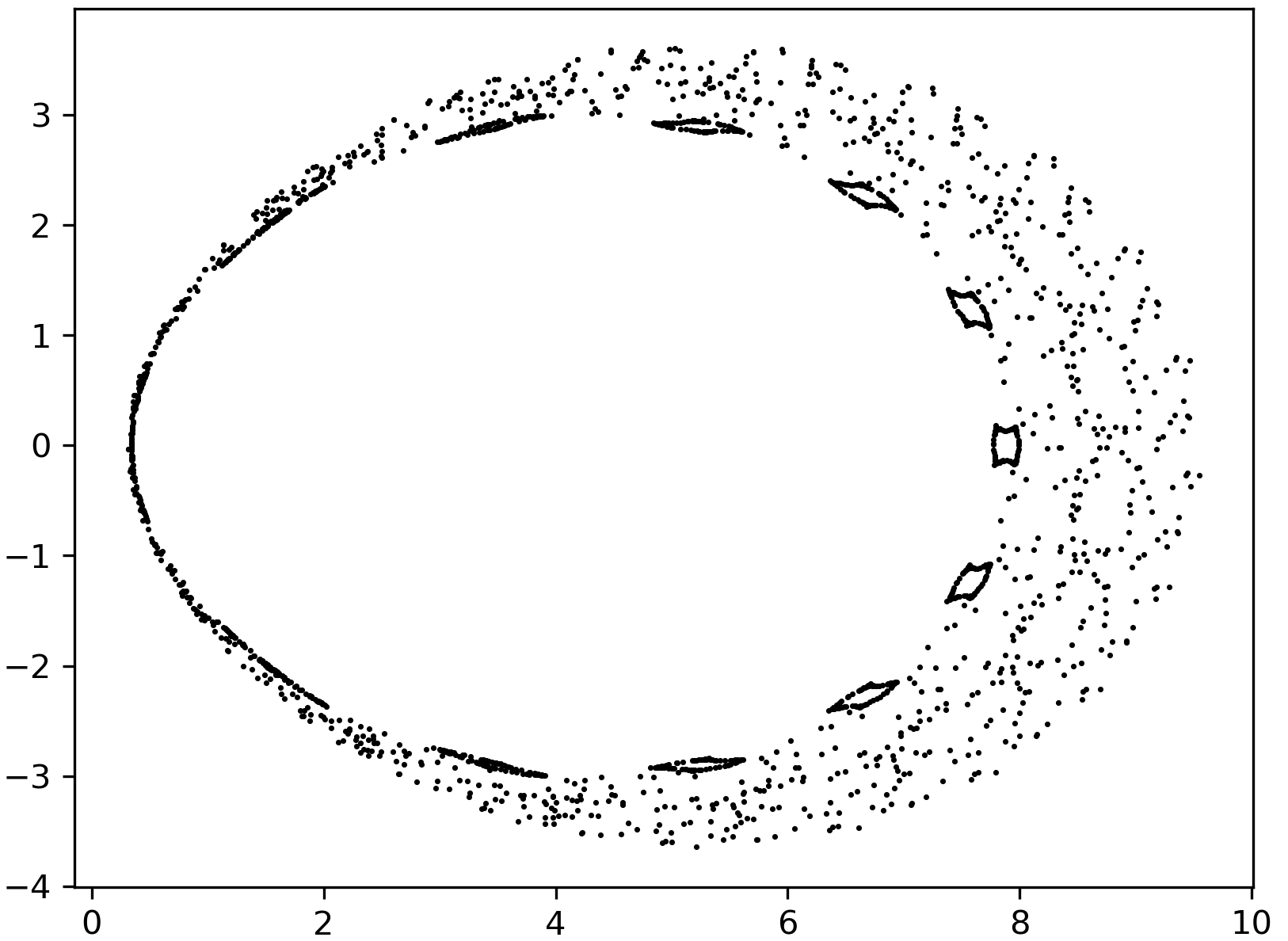}
    \caption{Chaotic orbit surrounding a $12$ island chain.} 
    \label{fig:StochasticLayerInternalIslands}
\end{figure}

However, if we actually attempt to directly use the $PH_1$ information to count the number of islands inside a chaotic layer we find that we usually under-count the number of islands. To see this, consider the chaotic orbit shown as Figure \ref{fig:StochasticLayerInternalIslands}. A chain of islands which the orbit passes very close to sticks out as the darkly outlined voids. As a note, there are 12 islands in this chain.

\begin{figure}
    \centering
        \includegraphics[width = 0.8\textwidth]{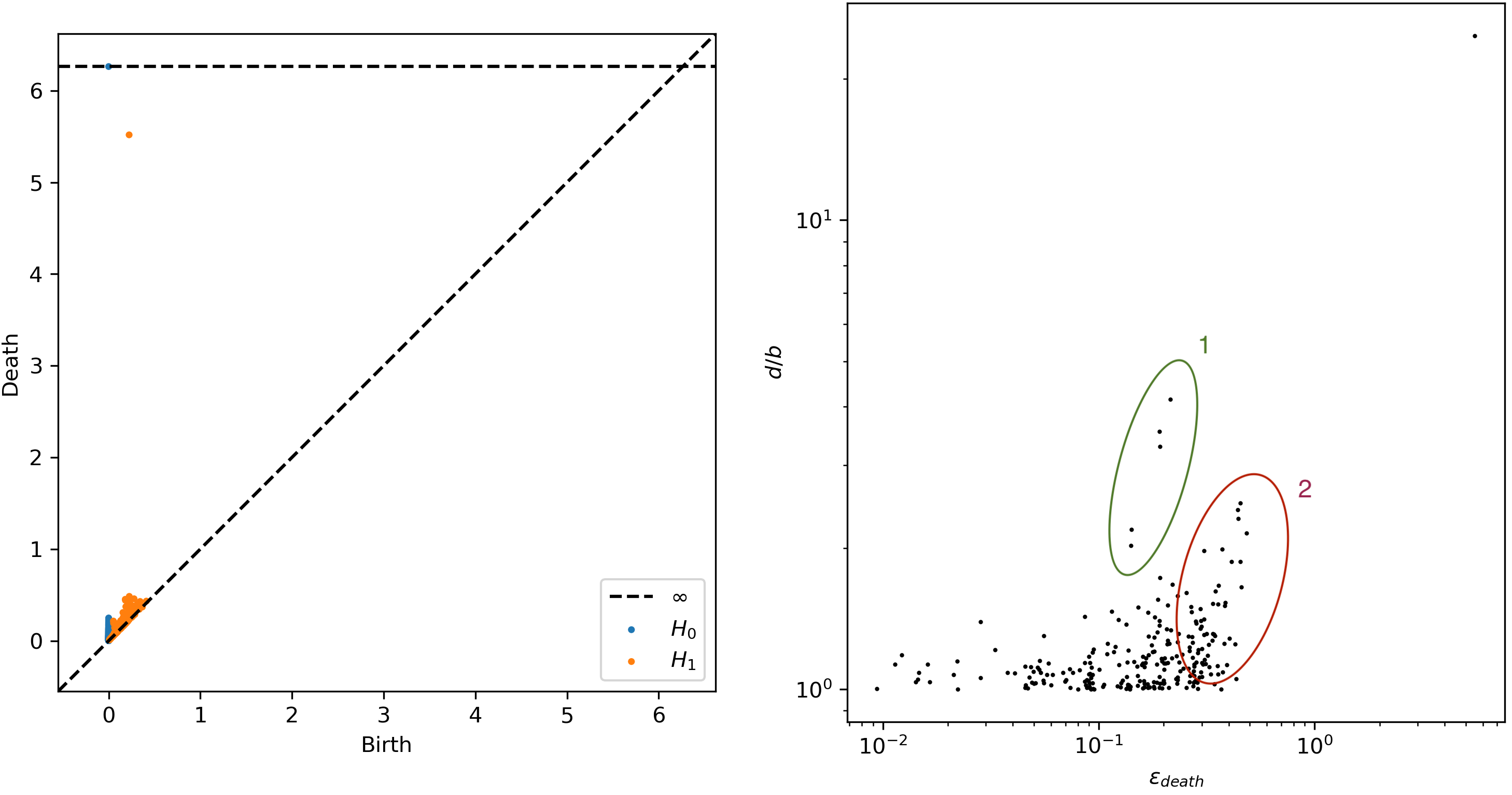}
    \caption{Topological descriptors for Figure \ref{fig:StochasticLayerInternalIslands}.}
    \label{fig:StochasticLayerInternalIslands_Data}
\end{figure}

Computing the $PH_1$ for this orbit we obtain the topological descriptors shown as Figure \ref{fig:StochasticLayerInternalIslands_Data}. Two clusters are visible in the MPD. Cluster 1 conatins five $PH_1$ classes with death time in $[0.08,0.1]$ and $\epsilon_{death}/\epsilon_{birth} > 2$ and it corresponds to the five islands on the outboard side. The other islands in the chain do not appear as classes separated from noise. So, we see that the $PH_1$ information is not accurately counting the number of islands in the chain the orbit bounds. Cluster 2 is associated to more subtle features and we will return to it later.

We can explain why the undercounting occurs by considering Figure \ref{fig:StochasticLayerInternalIslands} again. Note that the five islands which were detected correctly are the least elongated of the islands in the chain. The islands near to the inboard side are so deformed that they appear as almost straight lines. As a result, the homology classes corresponding to them die very early in the VR filtration, since classes die at $\epsilon$ on the order of the minor radius of the loop. Therefore they have small $\epsilon_{death}/\epsilon_{birth}$ and hence are not seen as statistically significant. If we want to use persistent homology to obtain an accurate count of the number of internal islands we need to solve this problem. 

One possible solution is to find a transform on the space which stretches out the elongated islands into more circular loops. Or more precisely, which transforms all of the islands in a chain into a single uniform shape. The persistent homology could then be computed in the transformed coordinates and a more accurate count on the number of islands would be obtained. The relevant question to ask then is ``does such a transformation exist?''. The answer to this is yes and finding it corresponds only to constructing the straight field line, or action-angle, coordinates for the integrable field of which ours is a perturbation.

\subsection{Straight field line coordinates}

Consider the case of the field generated by the currents shown in Figure \ref{fig:ToyTokamakModel} but with $\delta=0$. The current distribution, and hence the field is axisymmetric. As a result its field line Hamiltonian describes an integrable system and so action-angle coordinates exist for the field \cite{goldstein2011classical,hazeltine2003plasma,berry2020regular}. In a plasma physics context these are usually referred to as straight field line coordinates. 

\begin{figure}
    \centering
        \includegraphics[width = 0.6\textwidth]{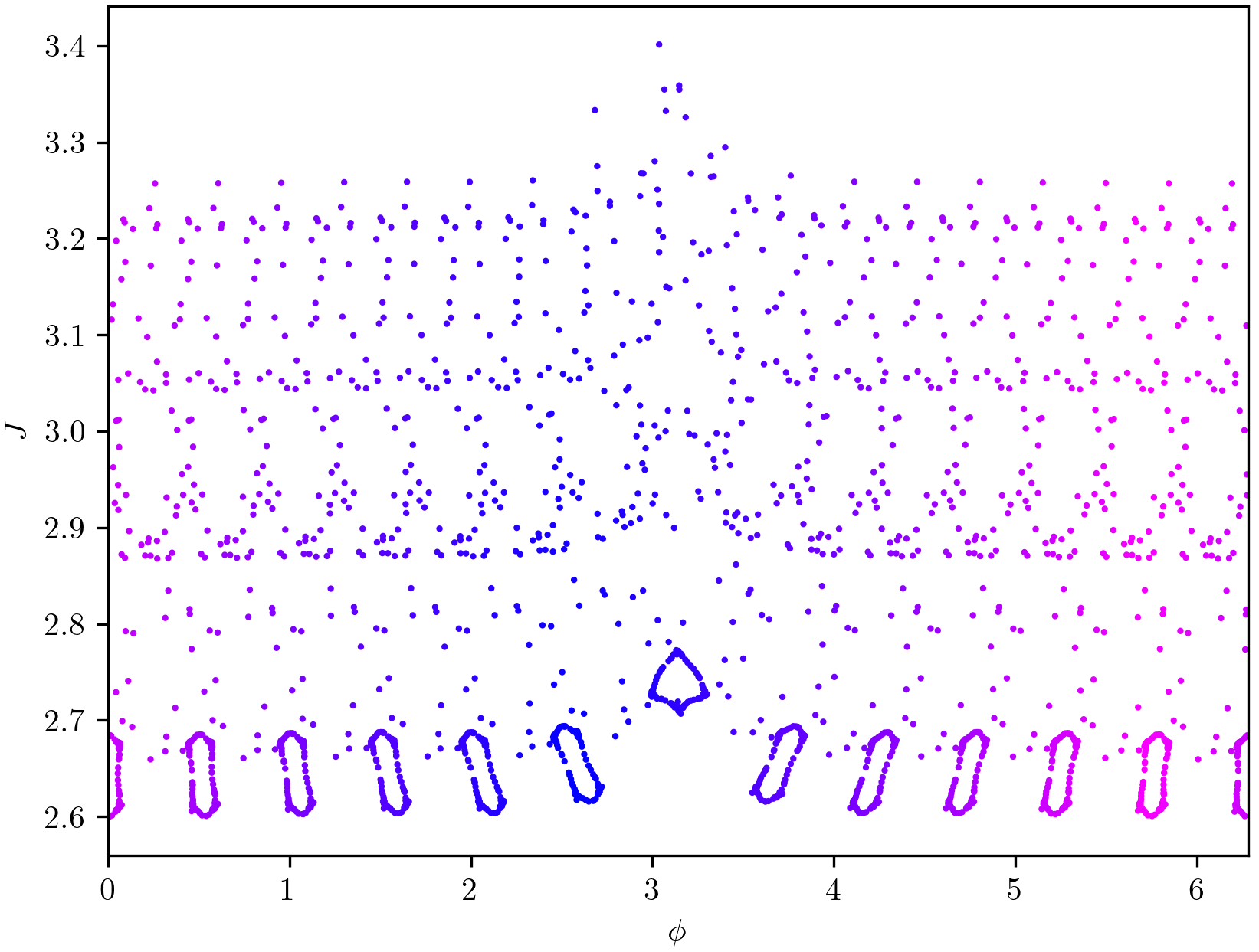}
    \caption{Chaotic orbit surrounding a $12$ island chain in straight field line coordinates. The colouring indicates the geodesic distance from to the initial point of the orbit. The periodicity is evident from the similarity of the colour on the left and right edges which are identified.}
    \label{fig:StochasticLayerInternalIslands_Straight}
\end{figure}

Since we have the exact for of the unperturbed field in cylindrical coordinates $(R,Z)$ it is possible to construct the transform into action-angle coordinates $(J,\phi)$ numerically. A brief discussion of this calculation is included in \ref{AppendixB}. Applying the transform to the point cloud in Figure \ref{fig:StochasticLayerInternalIslands} produces the cloud shown as Figure \ref{fig:StochasticLayerInternalIslands_Straight}. Note that, in the original coordinates the point cloud was a subset of $\mathbb{R}^2$, after the transformation it is a subset of a topologically cylindrical abstract phase space $\mathbb{R}\times S^1$. As a result Figure \ref{fig:StochasticLayerInternalIslands_Straight} should be interpreted with periodic boundary conditions, that is which the left and right edges identified.

Recall that, in order to calculate the VR Persistent Homology over a dataset it needs to be equipped with a distance function. In the $(R,Z)$ coordinates our dataset inherited the geometric distance on $\mathbb{R}^2$ itself because we were considering our Poincare section as a planar slice of Euclidean space. In contrast the $(J,\phi)$ space does not come equipped with a distance function as directly. This is common problem in TDA where we often want to measure the distance between points in abstract spaces which have no natural distance function, and the convention is to just artificially equip the space with a distance. Following the procedure, we use the trivial metric tensor $ds^2=dJ^2+d\phi^2$, which generates the same distance function as taking geodesic distances induced by embedding the space as a unit-radius cylinder in $\mathbb{R}^3$. 

\begin{figure}
    \centering
        \includegraphics[width = 0.8\textwidth]{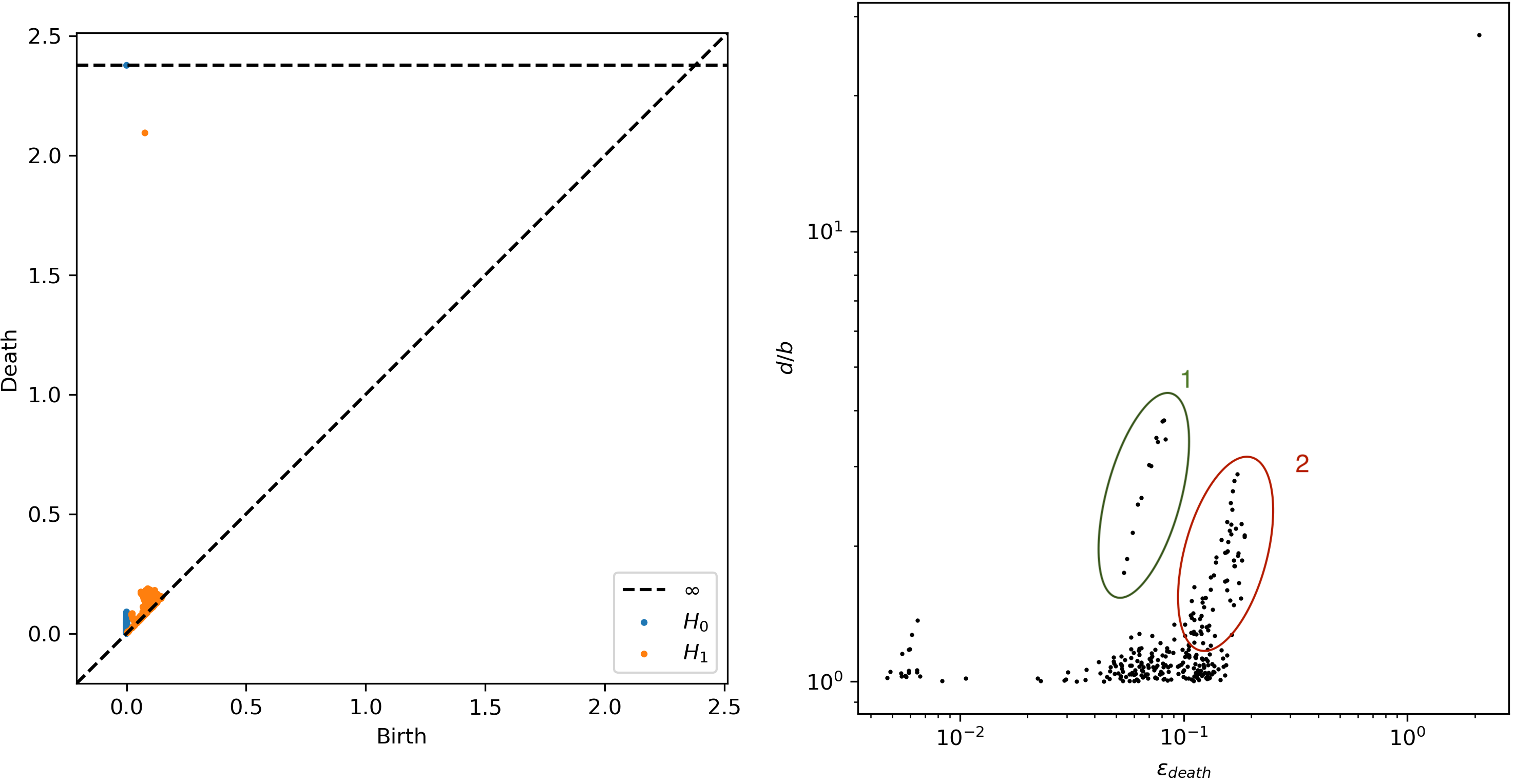}
    \caption{Topological descriptors for Figure \ref{fig:StochasticLayerInternalIslands_Straight}.}
    \label{fig:StochasticLayerInternalIslands_Data_Straight}
\end{figure}

Equipped with a distance metric we can compute the VR persistent homology of the point cloud shown in Figure \ref{fig:StochasticLayerInternalIslands_Straight}. This yields the topological descriptors presented as Figure \ref{fig:StochasticLayerInternalIslands_Data_Straight}. Three features are most notable. Firstly, on the PD there exists a persistent $H_1$ class with death time $\sim 2.1$, we will revisit this in the following section, and the largest multiplicative persistence of any class. It corresponds to the global topology of the cylindrical space and so analogous to the class used to define the enclosure of the orbits in Section \ref{EnclosureDefining}. Secondly, as in Figure \ref{fig:StochasticLayerInternalIslands_Data}, there is a cluster on the MPD, cluster $1$, corresponding to the primary island chain inside the chaotic orbit, but it is more prominent and now contains $12$ distinct classes, corresponding correctly to the $12$ islands in the chain. Finally, there is a prominent secondary cluster of classes, cluster $2$, corresponding to a second chain of $15$ islands which the orbit bounds. This island chain was effectively imperceptible to a human in the $(R,Z)$ coordinates but can be seen clearly in Figure \ref{fig:StochasticLayerInternalIslands_Straight} and appeared subtly as cluster $2$ in Figure \ref{fig:StochasticLayerInternalIslands_Data}. This cluster is not as well outlined by the chaotic orbit and as a result its homology classes are less statistically significant. For this reason the cluster overlaps substantially with the noise contribution to the MPD. 

The above numerical experiment provides evidence that the VR persistent homology is a more accurate tool for counting the number of islands embedded in a chaotic orbit when there exists a transformation into the straight field line coordinates of an integrable field near to the given chaotic field. In these coordinates, the islands necessarily all have a very similar shape because they exist near surfaces of constant $J$ and area enclosed by the islands must be the same since the Poincare map is area-preserving in canonical coordinates. As such, the homology classes associated to the islands have very similar birth and death times and so the island chain will correspond to a tight cluster on the multiplicative persistence scatter plot. This cluster can then be extracted, using any clustering algorithm of choice, and its points counted to give a more accurate count of the number of islands in the chain than was possible in the original physical coordinates. More generally, we can conclude that the VR persistent homology of the orbit in straight field line coordinates is a more useful topological descriptor for the field line orbit than that constructed using the physical coordinates.

\subsection{A global measure of non-integrability}

In this subsection we discuss the possibility of using the VR persistent homology of the Poincare map orbits of a field as a measure of the \textit{distance to integrability} of a given perturbed magnetic field. The measure we will construct is global, that is, it measures the failure of the whole Hamiltonian to be integrable throughout the whole phase space. This contrasts it to local measures of non-integrability such as measuring the time average Lyapunov exponents of individual orbits. 

\begin{figure}
    \centering
    \begin{subfigure}[b]{0.4\textwidth}
        \centering
        \includegraphics[width = \textwidth]{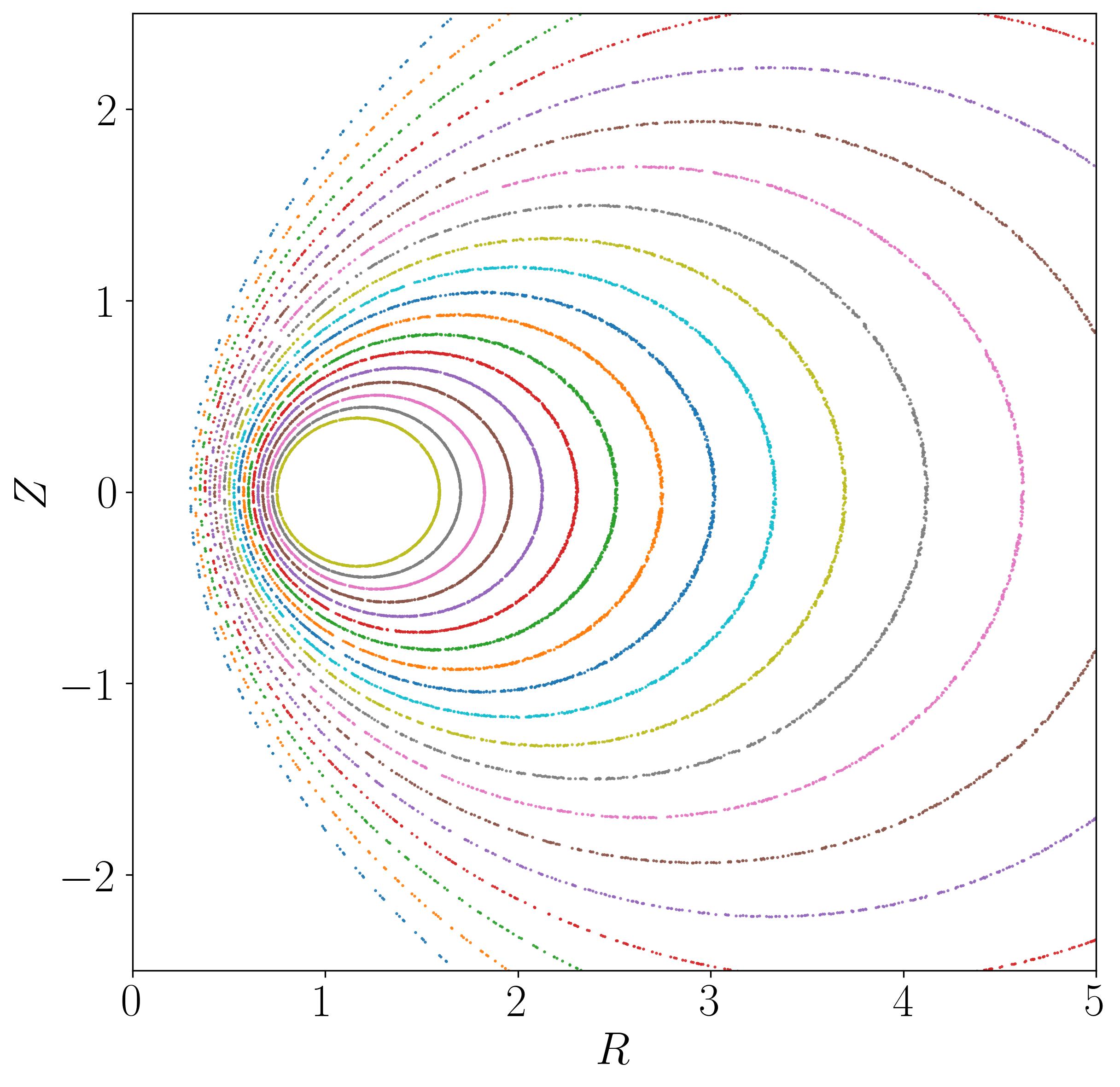}
    \end{subfigure}
    \begin{subfigure}[b]{0.4\textwidth}
        \centering
        \includegraphics[width = \textwidth]{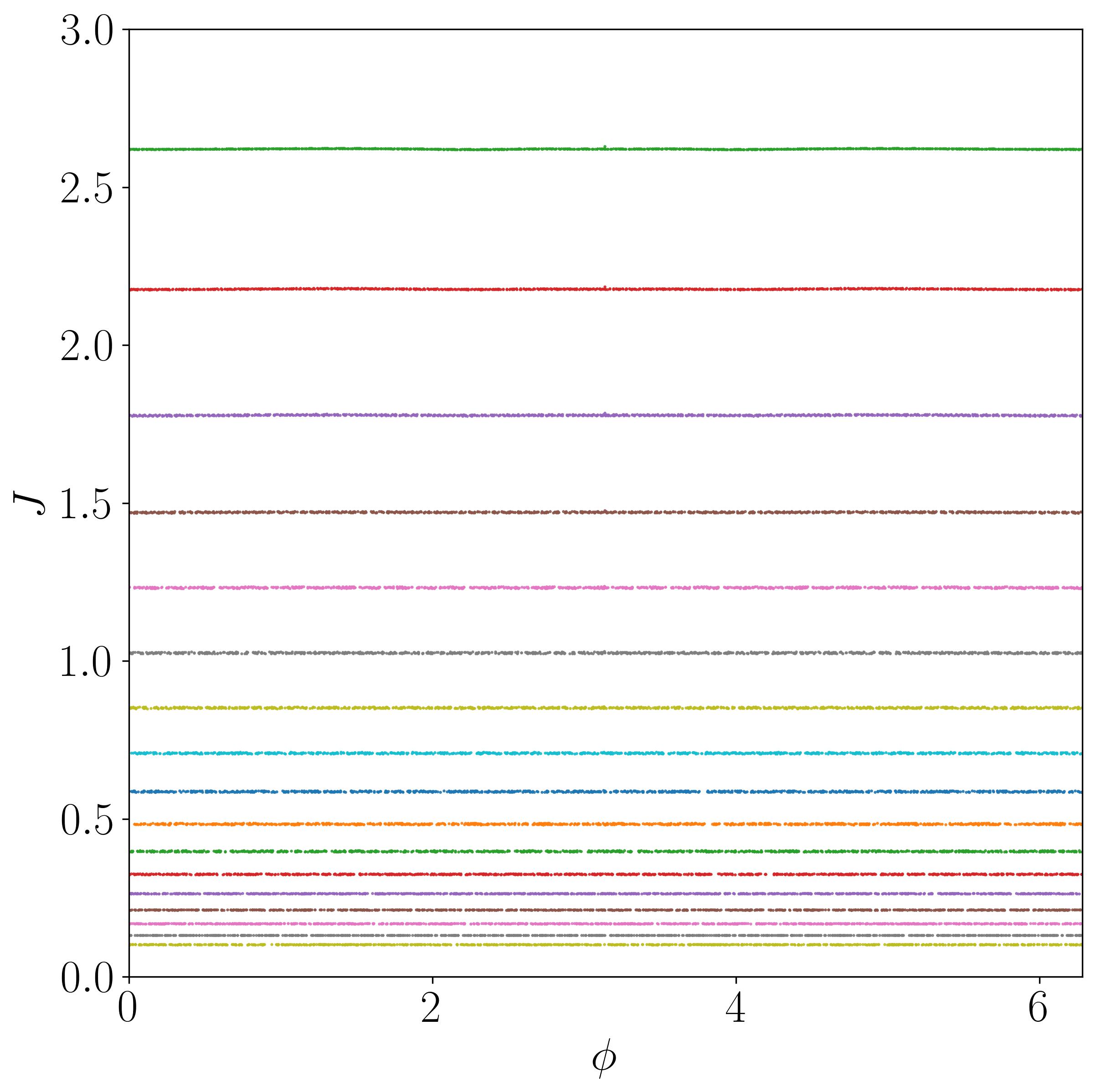}
    \end{subfigure}
    \caption{Poincare sections of the toy tokamak field when $\epsilon=0$ in both geometric and straight field line coordinates.}
    \label{fig:integrable_field}
\end{figure}

There are many definitions, or more accurately characterisations, of what defines an \textit{integrable} field. Consider Figure \ref{fig:integrable_field} which presents the Poincare map orbits of our toy tokamak field when $\delta=0$. In this case the field is axisymmetric and therefore all of the flux surfaces are nested. This is what is usually meant by an integrable field. In straight field line coordinates each orbit lies on a surface of constant $J$ and, ignoring periodic orbits, densely covers this surface. This is of course, the property the straight field line coordinates were constructed to have as action-angle variables. Further, we observe that the orbits in $(J,\phi)$ are identical circles which foliate the cylindrical space. This is our core observation and it implies that the VR persistent homology of each of the orbits in straight field line coordinates will be identical. 

\begin{figure}
    \centering
    \begin{subfigure}[b]{0.4\textwidth}
        \centering
        \includegraphics[width = \textwidth]{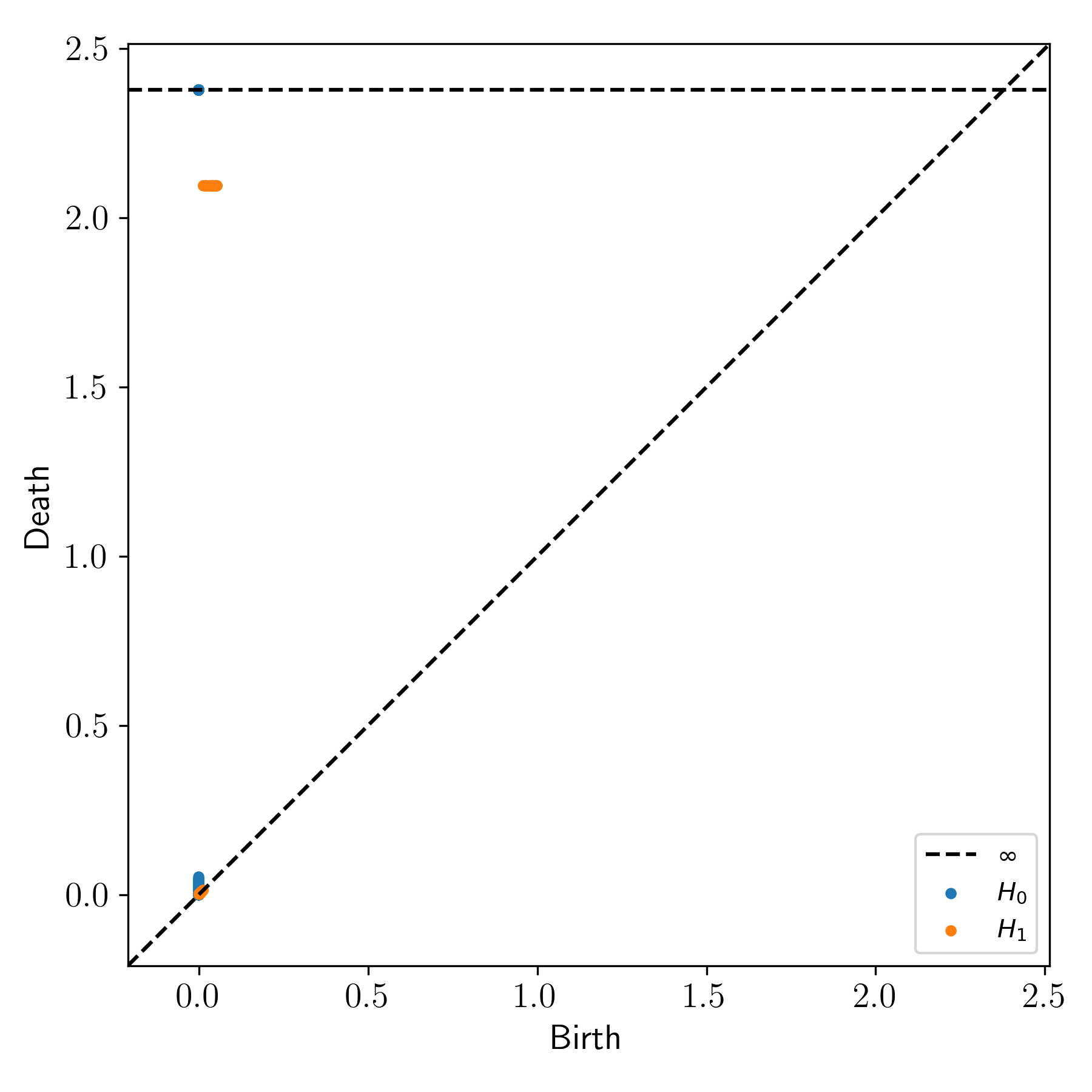}
        \caption{Integrable field PD.}
        \label{fig:integrable_field_StraightData}
    \end{subfigure}
    \begin{subfigure}[b]{0.4\textwidth}
        \centering
        \includegraphics[width = \textwidth]{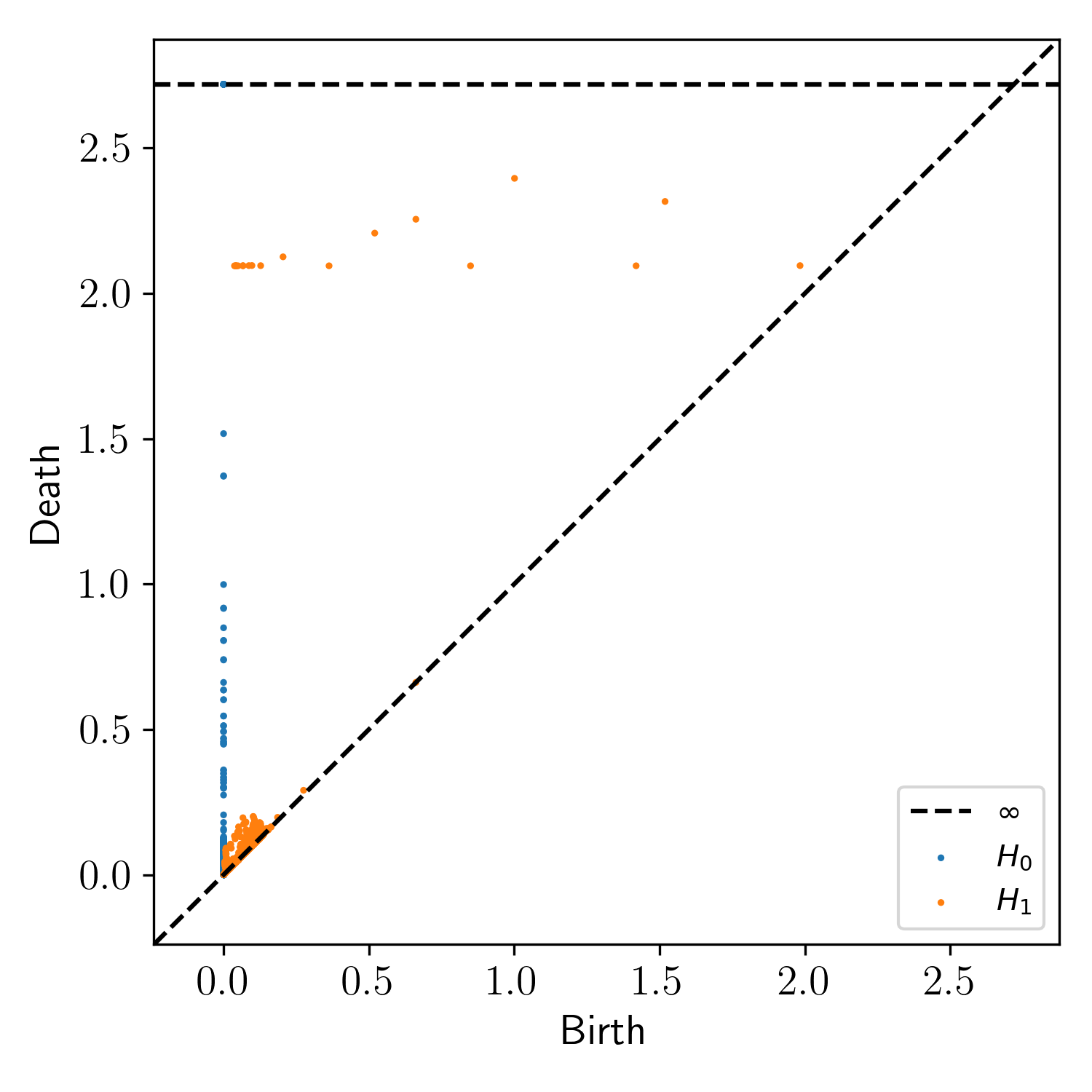}
        \caption{Non-integrable field PD.}
        \label{fig:nonintegrable_field_StraightData}
    \end{subfigure}
    \caption{Union of the persistence diagrams for each of straight field line coordinates representations of the orbits in Figures \ref{fig:integrable_field} and \ref{fig:nonintegrable_field}.}
\end{figure}

To confirm the above claim consider Figure \ref{fig:integrable_field_StraightData} which presents the union of the persistence diagrams of each of the $20$ straight field line orbits of our integrable system. There are $20$ orange dots with birth death pair $(b,d)\approx (0,2.1)$ each of which corresponds to the loop around the cylindrical phase space itself and one dot is contributed by each of the orbits. So, up to minor noise contributions we confirm that each of the orbits has identical $PH_1$ consisting of a single class. In the limit of infinite length orbits the birth and death of this class approaches $(0,\frac{2\pi}{3})$ as a consequence of the following lemma which is a special case of Corollary $7.7$ in \cite{adamaszek2017vietoris}. 

\begin{lem}
    The Vietoris-Rips $PH_1$ of a unit radius circle equipped with the circumferential distance metric is $\{(0,\frac{2\pi}{3})\}$.
\end{lem}

\begin{figure}
    \centering
    \begin{subfigure}[b]{0.45\textwidth}
        \centering
        \includegraphics[width = \textwidth]{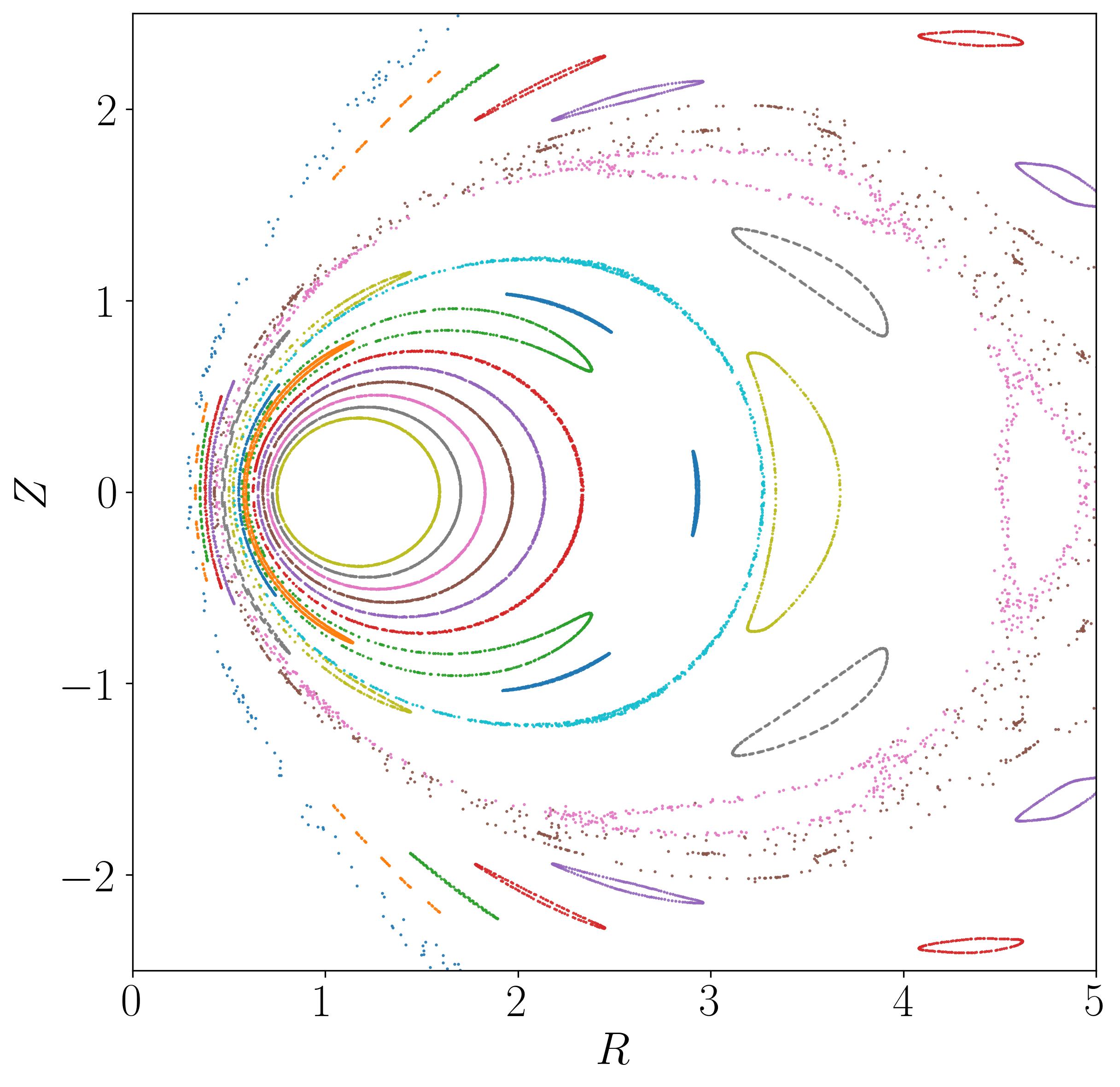}
    \end{subfigure}
    \begin{subfigure}[b]{0.45\textwidth}
        \centering
        \includegraphics[width = \textwidth]{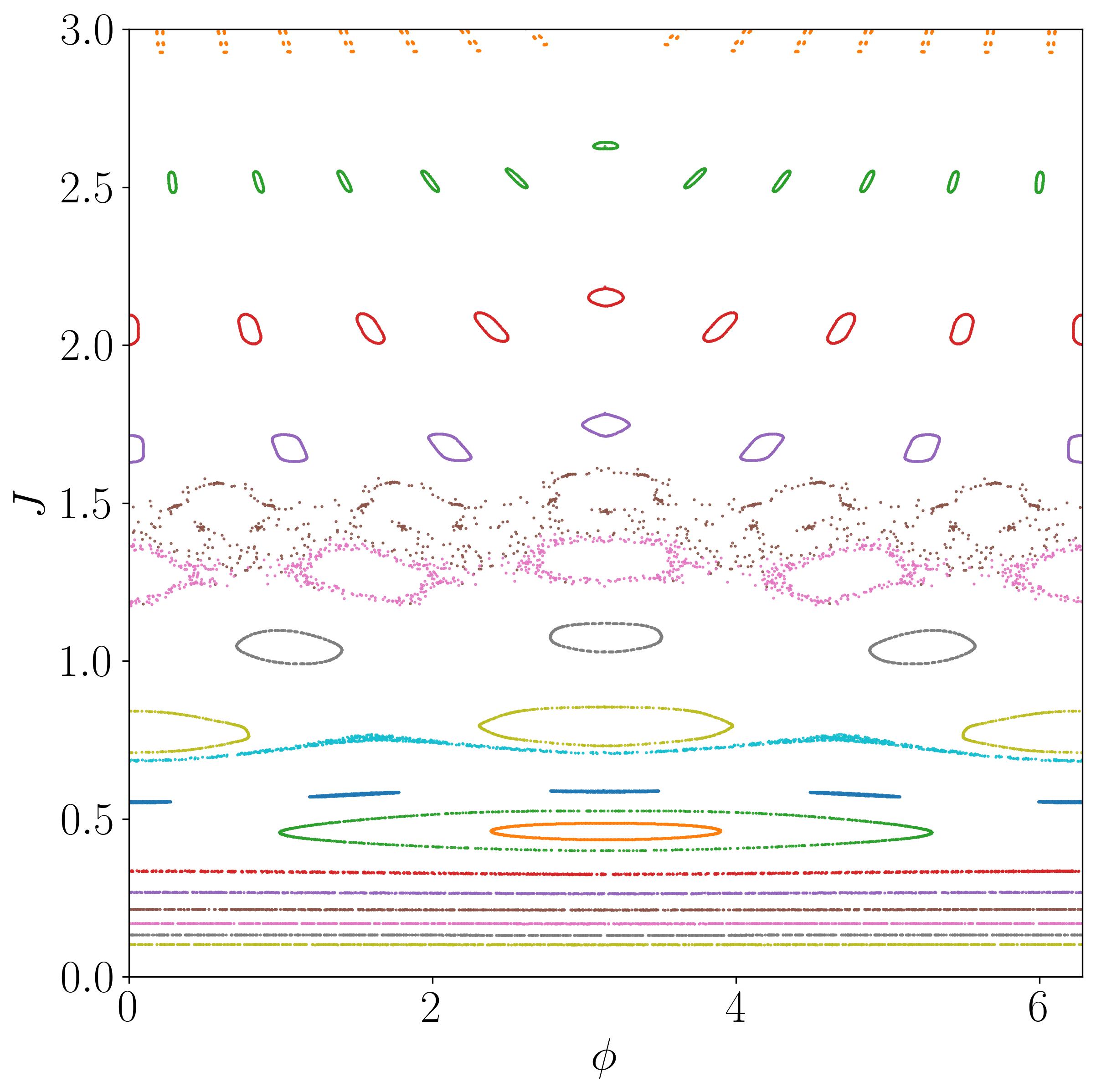}
    \end{subfigure}
    \caption{Poincare sections of the toy tokamak field when $\epsilon=0.006$ in both geometric and straight field line coordinates.}
    \label{fig:nonintegrable_field}
\end{figure}

This lemma guarantees for that any orbit which densely covers a circle of constant $J$ in the cylindrical phase space, and hence whose closure is the therefore circle, the VR $PH_1$ of the orbit will approach $\{(0,\frac{2\pi}{3})\}$ in the limit of infinite time. In an integrable field, ignoring periodic orbits, all orbits densely cover constant $J$ circles and so the $PH_1$ data will be identical for any integrable field, when computed in the straight field line coordinates associated to the field. 

In contrast we can consider the case of a non-integrable field such as Figure \ref{fig:nonintegrable_field}. For such a field many of the orbits do not densely cover constant $J$ circles and so the VR $PH_1$ of these trajectories will not be the same as the circle case. To see this we compute the VR $PH_1$ of $20$ orbits of the field and union to data as above to obtain Figure \ref{fig:nonintegrable_field_StraightData}. We see that the persistent homology is much more complicated and while there are still some $PH_1$ classes near to the point $(0,\frac{2\pi}{3})$ there are also many others which have emerged and which describe the more complicated topology of the orbit.

\begin{figure}
    \centering
        \includegraphics[width = \textwidth]{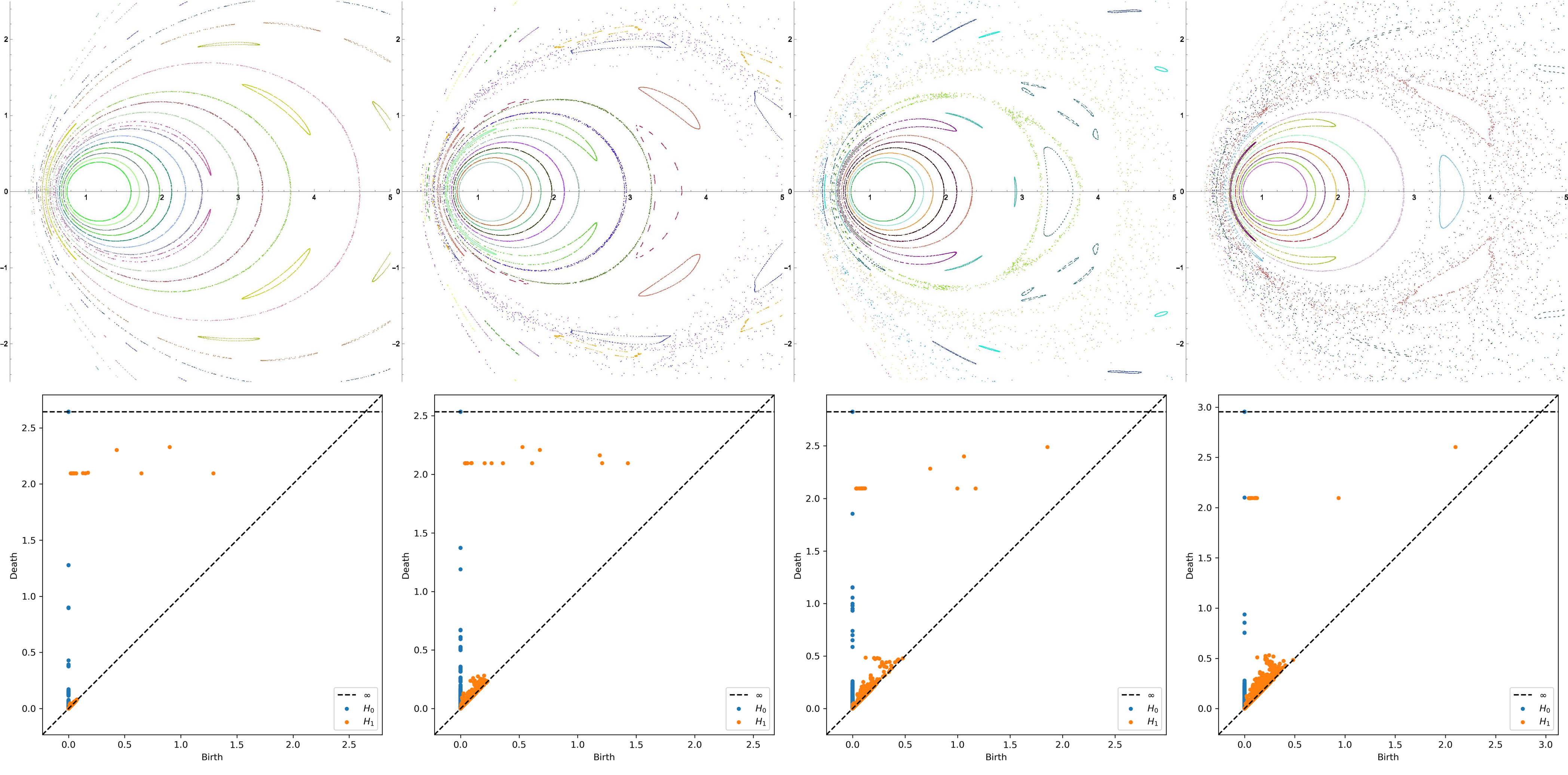}
    \caption{Combined persistence diagram of many non-integrable orbits. From left to right are shown $\delta=0.001,0.005,0.008,0.01$. Note that the axes are identical to the previously shown diagrams.}
    \label{fig:PerturbedFieldsData}
\end{figure}

We can now make the important observation that the union of persistence diagrams of our orbits changes ``continuously'' with respect to the perturbation parameter $\delta$ of the field. This occurs because the orbits of the points themselves change continuously, since the $\mathbf{B}$ field does, and persistence diagrams are continuous and stable under small changes in the underlying point cloud \cite{CohenSteiner2005StabilityOP,EDELSBRUNNER}. Figure \ref{fig:PerturbedFieldsData} presents some numerical evidence for this continuity. It shows the Poincare orbits and straight field line persistence diagrams for the toy tokamak field with increasing perturbation strength $\delta$. For larger $\delta$ the diagrams are ``further away'' from the integrable ideal. This suggests that if we can actually quantify how ``far apart'' two diagrams are then we use the distance between the union of persistence diagrams of the orbits of an integrable and a non-integrable one as a measure of how non-integrable the field is. We would expect that as the perturbation increases and the flux surfaces break-up the topology of the orbits will change and as a result the distance between the persistence diagrams will increase. However, at larger perturbations the amount of chaos will saturate and we would expect that the distance will not grow as quickly. Thus, we can quantify the sensitivity of our integrable field to a particular perturbation by looking at how quickly the topological distance between the fields grows. Note that this measure of non-integrability is \textit{global} in the sense defined above since we look at the union of persistence diagrams of many orbits rather than the local diagram of single orbits.

To make our topological measure of non-integrability rigorous we need to describe how to measure the distance between two persistence diagrams. The fact that this can be done is one of the core advantages of persistence diagrams as a method for visualising the persistent homology of a point cloud. Specifically, persistence diagrams form a metric space and there are several different choices of metric. One common metric on persistence diagrams is the Wasserstein-$q$ distance which is defined as
\begin{equation}
    W_q(PD,PD') = \left[ \underset{\eta:PD\rightarrow PD'}{\inf}\sum_{x\in PD}||x-\eta(x)||^q_q \right]^{1/q}\,,
\end{equation}
where $\eta:PD\rightarrow PD'$ is a matching function which assigns birth-death pairs in $PD$ either to birth-death pairs in $PD'$ or to the diagonal. Here, $q$ is referred to as the order or degree \cite{EDELSBRUNNER}. Also note $||\cdot||_q$ refers to the $L_q$ metric on the $(b,d)$ coordinates of the classes on a persistence diagram and that we are performing a $q$-th power summation. We will specifically utilise the case of $q=2$ below. 

We now describe a procedure for determining the topological distance between a perturbed magnetic field and its integral counterpart. Consider a sample of initial conditions $\{x_i\}\in \Sigma$ and their orbits $\{X_T(x_i)\}$, written in the straight field line coordinates of $\mathbf{B}_0$, for a particular time $T\in \N$ under the field line flow of $\mathbf{B}_\delta$. Calculate the VR $PH_1$ for each $X_T(x_i)$ and union them to form the Combined Persistence Diagram
\begin{equation}
    CPD=\bigcup_i{PH_1(X(x_i))}\,.
\end{equation}
Note that this union procedure is fairly natural because we are thinking of $\Sigma$ as foliated by the closure of the orbits. Where ``foliated'' means constructing $\Sigma$ as $\coprod_i \overline{X(x_i)}$ and the VR complex corresponding to it as $\coprod_i VR_\epsilon(X(x_i))$. The persistence diagram of which is $\bigcup_i{PH(X(x_i))}$, since simplices in the disjoint connected components of the complex never get connected to each other during the VR filtration.

\begin{figure}
    \centering
    \begin{subfigure}[b]{0.49\textwidth}
        \centering
        \includegraphics[width = \textwidth]{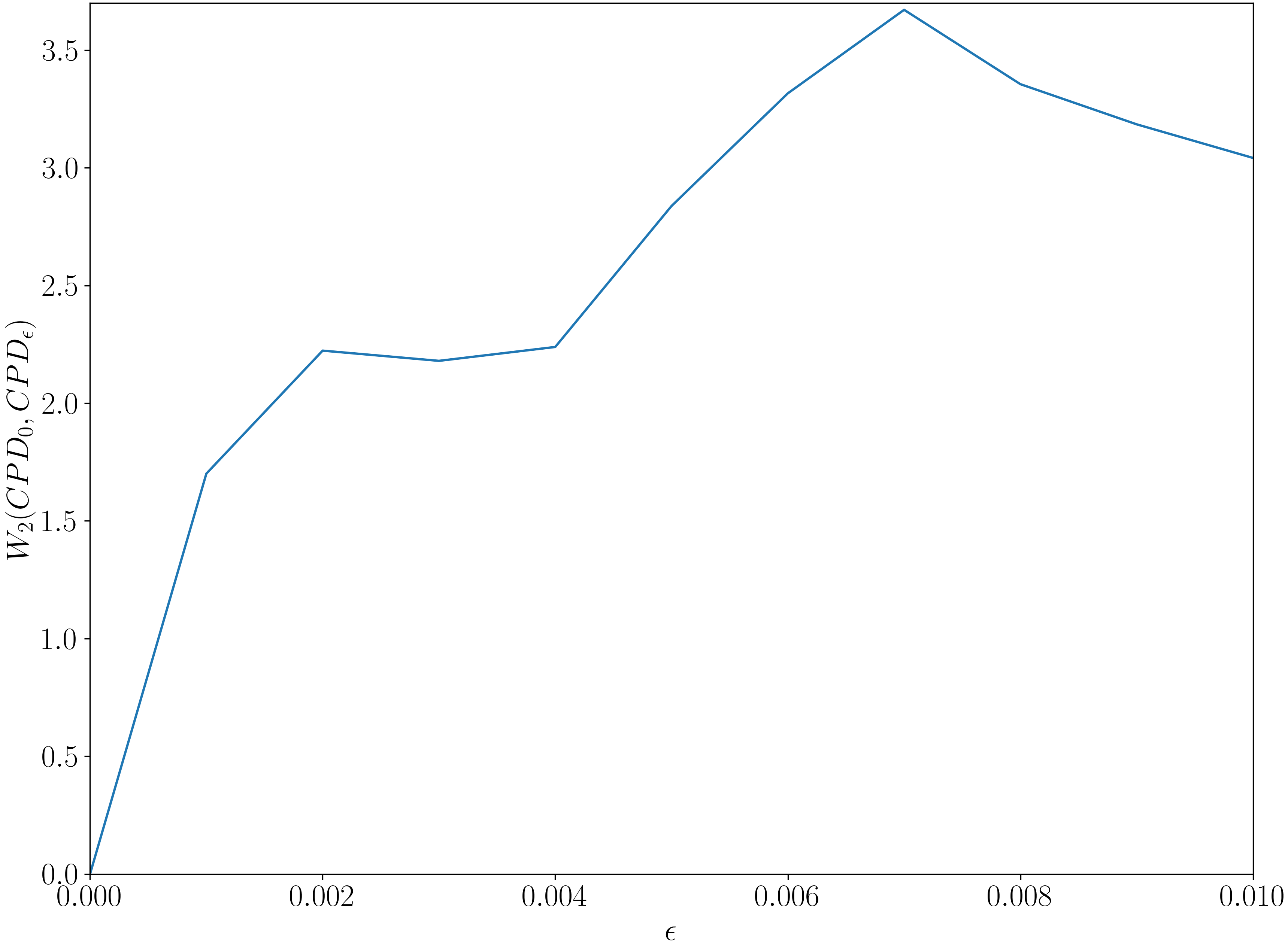}
        \subcaption{Incorrect placement case. \NB{Need bigger font and to change $\epsilon$ to $\delta$}}
        \label{subfig:Integrability_distance_plot_a}
    \end{subfigure}
    \begin{subfigure}[b]{0.49\textwidth}
        \centering
        \includegraphics[width = \textwidth]{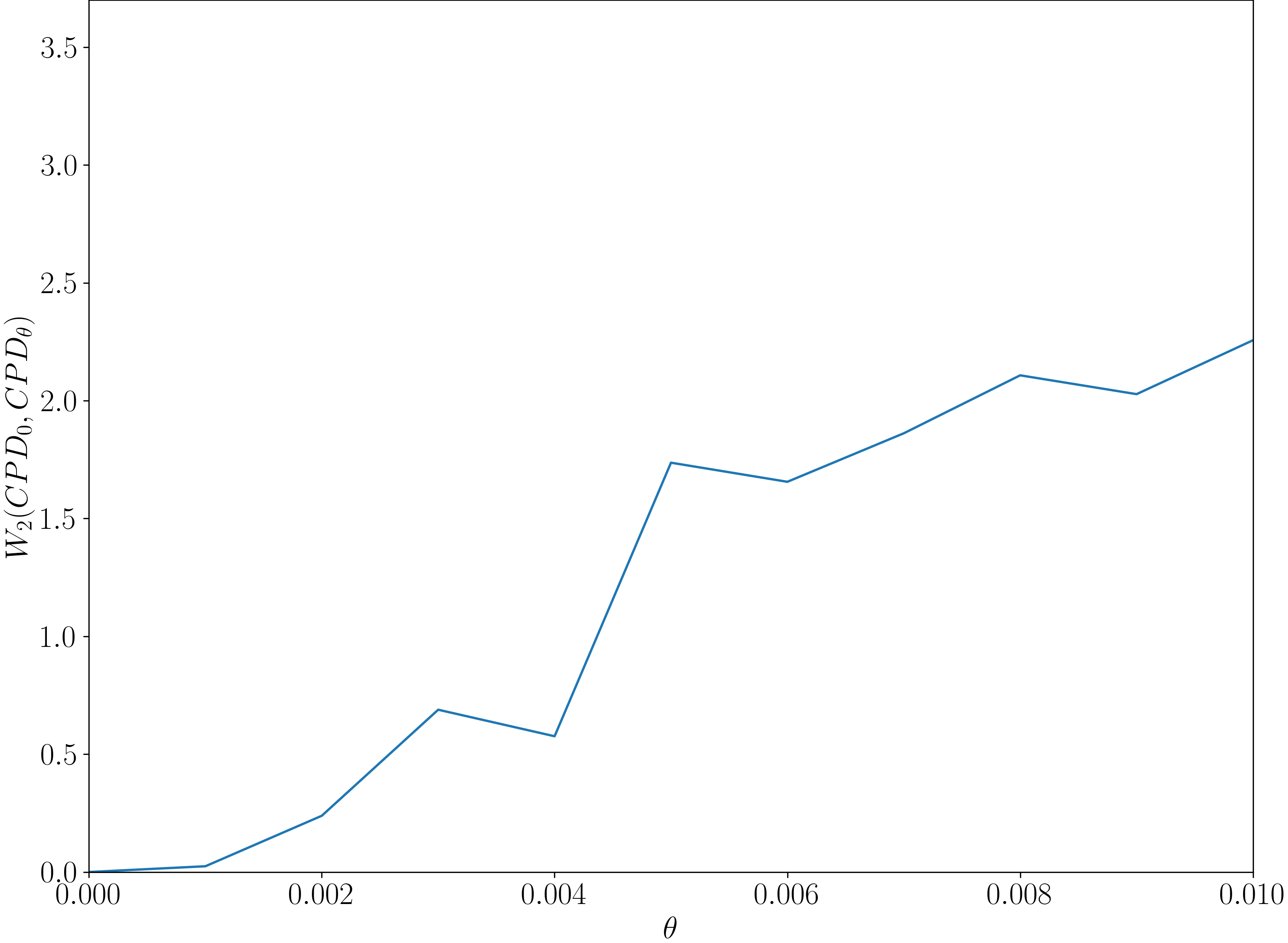}
        \subcaption{Incorrect alignment case.}
        \label{subfig:Integrability_distance_plot_b}
    \end{subfigure}
    \caption{Plot of the non-integrability measure as both $\delta$ and $\theta$ vary from $0$ to $0.01$.}
    \label{fig:Integrability_distance_plot}
\end{figure}

Fixing the initial conditions and varying the perturbation parameter $\delta$ yields a continuous one-parameter family of $CPD_\delta$. Since $CPD_0$ will always reproduce Figure \ref{fig:integrable_field_StraightData} we can define our global measure of non-integrability to be
\begin{equation}
    W_2(CPD_0,CPD_\delta)\,.
\end{equation}
As an example we computed $ W_2(CPD_0,CPD_\delta)$ for our toy tokamak field as a function of $\epsilon$ for the same sample of $20$ initial conditions which was used to generate Figure \ref{fig:PerturbedFieldsData}. The result of this calculation is presented as Figure \ref{subfig:Integrability_distance_plot_a}. We see that the integrability distance initially rises from zero very quickly as surfaces break up into island chains. This stops at $\delta\approx 0.002$ where $W_2(CPD_0,CPD_\delta)$ plateaus. This suggests that there is little change in the global topology between $0.002$ and $0.004$. Past $0.004$ the island chains begin to overlap, and the amount of chaos in the space quickly rises leading to another increase in the topological distance. This ceases at $\approx 0.007$ where the amount of chaos saturates and the magnetic field topology stabilises to a layer of nested flux surfaces around the magnetic axis surrounded by a large chaotic region. 

We now provide an example of using the non-integrability distance to perform an analysis of the relative sensitivity of the field to different perturbations. Whereas Figure \ref{subfig:Integrability_distance_plot_a} describes the case where the central line current is parallel to the $z$-axis but displaced, Figure \ref{subfig:Integrability_distance_plot_b} presents the case where the current passes through the centre of the circular coil but is angled at an angle of $\theta$ to the $z$-axis. That is, the current is misaligned rather than misplaced. We see that the distance to integrability grows faster when $\delta$ is increased than when $\theta$ is, at least in absolute terms. Also, $W_2$ varies more linearly with $\theta$ than with $\delta$, suggesting that as the misalignment angle is increased the surfaces break up slowly and in sequence rather than all breaking up simultaneously. So, we may tentatively conclude from the evidence that the integrability of our toy tokamak field is more sensitive to misplacement of the central current than its misalignment.  

Note that the above is not an exactly valid comparison as $\delta$ and $\theta$ have different scale lengths of $1$ and $\pi/2$ respectively. So by varying them both over the range $[0,0.01]$ we are not varying over the same proportion of the scale length. This is likely not a major issue those since $\pi/2$ is reasonably close to $1$.

Before concluding we would like to highlight that the discussion above is preliminary and there are many open questions about our non-integrability measure which need to be investigated further. Firstly, note that we choose a specific family of $N$ initial conditions $\{x_i\}$ which are used to compute $CPD$ for all values of the perturbation. The final Wasserstein distance clearly depends on the specific choice of initial conditions. We suspect that if these initial conditions are sampled randomly then the $CPD$ will converge, at least in a distributional sense, in the limit as $N\rightarrow \infty$ because the space of orbits will be sufficiently well-covered by the sample. However, we have not yet attempted to prove this and it is probably not trivial because it is likely sensitive to exactly what distribution the $\{x_i\}$'s are sampled from. Secondly, the rate of convergence as the length of each orbit is taken to $\infty$ has not been investigated. Thirdly, we do not know if the choice of $W_2$ for the distance function is the best choice, or even a particularly good choice. Finally, we have not yet carefully characterised how the non-integrability measure changes with the perturbation parameter of some familiar symplectic maps, such as Chirikov's standard map \cite{Morrison2000}. 


\section{Conclusion} \label{Chapter5}

Computational topology and TDA itself, were partially developed out of a need to produce new computational and mathematical tools which could provide insight into the geometry of chaotic dynamical systems. It is in this role that they are of interest in fusion science and general plasma dynamics. Specifically, as a tool for the characterisation of $3$d magnetic field geometry. In this paper we have provided some evidence that this geometry can be captured by persistent homology. 

We demonstrated numerically that by calculating VR persistent homology of the orbits of magnetic field lines under a Poincare map it is possible to automatically classify the field lines into their different orbit classes. We demonstrated the use of this classification to map the classes of field line orbits in a toy model of a tokamak. Further, we investigated limitations of this method arising from the use of non-canonical coordinates and showed that these limitations can be mitigated when straight field-line coordinates exist. Finally, we proposed a global approach to characterising the sensitivity of the topology of the magnetic surfaces to perturbations using the Wasserstein distance on PDs. 

There are possible avenues for further research into describing magnetic field geometry with TDA. 

Our classification procedures largely rely only on the $PH_1$ information of the orbits. In Section \ref{StochLayersWithIslands} we noted that we could incorporate some $PH_0$ information to automatically separate invariant tori from stochastic layers. It may be possible to incorporate more information from $PH_0$. This information is of value because the nearest neighbor distances between the points are contained in $PH_0$. However, more detailed analysis is needed to determine if this is feasible. 

Relatedly, throughout our discussion here we have not presented how to fit the thresholds of persistence at which we consider a feature to be statistically significant, except in Section \ref{StochLayersWithIslands}. This is because the standard approaches for this fitting in the TDA literature are largely ad-hoc and non-rigorous. However, in \cite{bobrowski2023universal}, Bobrowski and Skraba conjecture, and provide substantial supporting numerical evidence, that the distribution of multiplicative persistence from random point clouds is partially universal and so can be used to construct a rigorous hypothesis testing scheme which separates classes corresponding to noise from signal. That is, it provides rigorous automatic fitting of the thresholds for statistical significance. However, their universality claim is, at the time of writing, unproven and hence the hypothesis testing approach was not adopted here. Checking the universality for the noise generated from Hamiltonian orbits and applying it to separate out the signal classes from the noise in the MPDs is one possible future avenue for work.

We showed that in straight field-line coordinates the topological descriptors count the number of islands inside a chaotic layer and group them into clusters on the MPD associated to each island chain. However, this only provides insight into how many classes are present not where the features which generate them are located geometrically. That is, using the $PH_1$ information alone we can only determine how many islands exist, not what they represent. By finding explicit representatives of the homology classes it is actually possible to identify which class in the PD corresponds to which geometric feature. Using either \textit{optimal $1$-cycles} or \textit{stable volumes} this can be done is a statistically significant manner and to give tight bounding circles around islands \cite{obayashi2023stable}. This would allow for the creation of an automated phase space exploration procedure where the islands inside a candidate chaotic orbit are identified, bounded, and then further explored. We have yet to explore this concept in detail or develop software to perform the calculation as described.

Finally, note that while we considered only magnetic field lines in this paper, our methods should be generally applicable to chaotic Hamiltonian systems. It would be scientifically interesting to apply our automated classification procedure to the guiding centre Hamiltonian for particles in tokamak and stellarator fields and to specifically focus on how the topological descriptors change as the particle energy and pitch angle is varied. 

\section*{Acknowledgements}
The authors would like to thank the Australian National University's Mathematical Sciences Institute for the use of their facilities. This research was undertaken with the assistance of resources and services from the National Computational Infrastructure (NCI), which is supported by the Australian Government. 

\appendix

\section{Constructing the straight field-line coordinates by Hamilton-Jacobi theory}
\label{AppendixB}

In this appendix we both discuss some details regarding numerically constructing the field line Hamiltonian and the associated straight field-line coordinates. Specifically we will show that we can construct a Hamiltonian $\chi$ for a particular set of canonical coordinates $(\psi,\theta)$. We then explicitly compute the canonical transformation from $(\psi,\theta)$ into a set of action-angle variables $(J,\phi)$ associated to $\chi$.

We work in standard toroidal coordinates $(r,\theta,\zeta)$ with transformation equations
\begin{align}
    x&= (R+r\cos\theta)\cos\zeta\,,\nonumber\\
    y&= -(R+r\cos\theta)\sin\zeta\,,\nonumber\\
    z&= r\sin\theta\,,
\end{align}
and adopt the curvilinear calculus notation of \cite{hazeltine2003plasma} so that the Jacobian is 
\begin{equation}
    \mathcal{J} = \frac{1}{r(R+r\cos\theta)}\,.
\end{equation}

The $\mathbf{B}$ field for our toy tokamak configuration from Figure \ref{fig:ToyTokamakModel} with $\delta=0$ is then written in covariant components as 
\begin{equation}
    \mathbf{B} = B_r\nabla r + B_\theta \nabla \theta - \frac{\mu_0 I_z}{2\pi}\nabla \zeta\,,
\end{equation}
where the negative sign on $I_z$ accounts for the fact that the toroidal angle $\zeta$ is oriented oppositely to the standard cylindrical angle. $B_r$ and $B_\theta$ will not be written explicitly in what follows but they are linear combinations of elliptic functions. Define $\tilde{I}_z = \frac{\mu_0 I_z}{2\pi}$ for convenience. 

Finding the field line Hamiltonian $\chi$ and the flux $\psi$ amounts to solving the PDE
\begin{equation}
    \mathbf{B} = \nabla \psi \times \nabla \theta +  \nabla \zeta \times \nabla \chi\,, 
\end{equation}
from which \eqref{Field Line Hamiltonian} follows \cite{hazeltine2003plasma}. Here we are assuming that $\theta$ is a canonical position and attempting to find a canonical momentum $\psi$ with respect to which it is conjugate. Expanding and dotting with each to the gradient vectors yields
\begin{align}\label{HamiltonianEquations}
    &\pdv{\psi}{r} = \frac{B^\zeta}{J}\,,\nonumber\\
    &\pdv{\chi}{r} = \frac{B^\theta}{J}\,,\nonumber\\
    &\pdv{\psi}{\zeta}+\pdv{\chi}{\theta} = -\frac{B^r}{J}\,.
\end{align}
The general solution to which is
\begin{align}
    &\psi = \int^r \frac{B^\zeta}{J}dr'+f(\theta,\zeta)\,,\nonumber\\
    &\chi = \int^r \frac{B^\theta}{J}dr'+g(\theta,\zeta)\,,
\end{align}
where $f,g$ satisfy
\begin{equation}
    \pdv{f}{\zeta}+\pdv{g}{\theta} = -\int^r\left(\pdv{}{\zeta}\frac{B^\zeta}{J}+\pdv{}{\theta}\frac{B^\theta}{J}\right)dr'-\frac{B^r}{J}\,.
\end{equation}
Since we are free to select $f$ we can choose it them to be independent of $\zeta$. This is natural in an axisymmetric configuration such as ours.

We presented this argument in detail to highlight that this calculation can be partially performed analytically in our case where 
\begin{equation}
    \frac{B^\zeta}{J} =\frac{\tilde{I}_z r}{R+r\cos\theta}\,,
\end{equation}
can be integrated to yield
\begin{equation}\label{psi_equation}
    \psi = \frac{\tilde{I}_z}{\cos\theta}\left( r-\frac{R}{\cos\theta}\log(R+r\cos\theta)\right)+f(\theta)\,.
\end{equation}
We intend to use $\psi$ as a coordinate transform so prefer for it to exist everywhere. The above will not because it is generally singular where $\cos\theta=0$. We choose $f$ to cancel this singularity. Note that we have
\begin{equation}
    \lim_{\cos\theta\rightarrow 0}\psi = \lim_{\cos\theta\rightarrow 0}\frac{-\tilde{I}_zR\log R}{\cos^2\theta}+f(\theta) + \order{1}\,,
\end{equation}
so by choosing 
\begin{equation}
    f(\theta) = \frac{\tilde{I}_zR\log R}{\cos^2\theta}\,,
\end{equation}
we obtain a flux $\psi$ which is continuous globally
\begin{equation}\label{psi_cont}
     \psi(r,\theta) = \frac{\tilde{I}_z}{\cos\theta}\left( r-\frac{R}{\cos\theta}\log(\frac{R+r\cos\theta}{R})\right)\,.
\end{equation}

Note, that we are not actually interested in computing $\chi$ as a function of the fundamental toroidal coordinates $(r,\theta)$ and only of the canonical pair $(\psi,\theta)$. To do this directly we actually need $r(\psi,\theta)$, which is the inverse of $\psi(r,\theta)$ along rays of constant $\theta$. This can be done analytically and results in
\begin{equation}
    r(\psi,\theta) = -\frac{R(1+W_{\text{br}(\theta)}(e^{-\frac{\cos^2(\theta)}{\tilde{I}_z R}\psi+1}))}{\cos\theta}\,,
\end{equation}
where $W_k$ refers to the $k$-th branch of the Lambert $W$ function, and $\text{br}(\theta)$ is a branching function defined to be $0$ when $\theta \in [\pi/2,3\pi/2]$ and $-1$ otherwise.

Since $\theta$ is an angular coordinate we know that $\chi$ will be a periodic function of $\theta$ so we expand it as a Fourier series
\begin{equation}\label{fourierchi}
    \chi(\psi,\theta) = \sum_{m\in \Z} \chi_m(\psi)e^{im\theta}\,.
\end{equation}
Applying the chain rule to $\chi(\psi(r,\theta),\theta)$ yields
\begin{align*}
    \pdv{\chi}{r}\bigg)_\theta &= \pdv{\chi}{\psi}\bigg)_\theta\pdv{\psi}{r}\,,\\
    \pdv{\chi}{\theta}\bigg)_r &= \pdv{\chi}{\psi}\bigg)_\theta\pdv{\psi}{\theta}+\pdv{\chi}{\theta}\bigg)_\psi\,,
\end{align*}
which we can combine with both \eqref{fourierchi} and \eqref{HamiltonianEquations} to obtain the following integral expressions for the Fourier modes
\begin{align}
    \chi_0(\psi) &= \frac{1}{2\pi}\int^\psi d\psi'\int_0^{2\pi}d\theta\eval{\frac{B^\theta}{B^\zeta}}_{(r(\psi',\theta),\theta)}\,,\nonumber\\
    \chi_m(\psi) &=\frac{i}{2\pi m}\int_0^{2\pi}d\theta\eval{\left(\frac{B^r}{J}+\frac{B^\theta}{B^\zeta}\pdv{\psi}{\theta}\right)}_{(r(\psi,\theta),\theta)}\,.
\end{align}

It is using this integral form solution that the field line Hamiltonian associated to our integrable field was actually calculated numerically. Largely because it is very easy to implement when an analytic form for the contravariant components $B^r,B^\theta,B^\zeta$ are available, as in this case.

Recall from standard Hamiltonian mechanics that the action of the orbit with Hamiltonian value $\chi$ can be written as 
\begin{equation}
    J(\chi) = \frac{1}{2\pi}\oint \psi(\chi,\theta)\, d\theta\,,
\end{equation}
where $\psi(\chi,\theta)$ is the implicit inverse of $\chi(\psi,\theta)$ along rays of constant $\theta$. We can construct this inverse numerically, since $\chi$ is a monotonic function of $\psi$ said rays, at least for our particular case where there exists globally nested flux surfaces. We can then numerically compute each of the integrals over $\theta \in [0,2\pi]$ as that describes one full orbit. We then numerically invert $J(\chi)$ to obtain $\chi(J)$ and consequently $\psi$ as a function of $(J,\theta)$. Equipped with this we compute the angle as the derivative of the generating function
\begin{equation}
    \phi(J,\theta) =  \pdv{}{J}\int^\theta \psi\, d\theta = \int^\theta \pdv{\psi}{\chi}\dv{\chi}{J}\,d\theta\,.
\end{equation}

Therefore the full transform $(r,\theta)\rightarrow (J,\phi)$ can be calculated by the series of transforms below, where each transform only depends on the previous one and can be calculated by numerical integration or function inversion. 
\begin{center}
\begin{tikzcd}[column sep=3em, row sep = 3em,scale cd = 1]
    (r,\theta) \rightarrow (\psi,\theta) \rightarrow (\chi,
    \theta)\rightarrow (J,\theta) \rightarrow (J,\phi)
\end{tikzcd}
\end{center}

 \bibliographystyle{elsarticle-num} 
 \bibliography{cas-refs}





\end{document}